\documentclass[prb,aps,twocolumn,showpacs]{revtex4}
\usepackage{amsmath,amssymb,amsfonts,float}
\usepackage[dvips]{graphicx,color}
\usepackage{times}
\usepackage{pstricks,pst-node}
\newcommand{\bolS}{\text{\bf S}}
\newcommand{\bolJ}{\text{\bf J}}
\newcommand{\bolT}{\text{\bf T}}

\newcommand{\bolxi}{\mathbf{\xi}}
\newcommand{\bolchi}{\mathbf{\chi}}

\newcommand{\calU}{{\cal U}}
\newcommand{\VEV}[1]{\langle #1 \rangle}  % Vecuum Expectation Value

\newcommand{\be}{\text{e}}

\newsavebox{\dotdot}
\savebox{\dotdot}[3mm]{\shortstack{\circle*{0.8}\\ \\ \circle*{0.8}}}
\newcommand{\nord}{\usebox{\dotdot}}
\begin{document}
%%%%%%%%%%%%%%%%%%%%%%%%%%%%%%%%%%%%%%%%%%%%%%%%%%%%%%%
\title{Competing Orders and Hidden Duality Symmetries in 
Two-leg Spin Ladder Systems}
%%%%%%%%%%%%%%%%%%%%%%%%%%%%%%%%%%%%%%%%%%%%%%%%%%%%%%%
\author{P.~Lecheminant}
\affiliation{Laboratoire de Physique Th\'eorique et
Mod\'elisation, CNRS UMR 8089,
Universit\'e de Cergy-Pontoise, Site de Saint-Martin, 
2 avenue Adolphe Chauvin, 
95302 Cergy-Pontoise Cedex, France}
\author{K.~Totsuka}
\affiliation{Yukawa Institute for Theoretical Physics, 
Kyoto University, Kitashirakawa Oiwake-Cho, Kyoto 606-8502, Japan}
%%%%%%%%%%%%%%%%%%%%%%%%%%%%%%%%%%%%%%%%%%%%%%%%%%%%%%%%%%%%%%%%%%%%%%%%
\begin{abstract}
A unifying approach to competing quantum orders in generalized 
two-leg spin ladders is presented.    
Hidden relationship and quantum phase transitions among the competing 
orders are thoroughly discussed by means of a low-energy field theory
starting from an SU(4) quantum multicritical point.  
Our approach reveals that 
the system has a relatively simple phase structure in spite of 
its complicated interactions.  
On top of the U(1)-symmetry which is known from previous 
studies to mixes up antiferromagnetic 
order parameter with that of the $p$-type nematic, we find 
an emergent U(1)-symmetry which mixes order parameters dual to 
the above.    
On the basis of the field-theoretical- and variational analysis, 
we give a qualitative picture for the global structure of 
the phase diagram.  
Interesting connection to other models (e.g. bosonic $t$-$J$ model) 
is also discussed.  
\end{abstract}            
%%%%%%%%%%%%%%%%%%%%%%%%%%%%%%%%%%%%%%%%%%%%%%%%%%%%%%%%%%%%%%%%%%%%%%%
\pacs{75.10.Jm} 
\maketitle
%%%%%%%%%%%%%%%%%%%%%%%%%%%%%%%%%%%%%%%%%%%%%%%%%%%%%%%
% section 1 INTRODUCTION 
%%%%%%%%%%%%%%%%%%%%%%%%%%%%%%%%%%%%%%%%%%%%%%%%%%%%%%%
\section{Introduction}
\label{sec:intro}
%%%%%%%%%%%%%%%%%%%%%%%%%%%%%%%%%%%%%%%%%%%%%%%%%%%%%%%
In the past two decades, quantum magnetism has been serving not only 
as effective theories describing insulating phases of strongly-correlated 
electron systems but also as theoretical laboratories to look for 
and test new concepts.   
The discovery of high-temperature superconductors sparked 
the search for unconventional or exotic phases which are 
quite different from the ordinary ferromagnetic/antiferromagnetic 
phases.  Despite the effort of many researchers in searching for 
novelty, it is by now well-known that, in two- or higher dimensions, 
antiferromagnetic phases are found provided that spin frustration is 
not very strong \cite{misguich}.   In order to suppress antiferromagnetism and 
stabilize exotic phases\cite{Wen-book-04}, 
various mechanisms have been proposed.   
One realistic example of such mechanisms would be 
multispin-exchange interactions.  
Such interactions are expected to be
crucial for explaining unusual magnetic behavior in
${}^{3}\text{He}$ absorbed on graphite\cite{Roger-H-D-83}.
Moreover, it was reported that a certain amount of 
four-spin ring exchange would be necessary to account 
for neutron-scattering experiments for the parent compound 
of high-temperature superconductor\cite{Coldea-01} 
$\text{La}_{2}\text{CuO}_{4}$ and for 
a spin-ladder compound\cite{Brehmer-M-M-N-U-99,Matsuda-00} 
$\text{La}_{6}\text{Ca}_{8}\text{Cu}_{24}\text{O}_{41}$.  
Extensive numerical simulations carried out%
\cite{Misguich-B-L-W-98,Lauchli-D-L-S-T-05} for 
the two-dimensional Heisenberg antiferromagnets with 
four-spin ring exchange found various phases with unconventional 
orders: a spin-nematic phase \cite{Lauchli-D-L-S-T-05}  
and a spin-liquid phase with topological ordering \cite{Misguich-B-L-W-98}.

On the other hand, various approaches have been proposed 
in electron systems to unify several (and sometimes 
quite different) competing orders%
\cite{Zhang-97,Demler-H-Z-04,demler,Hermele-S-F-05} and succeeded 
in clarifying the nature of the quantum phase transitions 
among them \cite{sachdev,Senthil-V-B-S-F-04}. 
Usually, in those approaches, extended symmetries are adopted 
so that mutually competing order parameters may be transformed to 
each other.  
A typical example would be the SO(5) theory\cite{Zhang-97,Demler-H-Z-04} 
for the competition between $d$-wave superconductivity and 
antiferromagnetism, where the order parameters of $d$-wave 
superconductivity and those of antiferromagnetism are combined 
to form a unified order parameter quintet.  
For a one-dimensional geometry (two-leg ladders), it is known 
that even larger symmetries SO(8) \cite{Lin-B-F-98,saleur} and 
SO(6) \cite{schulz,boulat,controzzi} can emerge 
at low energies and be useful for the description of 
the electronic phases at half-filling and away from half-filling, 
respectively. 
In particular, the existence of an extended symmetry 
might provide a route to classify one-dimensional gapped phases
\cite{boulatpreprint}.

\begin{figure}[ht]
\begin{center}
\includegraphics[scale=0.4]{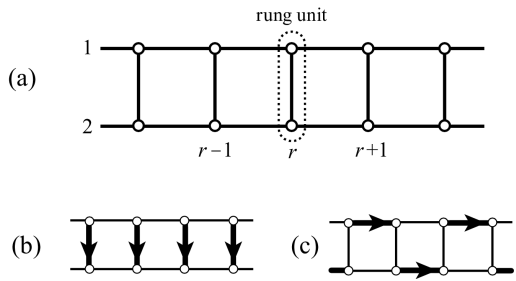}
\caption{(a): Two-leg spin ladder.  The unit of constructing 
the model Hamiltonian (see section \ref{sec:Model_and_Sym}) 
is a pair of spin-1/2s enclosed in 
a broken line (called {\em rung} in the text).  
Any site can be labeled by the chain index (1,2) and the rung 
index $r$.  (b) and (c): Two typical states appearing 
in two-leg ladders.  The usual rung-singlet phase (b) and 
the staggered-dimer phase (c) to be discussed in the text.  
Arrows denote spin-singlet pairs.
\label{fig:2leg_ladder}}
\end{center}
\end{figure}

Unfortunately, no systematic approach based on extended symmetries 
is known for unconventional phases found in spin ladders and 
it is desirable to construct such theories.  
As the first step along this line, 
we investigate here two-leg spin ladders with four-spin interactions, 
since they possess high enough symmetry to unify various competing orders.  
Without the four-spin interactions, the two-leg spin ladder 
has a finite spin gap in magnetic excitations and short-ranged 
magnetic correlations\cite{Dagotto-R-96}.  
Basic physics of the two-leg ladder can be understood by 
considering a ground state consisting of 
almost localized dimer singlets on antiferromagnetic rung bonds 
(see Fig. \ref{fig:2leg_ladder}) and low-energy excitations 
over it.   For this reason, the spin-gap phase in the usual 
two-leg spin ladder sometimes is dubbed as rung-singlet- or 
rung-dimer phase.  If we change the rung coupling to ferromagnetic, 
the rung bonds will be dominantly occupied not by singlets but by 
triplets. Then, the system is effectively equivalent 
to the spin-1 systems and the knowledges in those systems may 
apply \cite{affleckrev,Mikeska,revphle}.  
When the four-spin interactions (say, ring-exchange) are switched on, 
the size of the spin gap will be reduced and finally at a certain 
critical strength it even 
vanishes\cite{Nersesyan-T-97,Muller-V-M-02,Hijii-Q-N-03,gritsev}.  
Large-scale numerical simulations\cite{Lauchli-S-T-03,Hikihara-M-H-03} 
suggested that the model has a rather rich phase diagram.   
In particular, a spin liquid phase with scalar chirality ordering
was found \cite{Lauchli-S-T-03}. Such a phase breaks both time-reversal 
and parity symmetries and 
has been discussed previously in the context of anyon superconductivity
\cite{Wen-W-Z-89,laughlin}.  
A hallmark of the unconventional phases in the phase diagram of the
two-leg spin ladder with ring-exchange interaction 
is that neither singlets nor triplets dominate over the others.  
Hence, the conventional approaches starting from the limit of 
strong rung couplings (whether ferromagnetic or antiferromagnetic) 
is not very convenient to explore the nature of novel phases 
stabilized by four-spin interactions.  The main goal of this paper 
is to fully describe the nature of the unconventional phases 
and quantum phase transitions among them by an approach 
based on an extended symmetry.    
%%%%%%%%%%%%%%%%%%%%%%%%%%%%%%%%%%%%%%%%%%%%%%%%%%%%%%
%%%%%%%%%%%%%%%%%%%%%%%%%%%%%%%%%%%%%%%%%%%%%%%%%%%%%%

The organization of the present paper reads as follows. 
In section \ref{sec:Model_and_Sym}, we construct the 
ladder Hamiltonian by requiring rotational invariance and 
equivalence of two constituent chains.  
Our goal is to unify several unconventional phases 
stabilized by four-spin interactions.  To this end, 
we shall use an enlarged symmetry SU(4) which contains 
the ordinary (spin) SU(2).   
We shall also pay particular attention to an interesting symmetry 
({\em spin-chirality transformation}) 
which is a special case of the above SU(4) and commutes with 
the spin rotation.   
This symmetry, which has been introduced\cite{Hikihara-M-H-03,Momoi-H-N-H-03} 
in the context of the ring-exchange two-leg spin ladder,  
exchanges the N{\'e}el order parameter with the vector chirality 
(the order parameter of a $p$-type nematic \cite{Andreev-G-84}).  
It will play a crucial role in our analysis and give 
an important clue to understand the global structure 
of the phase diagram.  

A low-energy field-theory analysis will be developed 
in section \ref{sec:cont_limit}.  
Although various field-theory approaches are known for two-leg 
ladder systems\cite{Dagotto-R-96,Gogolin-N-T-book,Giamarchi-book,cabra}, 
most of them start from the limit of two decoupled chains 
and will not be suited to investigating the phase 
structure when four-spin interactions are by no means small.  
Instead of starting from two weakly-coupled 
$S=1/2$ chains, we shall take the SU(4)-invariant point,  
which is a special case of the lattice Hamiltonian, 
as the starting point.  
The rotational- and the spin-chirality symmetry are beautifully 
incorporated into our low-energy effective action.  

In section \ref{sec:RG-analysis}, 
the phase structure and unexpected high symmetry among 
unconventional phases will be discussed with the help of 
one-loop renormalization group (RG) calculation.   
In particular, we shall find four dominant phases where 
SU(4) (SO(6), more precisely) symmetry is approximately restored 
in the low-energy limit.   
The quantum phase transitions among these dominant phases 
will be investigated in sections \ref{sec:physics_selfdual} 
and \ref{sec:other_QCP}. 
The crucial role played by 
the spin-chirality symmetry in transitions 
among spin-singlet phases will be revealed.  In this respect, 
a bosonization scheme based on 
U(1)$\times$SU(3) symmetry will be introduced to extract 
relevant low-energy degrees of freedom which govern the phase transition.  
By using an effective theory for the low-energy fluctuations of  
a vector doublet composed of competing order parameters, 
we shall clarify how time-reversal symmetry is broken 
in one of these unconventional phases. 

Readers who are not interested in the detail of 
the field-theoretical analysis may skip sections 
\ref{sec:cont_limit}, \ref{sec:RG-analysis}, 
and \ref{sec:other_QCP}.  
Our main results have been already 
published in part as a short communication \cite{Lecheminant-T-05}.

In order to supplement the field-theoretical analysis, 
we shall carry out a variational- and a strong-coupling 
analysis in section \ref{sec:global-str}.  
As has been described above, the usual strong-coupling 
expansion starting from the limit of isolated rungs 
is not very convenient.  Instead, we shall start from 
the limit of isolated {\em plaquettes} to successfully 
describe the competition among several quantum phases.  
On the basis of results obtained from these analyses, 
we map out the global phase diagram and discuss
a connection to a spin-1 bosonic $t$-$J$ model.   
This relationship might be useful for 
the possible experimental realization of the
exotic gapped phases stabilized by four-spin 
exchange interactions.  
Indeed, in standard two-leg ladder compounds discovered so far, 
four-spin exchange interactions (10--20\% of leg/rung interactions) 
are not strong enough to stabilize the spin liquid phase 
with scalar chirality ordering in these systems.  
However, with the connection to spin-1 bosonic $t$-$J$ model
presented in section \ref{sec:global-str}, 
one can expect that this phase (together with other unconventional 
ones) might be feasible in ultra-cold bosonic atomic gases in optical 
lattices\cite{lewenstein}.   
Finally, section VIII presents our concluding remarks
and technical details will be presented in the Appendixes.  
%%%%%%%%%%%%%%%%%%%%%%%%%%%%%%%%%%%%%%%%%%%%%%%%%%%%%%%
\section{The Model and Its Symmetries}
%%%%%%%%%%%%%%%%%%%%%%%%%%%%%%%%%%%%%%%%%%%%%%%%%%%%%%%
\label{sec:Model_and_Sym}
%%%%%%%%%%%%%%%%%%%%%%%%%%%%%%%%%%%%%%%%%%%%%%%%%%%%%
\subsection{Building Blocks of a Hamiltonian}
%%%%%%%%%%%%%%%%%%%%%%%%%%%%%%%%%%%%%%%%%%%%%%%%%%%%%
As it has been discussed in the introduction, enlarged symmetries 
are very powerful in unifying various competing orders.  
Therefore, it is necessary first to identify 
the enlarged symmetry in our problem. 
Since we are considering spin-1/2 two-leg ladders, the maximal 
symmetry would be SU(4).  The appearance of SU(4) in our 
problem is easily understood by noting that four states 
($|\uparrow\uparrow\rangle$, $|\uparrow\downarrow\rangle$, 
$|\downarrow\uparrow\rangle$, $|\downarrow\downarrow\rangle$) 
on a single rung (say, $r$-th rung in Fig. \ref{fig:2leg_ladder}(a)) 
span the four-dimensional 
defining representation ({\bf 4}) of SU(4) 
\cite{Li-M-S-Z-98,Yamashita-S-U-98}.  
Let us denote $S=1/2$ spins on the first- (upper) and the second 
(lower) chains by $\bolS_{1,r}$ and $\bolS_{2,r}$, respectively.  
Then, one of the standard choices of the 15 generators $X_{i}$ 
on the $r$-th rung is 
\begin{equation}
\begin{split}
& X^{i}(r) = S_{1,r}^{i} \; , \; 
X^{i+3}(r) =S_{2,r}^{i}\quad (i=1,2,3)\; , \\ 
& X^{i}(r) =2\, S_{1,r}^{a}S_{2,r}^{b} \equiv G_{ab}(r) 
\; (i=7,\ldots,15; \; a,b,=1,2,3) \; .
\end{split}
\label{su4gen}
\end{equation} 
The explicit matrix expressions of these generators are given 
in Appendix \ref{sec:SU4_generators}.     

Now we are at the point of constructing the Hamiltonian by 
requiring the invariance under 
(i) SU(2), (ii) $\mathbb{Z}_{2}$ ($ 1\leftrightarrow2$: 
interchange of the two chains. See Fig. \ref{fig:2leg_ladder}) 
and (iii) reflection with respect to horizontal links ({\em link parity}).   
Since we are considering $S=1/2$, we can safely restrict ourselves 
to second-order polynomials in $S_{1}^{a}$ and $S_{2}^{b}$.  
We can divide them into (a) a scalar, (b) (SO(3)-)vectors and 
(c) rank-2 symmetric tensors.   
Of course, the scalar is given by $\bolS_{1,r}{\cdot}\bolS_{2,r}$  
and the followings are all vectors:
\begin{equation}
\begin{split}
& \bolS_{1,r}\!+\!\bolS_{2,r} \qquad \ldots \mathbb{Z}_{2}\text{-even} \\
& \bolS_{1,r}\!-\!\bolS_{2,r} \qquad \ldots \mathbb{Z}_{2}\text{-odd} \\
& \bolS_{1,r}\!\times\!\bolS_{2,r} \qquad \ldots \mathbb{Z}_{2}\text{-odd} \; .
\end{split}
\end{equation}
Although the last one seems an antisymmetric tensor, it behaves like 
a (pseudo)vector in the spin space and serves as the order parameter 
of the $p$-type spin nematic\cite{Andreev-G-84}.   
On top of them, we have a symmetric tensor whose components are 
given essentially by $G_{ab}$ \cite{comment1}:
\begin{equation}
Q_{ab}(r) \equiv S_{1,r}^{a}S_{2,r}^{b}+S_{1,r}^{b}S_{2,r}^{a} 
= \frac{1}{2}\left(G_{ab}(r)+G_{ba}(r)\right)  \; ,
\end{equation}
which reduces to a set of order parameters of 
the $n$-type spin nematic\cite{Andreev-G-84} 
when the dimer bonds are occupied by triplets.  
The point here is that all these operators are essentially 
the SU(4) generators 
(see Appendix \ref{sec:SU4_generators}.2 for more details).   

Now we proceed to constructing a Hamiltonian.  
If we consider only interactions involving two neighboring rungs 
(i.e. four spins), the SU(2)- and $\mathbb{Z}_{2}$-invariance 
strongly restrict the possible interactions and  
we are left with the following seven ones:
%%%%%%%%%%%%%%%%%%%%%%%%%
%\begin{widetext}
\begin{subequations}
%%%%%%%%%%%%%%%%%%%%%%%%%
%%%%%%%%%%%%%%%%%%%%%%%%%
\begin{align}
{\cal H}_{1} &=\sum_{r}
(\bolS_{1,r}+\bolS_{2,r}){\cdot}(\bolS_{1,r+1}+\bolS_{2,r+1}) 
\label{eqn:int_1}
\\
%%%%%%%%%%%%%%%%%%%%%%%%%%
%\underline{leg$-$diagonal-2}:
{\cal H}_{2}&=
\sum_{r} (\bolS_{1,r}-\bolS_{2,r}){\cdot}(\bolS_{1,r+1}-\bolS_{2,r+1}) 
\label{eqn:int_2}
\\
%%%%%%%%%%%%%%%%%%%%%%%%%
%\underline{4-body (a)}:
\begin{split}
{\cal H}_{3}&= 
2 \sum_{r}\sum_{a,b=1}^{3}Q_{ab}(r)Q_{ab}(r+1)  \\
&= 4 \sum_{r} \bigl[
(\bolS_{1,r}{\cdot}\bolS_{1,r+1})(\bolS_{2,r}{\cdot}\bolS_{2,r+1}) \\
& \qquad \qquad \qquad 
+(\bolS_{1,r}{\cdot}\bolS_{2,r+1})(\bolS_{2,r}{\cdot}\bolS_{1,r+1})
\bigr]
\end{split}
\label{eqn:int_3}
\\
%%%%%%%%%%%%%%%%%%%%%%%%
%\underline{4-body (b)}:
\begin{split}
{\cal H}_{4}&= \sum_{r}
4(\bolS_{1,r}\times\bolS_{2,r}){\cdot}(\bolS_{1,r+1}\times\bolS_{2,r+1})
\\
&= 4 \sum_{r} \bigl[
 (\bolS_{1,r}{\cdot}\bolS_{1,r+1})(\bolS_{2,r}{\cdot}\bolS_{2,r+1}) \\
& \qquad \qquad \qquad 
-(\bolS_{1,r}{\cdot}\bolS_{2,r+1})(\bolS_{2,r}{\cdot}\bolS_{1,r+1})
\bigr]
\end{split}
\label{eqn:int_4}
\\
%%%%%%%%%%%%%%%%%%%%%%%
%\underline{3-body (time-reversal odd)}:
\begin{split}
{\cal H}_{5} &= 
2\sum_{r} \bigl[
   (\bolS_{1,r}-\bolS_{2,r}){\cdot}(\bolS_{1,r+1}\times\bolS_{2,r+1})  
\\
& \qquad \qquad \qquad 
+ (\bolS_{1,r}\times\bolS_{2,r}){\cdot}(\bolS_{1,r+1}-\bolS_{2,r+1})
\bigr]
\end{split}
\label{eqn:int_5}
\\
%%%%%%%%%%%%%%%%%%%%%%
%\underline{rung}:
{\cal H}_{6} &= \frac{1}{2}\sum_{r} (\bolS_{1,r}{\cdot}\bolS_{2,r}+
\bolS_{1,r+1}{\cdot}\bolS_{2,r+1})
\label{eqn:int_6}
\\
%%%%%%%%%%%%%%%%%%%%%
%\underline{rung-rung 4-body}:
{\cal H}_{7} &= \sum_{r} (\bolS_{1,r}{\cdot}\bolS_{2,r})
(\bolS_{1,r+1}{\cdot}\bolS_{2,r+1}) \; ,
\label{eqn:int_7}
%%%%%%%%%%%%%%%%%%%%%%%%%%%%%%
\end{align}
\end{subequations}
%\end{widetext}
%%%%%%%%%%%%%%%%%
where the summation $\sum_{r}$ is taken over all rungs of the ladder.  
Aside from the four-spin terms ${\cal H}_{3}$, ${\cal H}_{4}$, 
and ${\cal H}_{7}$, we have a three-spin term ${\cal H}_{5}$ 
which {\em explicitly} breaks time-reversal symmetry.   
If $S=1/2$ comes from the electron spin (more generally, 
magnetic moment of charged particles), three-spin term ${\cal H}_{5}$ 
may result from the electron hopping on each plaquette. 

It would be useful to rewrite ${\cal H}_{1}$ and ${\cal H}_{3}$ 
in terms of spin-1 operator $\bolT_{r}$ defined on the $r$-th rungs: 
\begin{equation}
\begin{split}
& {\cal H}_{1}=\sum_{r}\bolT_{r}{\cdot}\bolT_{r+1} \; , \\ 
& {\cal H}_{3} = \sum_{r}\left[
\bolT_{r}{\cdot}\bolT_{r+1}
+2\left(\bolT_{r}{\cdot}\bolT_{r+1}\right)^{2} 
-4 {\cal H}_{6} - \frac{3}{2}\right]   \; .
\label{eqn:H1-H3-BBQ}
\end{split}
\end{equation}
In the above equations, the projection operators onto 
the triplet subspace 
$P_{\text{triplet}}(r)\equiv\bolS_{1,r}{\cdot}\bolS_{2,r}+3/4$ 
are implied, {\em i.e.}, 
\[
 \bolT_{r}\equiv P_{\text{triplet}}(r)(\bolS_{1,r}+\bolS_{2,r})
P_{\text{triplet}}(r) \ .
\]
These two blocks describe the interaction between effective spin-1 
objects,  that is, they dictate the {\em magnetic} part 
of the Hamiltonian.  
All the above seven interactions are used to construct the following 
general Hamiltonian: 
\begin{equation}
{\cal H} = A {\cal H}_{1}+B {\cal H}_{2}+C{\cal H}_{3}+D{\cal H}_{4}
+E{\cal H}_{5}+F{\cal H}_{6}+G{\cal H}_{7} \; .
\label{eqn:family}
\end{equation} 

Of course, another set of interactions could have been used.  
For example, basis ${\cal H}_{1}\pm{\cal H}_{2}$ and 
${\cal H}_{3}\pm{\cal H}_{4}$ could have been chosen instead of 
${\cal H}_{1}$, ${\cal H}_{2}$, ${\cal H}_{3}$, and 
${\cal H}_{4}$.   In fact, the latter choice is convenient 
when discussing the systems with SU(2)$\times$SU(2)-symmetry 
(e.g. the spin-orbital model \cite{Kugel-K-82,Pati-S-K-98,Yamashita-S-U-98}), 
while our choice here is useful when dealing 
with the spin-chirality transformation 
which will be introduced in the next subsection.    
%%%%%%%%%%%%%%%%%%%%%%%%%%%%%%%%%%%%%%%%%%%%%%%%%%%%%%%%%%%%%
\subsection{Spin-chirality Transformation}
%%%%%%%%%%%%%%%%%%%%%%%%%%%%%%%%%%%%%%%%%%%%%%%%%%%%%%%%%%%%%
%%%%%%%%%%%%%%%%%%%%%%%%%%%%%%%%%%%%%%%%%%%%%%%%%%%%%%%%%%%%%
\subsubsection{Construction}
%%%%%%%%%%%%%%%%%%%%%%%%%%%%%%%%%%%%%%%%%%%%%%%%%%%%%%%%%%%%%
As mentioned in the previous sections, the largest symmetry 
of the problem is SU(4), whose defining representation 
is spanned by the four states (singlet and triplet) on a single rung.  
As a subgroup of SU(4), there is an interesting 
U(1) symmetry\cite{Momoi-H-N-H-03} 
called the spin-chirality transformation. 

Let us consider two spins ($\bolS_{1,r}$ and $\bolS_{2,r}$) 
on the $r$-th rung and look for a local unitary transformation 
${\cal U}_{r}(\theta)$ which commutes with the SU(2)-rotation 
generated by $\bolS_{1,r}+\bolS_{2,r}$.  
The commutation relations 
\begin{equation}
 \biggl[\,  \bolS_{1,r}+\bolS_{2,r}\, , \, \sum_{i=1}^{15}x_{i} 
X_{r}^{i} \, \biggr]=\text{\bf 0}
\label{eqn:requirement}
\end{equation}
satisfied by the generator of 
${\cal U}_{r}(\theta)$ {\em uniquely} (up to a constant phase) 
determine the following form:
\begin{equation}
\begin{split}
{\cal U}_{r}(\theta) &= \exp \left[ i \theta \, 
P_{\text{triplet}}(r) \right]  \\
& \equiv \exp \left[ i \theta \left(\frac{3}{4}+
\bolS_{1,r}{\cdot}\bolS_{2,r} \right)\right]  \\
&= 
\frac{1}{4}( 1 + 3\be^{i\theta})+
(\be^{i\theta}-1)\, \bolS_{1,r}{\cdot}\bolS_{2,r} \; . 
\end{split}
\end{equation}
In the above expression, we have adopted a slightly different 
definition from the original one in Ref. \onlinecite{Momoi-H-N-H-03} 
for reasons which will become clear later.    
By construction, it is obvious that 
${\cal U}_{r}(\theta)$ is the {\em only} U(1) transformation which 
commutes with the spin-rotation symmetry.  
The U(1) transformation ${\cal U}_{r}(\theta)$ has remarkable 
properties.  
By fully utilizing the properties of $S=1/2$, we can show that 
the following equations hold:
\begin{subequations}
\begin{align}
\begin{split}
& {\cal U}_{r}(\theta)(\bolS_{1,r}\!+\!\bolS_{2,r})\,
{\cal U}_{r}^{\dagger}(\theta) = \bolS_{1,r}\!+\!\bolS_{2,r}  \\
& \qquad \qquad (\text{total-spin conservation}) \label{eqn:transf1}
\end{split}
\\
\begin{split}
& {\cal U}_{r}(\theta)(\bolS_{1,r}\!-\!\bolS_{2,r})\, 
{\cal U}_{r}^{\dagger}(\theta) \\
& \qquad = (\bolS_{1,r}\!-\!\bolS_{2,r})\cos\!\theta
-2 (\bolS_{1,r}\!\times\!\bolS_{2,r})\sin\!\theta \\
& {\cal U}_{r}(\theta)(\bolS_{1,r}\!\times\!\bolS_{2,r})\, 
{\cal U}_{r}^{\dagger}(\theta)  \\ 
& \qquad = \frac{1}{2}(\bolS_{1,r}\!-\!\bolS_{2,r})\sin\!\theta +  
(\bolS_{1,r}\!\times\!\bolS_{2,r})\cos\!\theta \; .
\end{split}
\label{eqn:transf2}
\end{align}
\end{subequations}
The first line (\ref{eqn:transf1}) is a trivial consequence 
of the requirement (\ref{eqn:requirement}).  
The local U(1)-transformation ${\cal U}_{r}(\theta)$ 
can be readily generalized to the whole lattice:
\begin{equation}
{\cal U}(\theta) \equiv 
\prod_{r=\text{rungs}} {\cal U}_{r}(\theta) 
\label{eqn:duality2}
\end{equation} 
and all the above properties are preserved for ${\cal U}(\theta)$ 
as well.  
In what follows, we shall call ${\cal U}(\theta)$ 
{\em spin-chirality transformation}, since, as can be seen   
in the above equations (\ref{eqn:transf2}), it mixes up 
the antiferromagnetic order parameters $\bolS_{1}-\bolS_{2}$ and 
the vector chirality (or the order parameter of the $p$-type spin 
nematic\cite{Andreev-G-84}) $\bolS_{1}\,\times\, \bolS_{2}$.  

For our purpose, it is helpful to view ${\cal U}(\theta)$ as 
an SU(4) transformation rather than as a non-linear transformation 
for two spin operators $\bolS_{1}$ and $\bolS_{2}$.  
The latter two lines (\ref{eqn:transf2}) suggest that two antisymmetric 
(in $1\leftrightarrow 2$) quantities $\bolS_{1}-\bolS_{2}$ and 
$2(\bolS_{1}\times\bolS_{2})$ behave as an O(2)-doublet 
under the spin-chirality transformation ${\cal U}(\theta)$:
\begin{equation}
 \begin{pmatrix} 
\tilde{\bolS}_{1}-\tilde{\bolS}_{2} \\
2\tilde{\bolS}_{1}\times\tilde{\bolS}_{2}
\end{pmatrix}
= \begin{pmatrix} \cos\! \theta & -\sin\! \theta \\
\sin\! \theta & \cos \!\theta 
\end{pmatrix}
 \begin{pmatrix} 
\bolS_{1}-\bolS_{2} \\
2\bolS_{1}\times \bolS_{2}
\end{pmatrix}  \; .
\label{eqn:O2_doublet}
\end{equation}
%%%%%%%%%%
The remaining 9 generators ($S_{1}^{a}+S_{2}^{a}$ and $G_{ab}$) 
are invariant under ${\cal U}(\theta)$.  
%%%%%%%%%%%%%%%%%%%%%%%%%%%%%%%%%%%%%%%%%%%%%%%%%%%%%%%%%%%
\subsubsection{Duality}
%%%%%%%%%%%%%%%%%%%%%%%%%%%%%%%%%%%%%%%%%%%%%%%%%%%%%%%%%%%
Now we are at the position to discuss on the effect of 
spin-chirality transformation on our building blocks.  
After some algebra, we obtain the following rules:
\begin{subequations}
\begin{align}
& {\cal H}_{1} \mapsto {\cal H}_{1}  
\label{eqn:duality_rule1} \\
& {\cal H}_{2} \mapsto 
\frac{1}{2}({\cal H}_{2}+{\cal H}_{4}) +\frac{1}{2}\cos2\theta 
\,({\cal H}_{2}-{\cal H}_{4})
-\frac{1}{2}\sin\!2\theta \,{\cal H}_{5}  \\
& {\cal H}_{3} \mapsto {\cal H}_{3}  \\
& {\cal H}_{4} \mapsto 
\frac{1}{2}({\cal H}_{2}+{\cal H}_{4}) -\frac{1}{2}\cos2\theta 
\,({\cal H}_{2}-{\cal H}_{4})
+ \frac{1}{2}\sin\!2\theta \,{\cal H}_{5}  \\
& {\cal H}_{5} \mapsto \sin\!2\theta ({\cal H}_{2}-{\cal H}_{4}) 
+ \cos \!2\theta \, {\cal H}_{5} \\
& {\cal H}_{6} \mapsto {\cal H}_{6}   \\
& {\cal H}_{7} \mapsto {\cal H}_{7}   \; .
\label{eqn:duality_rule2} 
\end{align}
\end{subequations}
%%%%%%%%%%%%%%%%%%%%%%%%%%%%%%%%%%%%%%%%%%%%%%%%%%%%%%%%%%
Now the reason why we have decomposed the Hamiltonian into 
${\cal H}_{1},\ldots,{\cal H}_{7}$ can be easily understood 
from the above equations.  
The spin-chirality transformation for $\theta=\frac{\pi}{2}$ is 
particularly simple:
\begin{equation}
{\cal H}_{2} \leftrightarrow {\cal H}_{4} \quad , \quad 
{\cal H}_{5} \mapsto - {\cal H}_{5} 
\quad \text{(all the others are invariant)} \; .
\label{eqn:duality3}
\end{equation}
Hereafter, this special case 
\begin{equation}
{\cal D} \equiv {\cal U}(\theta=\pi/2)
\end{equation}
will be called {\em duality transformation}\cite{Hikihara-M-H-03,%
Momoi-H-N-H-03}, although it is rather different from 
the standard `duality' which maps local objects onto 
non-local ones and {\em vice versa}.  
It readily follows from 
Eqs. (\ref{eqn:duality_rule1}-\ref{eqn:duality_rule2}) and 
(\ref{eqn:duality3}) that  
a model with $B=D$ and $E=0$ is invariant (or {\em self-dual}) not only for 
$\theta=\pi/2$ but also for {\em arbitrary} values of $\theta$ 
\cite{Momoi-H-N-H-03}.  
In what follows, we will see that this enhanced U(1)-symmetry 
at $B=D$ will play a crucial role. 
%%%%%%%%%%%%%%%%%%%%%%%%%%%%%%%%%%%%%%%%%%%%%%%%%%%%%%%%%%%%%%
\subsection{Special Cases}
%%%%%%%%%%%%%%%%%%%%%%%%%%%%%%%%%%%%%%%%%%%%%%%%%%%%%%%%%%%%%%
Before discussing the duality property of them, we mention 
interesting special cases of Hamiltonian (\ref{eqn:family}).  

%%%%%%%%%%%%%%%%%%%%%%%%%%%%%%%%%%%%%%%%
\underline{\em Ordinary two-leg ladder}: 
Much is known\cite{Dagotto-R-96,Gogolin-N-T-book,Giamarchi-book} about the ordinary two-leg ladder 
defined by the following Hamiltonian:
\begin{equation}
\begin{split}
 {\cal H} &= J\sum_{r}\left(
\bolS_{1,r}{\cdot}\bolS_{1,r+1}+\bolS_{2,r}{\cdot}\bolS_{2,r+1}
\right)
+ J_{\perp}\sum_{r}\bolS_{1,r}{\cdot}\bolS_{2,r}  \\
&= \frac{1}{2}J\left( {\cal H}_{1}+{\cal H}_{2}\right) 
+ J_{\perp}{\cal H}_{6} \; .
\end{split}
\end{equation}
The basic picture of the ground state is provided by putting dimer singlets 
on rung (i.e. $J_{\perp}$) bonds and low-lying excitations 
may be understood as propagating dimer triplets \cite{barnes}.  

%%%%%%%%%%%%%%%%%%%%%%%%%%%%%%%%%%%%%%%%
%\noindent%
\underline{\em Spin-orbital model}: 
The Hamiltonian of the spin-orbital model 
is defined by \cite{Kugel-K-82,Pati-S-K-98,Yamashita-S-U-98} 
\begin{equation}
\begin{split}
 {\cal H}_{\text{SO}} &= J\sum_{r}\left(
\bolS_{1,r}{\cdot}\bolS_{1,r+1}+\bolS_{2,r}{\cdot}\bolS_{2,r+1}
\right) \\
& \qquad + K \sum_{r}\left(\bolS_{1,r}{\cdot}\bolS_{1,r+1}\right)
\left(\bolS_{2,r}{\cdot}\bolS_{2,r+1}\right)
\end{split}
\label{eqn:spin-orb-ham}
\end{equation}
and is obtained by choosing
\[
 A=B=\frac{1}{2}J \, , \quad C=D=\frac{K}{8}\, , \quad E=F=G=0 \; . 
\]
It was shown by weak-coupling analysis\cite{Nersesyan-T-97} 
and later by explicitly constructing the exact ground 
state\cite{Kolezhuk-M-98} that the model displays a staggered 
dimer (SD) ordering in a certain region ($0<K<4J$) 
of the parameter space.   

If we further impose the restriction
\[
 A=B=C=D\left(=\frac{1}{2}J \right) \; , 
\]
the model reduces to the so-called 
{\em SU(4)-spin-orbital model}%
\cite{Sutherland-75,Li-M-S-Z-98,Yamashita-S-U-98}. 
In this particular case, the Hamiltonian can be written as 
\begin{equation}
\begin{split}
& {\cal H}_{\text{SU(4)}} \\
&= J \sum_{r\in \text{rung}}\left[
\left( \frac{1}{2}+2 \bolS_{1,r}{\cdot}\bolS_{1,r+1}\right) 
\left( \frac{1}{2}+2 \bolS_{2,r}{\cdot}\bolS_{2,r+1}\right) 
-\frac{1}{4} \right] \\
&= J \sum_{r\in \text{rung}}\left(
P(\bolS_{1,r},\bolS_{1,r+1})P(\bolS_{2,r},\bolS_{2,r+1}) 
-\frac{1}{4} \right) \\
&= J \sum_{a=1}^{15}X^{a}_{r}X^{a}_{r+1}  \; ,
\end{split}
\label{eqn:SU4_Heisenberg}
\end{equation}
where $P(\bolS,\bolT)$ denotes a permutation operator for 
two $S=1/2$ modules (corresponding to $\bolS$ and $\bolT$).  
This is an SU(4) generalization of the $S=1/2$ Heisenberg model 
since $P(\bolS_{1,r},\bolS_{1,r+1})P(\bolS_{2,r},\bolS_{2,r+1})$  
is nothing but the SU(4)-permutation operator if we regard 
the four states on a rung as {\bf 4}.  
The model ${\cal H}_{\text{SU(4)}}$ is integrable\cite{Sutherland-75} 
and will be used as a starting point of the following analysis. 

As shown in Ref. \onlinecite{Wang-99b}, a term $G {\cal H}_{6}$ 
can be added to ${\cal H}_{\text{SU(4)}}$ without 
spoiling integrability.  According to the Bethe ansatz 
results\cite{Wang-99b}, we have two critical values of $G$; 
when $G<G_{\text{c},1}=\pi/(2\sqrt{3})+\ln 3/2$,  
all rungs are occupied by triplets and 
the system is described by two gapless bosons 
while for $G>G_{\text{c},2}=4$ the system is in the 
so-called rung-singlet phase and all excitations are gapped.     
%%%%%%%%%%%%%%%%%%%%%%%%%%%%%%%

%\noindent%
\underline{\em Self-dual models}: 
An important class of models is defined by the following choice 
of parameters:
\[
 B=D,\, E=0, \, A,C,F,G=\text{arbitrary}  \, . 
\]
This defines a family of models which are invariant under 
the full spin-chirality rotation ${\cal U}(\theta)$.  
Hereafter, we call this family of models {\em self-dual models} and 
the manifold characterized by the above set of parameters 
{\em self-dual manifold}.    
Obviously, the self-dual models have $\text{SU(2)}_{\text{spin}}\times
\text{U(1)}_{\text{spin-chiral}}$ symmetry.  
%%%%%%%%%%%%%%%%%%%%%%%%%%%%%%%

%\noindent%
\underline{\em SU(3)$\times$U(1)-models}: 
On a special sub-manifold
\[
 A=C=\frac{1}{2}J_{1} \, , \, B=D=\frac{1}{2}J_{2}\, , \, E=0 \, , \,
F, \, G=\text{arbitrary}
\]   
of the self-dual manifold (obtained by setting $A=C$ in 
the self-dual models), 
the spin-SU(2) gets enlarged to SU(3) (see Appendix 
\ref{sec:U1_SU3}).  
The conditions $A=C$ and $B=D$ are crucial for the SU(3)-invariance.  
In fact, these SU(3) and U(1) are broken {\em simultaneously} 
($\text{SU(3)}\mapsto \text{SU(2)}$, 
$\text{U(1)}\mapsto \mathbb{Z}_{2}$) when we move away 
from the self-dual manifold ($B=D$).  

%%%%%%%%%%%%%%%%%%%%%%%%%%%
%\noindent%
\underline{\em composite-spin model}: 
The so-called composite-spin model is defined by%
\cite{Timonen-L-85,Schulz-86} 
\begin{equation}
\begin{split}
{\cal H}_{\text{composite}} & = \sum_{r}\bigl[
\left( \bolS_{1,r}+\bolS_{2,r}\right){\cdot}
\left( \bolS_{1,r+1}+\bolS_{2,r+1}\right)  \\
& \qquad -\beta  \left[\left( \bolS_{1,r}+\bolS_{2,r}\right){\cdot}
\left( \bolS_{1,r+1}+\bolS_{2,r+1}\right)  \right]^{2} \bigr]  \; ,
\end{split}
\end{equation}
which can be rewritten as 
\[
{\cal H}_{\text{composite}}=
 \left(1+\frac{1}{2}\beta \right){\cal H}_{1} 
-\frac{1}{2}\beta {\cal H}_{3} -2\beta {\cal H}_{6} \; . 
\]
This preserves the total spin $\bolS_{1,r}+\bolS_{2,r}$ on each rung 
and the problem reduces essentially to that of a collection of 
finite chain segments.  

Note that this is a special case ($B=D=0$) of the self-dual models.  
If we choose $\beta=-1$, we get $A=C=1/2$ and as a consequence 
we have SU(3) symmetry.   This is in agreement with the well-known 
fact that the $\beta=-1$ bilinear-biquadratic chain is SU(3)-invariant 
(the SU(3) Uimin-Lai-Sutherland model \cite{uimin,Sutherland-75}). 
%%%%%%%%%%%%%%%%%%%%%  

%\noindent%
\underline{\em spin-ladder with four-body cyclic exchange}: 
The four-body cyclic exchange on elementary plaquettes made up of 
two rungs $r$ and $r+1$ can be recasted as 
%\[
\begin{equation}
\begin{split}
{\cal H}_{\text{cyc}} & \equiv 
\sum_{r}\left(
P_{4}(r,r+1)+P^{-1}_{4}(r,r+1)\right)  \\
& = {\cal H}_{1}+{\cal H}_{4}+2{\cal H}_{6}
+4 {\cal H}_{7}+\text{const.} \; ,
\end{split}
\end{equation}
%\]
where $P_{4}(r,r+1)$ and $P_{4}^{-1}(r,r+1)$ respectively 
make a cyclic permutation 
and its inverse on a plaquette formed by rungs $r$ and $r+1$.  
Using this, the Hamiltonian for a two-leg ladder with four-spin exchange 
is given as\cite{Sakai-H-99,Brehmer-M-M-N-U-99}:
\begin{equation}
\begin{split}
&{\cal H}_{\text{ladder+4-spin}} \\
& = J \sum_{r}\left(\bolS_{1,r}{\cdot}\bolS_{1,r+1}
+\bolS_{2,r}{\cdot}\bolS_{2,r+1}\right) + 
K_{4}{\cal H}_{\text{cyc}} \\
& \qquad \qquad + J_{\text{R}}\sum_{r}\bolS_{1,r}{\cdot}
\bolS_{2,r} \\
&= \left( \frac{1}{2}J + K_{4}\right){\cal H}_{1}
+\frac{1}{2}J{\cal H}_{2}+K_{4}{\cal H}_{4}  \\
& \phantom{\left( \frac{1}{2}J + K_{4}\right){\cal H}_{1}}
+(J_{\text{R}}+2K_{4}){\cal H}_{6} + 4K_{4}{\cal H}_{7} \; .
\end{split}
\label{eqn:4-spin-ladder}
\end{equation}
Note that the model with 
$K_{4}=J/2$ is self-dual ($B=D$) and we can find not only 
the exact rung-singlet ground state but also the exact one-magnon state%
\cite{Muller-V-M-02} (in notations used in Ref. \onlinecite{Muller-V-M-02}, 
$J_{\text{ring}}\equiv 2K_{4}$) for certain choices of parameters.  
The phase diagram for $J_{\text{R}}=1$ has been mapped out in 
Refs. \onlinecite{Muller-V-M-02,Hijii-Q-N-03,Lauchli-S-T-03,Hikihara-M-H-03}.  

\begin{figure}[h]
\begin{center}
\includegraphics[scale=0.45]{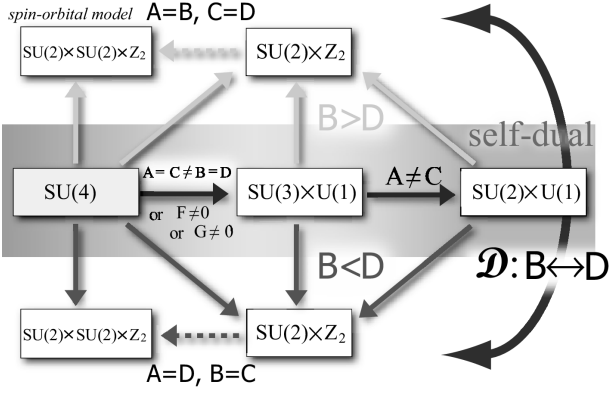}
\end{center}
\caption{Changes in the symmetry as the parameters are varied. 
Our starting point denoted by `SU(4)' corresponds to the choice 
$A=B=C=D$, $E=F=G=0$.  `Duality' ${\cal D}$ maps a model with 
$B>D$ onto another with $B<D$ and vice versa.  
Note that the full symmetry for generic cases (including the 
spin-orbital case) is $\text{SU(2)}_{\text{spin}}
\times\mathbb{Z}_{2}$ .\label{fig:sym}}
\end{figure}

We summarize the relation between parameters 
and the symmetries of Hamiltonian (\ref{eqn:family}) in Fig. \ref{fig:sym}.   
First of all, in most cases perturbation around SU(4)-point 
explicitly breaks the SU(4) symmetry.  
The fate of the system after the breaking of SU(4) is 
different according to whether or not the system is invariant under 
the spin-chirality rotation ${\cal U}(\theta)$;   
if the system is invariant (i.e. $B=D$), 
then we can have a high symmetry like 
SU(3)$\times$U(1) (when $A=C$) or SU(2)$\times$U(1) (when $A\neq C$).   
These cases will be treated in section \ref{sec:physics_selfdual}.  
Otherwise, the system generically assumes the lowest possible symmetry 
$\text{SU(2)}\times\mathbb{Z}_{2}$.     
%%%%%%%%%%%%%%%%%%%%%%%%%%%%%%%%%%%%%%%%%%%%%%%%%%%%%
%\begin{widetext}
%%%%%%%%%%%%%%%%%%%%%%%%%%%%%%%%%%%%%%%%%%%%%%%%%%%%%%
%%%%%%%%%%%%%%%%%%%%%%%%%%%%%%%%%%%%%%%%%%%%%%%%%%%%%
%\end{widetext}
%%%%%%%%%%%%%%%%%%%%%%%%%%%%%%%%%%%%%%%%%%%%%%%%%%%%%%%%%%%%%%%%%%
\subsection{Useful Analogies}
%%%%%%%%%%%%%%%%%%%%%%%%%%%%%%%%
\label{sec:useful_analogy}
%%%%%%%%%%%%%%%%%%%%%%%%%%%%%%%%%%%%%%%%%%%%%%%%%%%%%%%%%%%%%%%%%%
\subsubsection{Pseudospin-1/2 model}
%%%%%%%%%%%%%%%%%%%%%%%%%%%%%%%%%%%%%%%%%%%%%%%%%%%%%%%%%%%%%%%%%%
Although the Hamiltonian (\ref{eqn:family}) looks complicated, there is a useful 
analogy to a more familiar model--$S=1/2$ XXZ model.  
To understand this, we note that the quantity 
\begin{equation}
{\cal S}^{z}_{r} = \bolS_{1,r}{\cdot}\bolS_{2,r} + \frac{1}{4}
\end{equation}
formally plays a role of $S^{z}$ in the $S=1/2$ XXZ problem.   
Indeed, it is not difficult to show 
\begin{align}
& [\, {\cal S}^{z} \, , \, \left(\bolS_{1}-\bolS_{2}\right)^{a}\,] = 
2i \left( \bolS_{1}\times\bolS_{2}\right)^{a} \notag \\
& [\, {\cal S}^{z} \, , \, 
2 \left( \bolS_{1}\times\bolS_{2}\right)^{a}\,] = 
-i \left(\bolS_{1}-\bolS_{2}\right)^{a}  
\end{align}
and 
\begin{align}
& [\, {\cal S}^{z} \, , \, \left(\bolS_{1}+\bolS_{2} \right)^{a} \,]
=0  \notag \\ 
& [\, {\cal S}^{z} \, , \, Q_{ab} \,] = 0 \; .
\end{align}
The first two equations imply that $\bolS_{1}-\bolS_{2}$ and 
$2\left(\bolS_{1}\times\bolS_{2}\right)$ rotate like 
$S^{x}$ and $S^{y}$, respectively, as is expected 
from Eq. (\ref{eqn:O2_doublet}). 
Of course, $\left(\bolS_{1}-\bolS_{2}\right)^{a}\pm 2i 
\left(\bolS_{1}\times\bolS_{2}\right)^{a}$ 
($a=x,y,z$) give rise to singlet-triplet transitions and play a role of 
raising/lowering operators of the pseudo-spins.  

Therefore two building blocks 
\begin{align*}
{\cal H}_{2} &= \sum_{r}\sum_{a=x,y,z}\left(\bolS_{1,r}-\bolS_{2,r}
\right)_{a}
\left(\bolS_{1,r+1}-\bolS_{2,r+1}\right)_{a} \\
{\cal H}_{4} &=  \sum_{r}\sum_{a=x,y,z}
4\left(\bolS_{1,r}\times \bolS_{2,r}\right)_{a}
\left(\bolS_{1,r+1}\times \bolS_{2,r+1}\right)_{a}
\end{align*}
behave like $\sum {\cal S}^{x}_{r}{\cal S}^{x}_{r+1}$ and 
$\sum {\cal S}^{y}_{r}{\cal S}^{y}_{r+1}$ with respect to U(1) (SO(2)) 
generated by $\sum {\cal S}^{z}_{r}$ except that they have 
additional degeneracy 
coming from the SU(2)-invariance \cite{comment2}.
Of course, as far as the pseudo-spin degrees of freedom are concerned, 
the roles of $-h \sum {\cal S}^{z}$ and 
$\sum {\cal S}^{z}_{r}{\cal S}^{z}_{r+1}$ are played by the rung 
interaction ${\cal H}_{6}$ and the rung-rung 
4-body term ${\cal H}_{7}$, respectively.  
Although the Hamiltonian 
$B({\cal H}_{2}+{\cal H}_{4})+F{\cal H}_{6}+G{\cal H}_{7}$ 
may formally look like a collection of three $S=1/2$ XXZ chains
($a=x,y,z$) in a finite magnetic field, 
it is not true because $S_{1}^{a}-S_{2}^{a}$ and 
$S_{1}^{b}-S_{2}^{b}$ ($a\neq b$) do {\em not} commute with each other.  
(In fact, they obey the SU(4) commutation relations. 
See Appendix \ref{sec:SU4_generators}.)  
Nevertheless, an analogy to $S=1/2$ XXZ chain is still useful 
because the model Hamiltonian decomposes into (spin-chirality) 
pseudo-spin $S=1/2$ XXZ part and a (real) spin part.   
That is, the effective XXZ part 
(${\cal H}_{2}+{\cal H}_{4}+F{\cal H}_{6}+G{\cal H}_{7}$) has non-zero 
off-diagonal elements only for singlet-triplet transitions (pseudo-spin 
flipping), 
while the magnetic part (${\cal H}_{1}$ and ${\cal H}_{3}$) gives rise 
only to triplet-triplet transitions.  
Therefore, we may expect that, if for some reasons the spin sector 
gets gapped and the magnetic dynamics is frozen, the low-energy 
part of the full dynamics will be described by the above effective  
(pseudospin) XXZ model.   
%%%%%%%%%%%%%%%%%%%%%%%%%%%%%%%%%%%%%%%%%%%%%%%%%%%%%%%%%%%%%%%%
\subsubsection{spin-1 Bose system}
\label{sec:spin-1-bose}
%%%%%%%%%%%%%%%%%%%%%%%%%%%%%%%%%%%%%%%%%%%%%%%%%%%%%%%%%%%%%%%%
Another formulation of our system facilitates us capturing 
the physical meaning of the spin-chirality transformation 
${\cal U}(\theta)$.   
As a first step, we note that the asymmetric part 
($\bolS_{1}-\bolS_{2}$ and $\bolS_{1}{\times}\bolS_{2}$) 
of the Hamiltonian $B{\cal H}_{2}+D{\cal H}_{4}$ can be written 
as a hopping term of spin-1 bosons:
%%%%%%%%%%%%%%%%%%%%%%%%%%%%%%%%%%%%%%%%%%%%%%%%
\begin{multline}
B{\cal H}_{2}+D{\cal H}_{4} \\
= (B+D) \sum_{r,a}
\left(b_{r,a}^{\dagger}b_{r+1,a}+b_{r+1,a}^{\dagger}b_{r,a}\right) \\
+(B-D)\sum_{r,a}
\left(b_{r,a}^{\dagger}b^{\dagger}_{r+1,a}+b_{r+1,a}b_{r,a}\right) \\
\quad (a =x,y,z) \; , \label{eqn:tJ_hop}
\end{multline}
where the operator $b^{\dagger}_{r,a}$ creates a spin-1 (triplet) 
boson with spin index $a$ on the site $r$.  
Since, in the spin language, the $r$-th rung is occupied either by 
a singlet or by a triplet, the bosons $b_{a}$ should be thought of
as hardcore particles. 
Namely, the following local constraints on the boson 
occupation numbers should be imposed:
\begin{equation}
 n^{\text{B}}_{r} \equiv \sum_{a=x,y,z} b_{r,a}^{\dagger}b_{r,a}
=\bolS_{1,r}{\cdot}\bolS_{2,r}+3/4 = {\cal S}_{r}^{z}+1/2 = 0,1  \; .
\end{equation}
Note that  these particles obey non-standard commutation relations:
\[
 [\, b_{r,a}\, , \, b^{\dagger}_{r^{\prime},b}\,] = 
\left\{
\delta_{ab}\!\left(1-n^{\text{B}}_{r} \right)
-b_{r,a}^{\dagger}b_{r,b}
\right\} \delta_{r,r^{\prime}} \; .
\]   
One of the greatest merits of this mapping is that 
the above bosons are directly related to the order parameters 
$\bolS_{1}-\bolS_{2}$ (antiferromagnetic) 
and $\bolS_{1}\times\bolS_{2}$ ($p$-type nematic) as 
\begin{equation}
b^{\dagger}_{a,r} = \frac{1}{2}(\bolS_{1,r}-\bolS_{2,r})^{a}
+ i (\bolS_{1,r}\!\times\!\bolS_{2,r})^{a} \quad 
(a=x,y,z) \; .
\end{equation} 
The spin-chirality 
transformation (\ref{eqn:duality2}) is simply expressed as 
a gauge transformation of these bosons:
\begin{equation}
 b^{\dagger}_{a,r} \mapsto \be^{i\theta}\, b^{\dagger}_{a,r} \; .
\end{equation}

As has been mentioned above, 
${\cal H}_{6}$ and ${\cal H}_{7}$ have simple interpretation 
in terms of an effective spin ${\cal S}^{z}_{r}$: 
\begin{align}
F{\cal H}_{6}& = F \sum_{r}n_{r}^{\text{B}} + \text{const.}  \\
\begin{split}
G{\cal H}_{7}& = G \sum_{r}\left({\cal S}_{r}^{z}-\frac{1}{4}\right)
\left({\cal S}_{r+1}^{z}-\frac{1}{4}\right) \\
&= G\sum_{r}n^{\text{B}}_{r}n^{\text{B}}_{r+1}
-\frac{3}{2}G {\cal H}_{6}
+\text{const.}  \; .
\end{split}
\end{align}
Last, the symmetric part of the Hamiltonian 
$A {\cal H}_{1}+C {\cal H}_{3}$ can be recasted as 
(see Eq.(\ref{eqn:H1-H3-BBQ}))
\begin{equation}
\begin{split}
& A {\cal H}_{1}+ C {\cal H}_{3} \\
& = \sum_{r}\left[
(A+C)\bolT_{r}{\cdot}\bolT_{r+1}+2C \left(\bolT_{r}{\cdot}
\bolT_{r+1}\right)^{2}-4C {\cal H}_{6}
\right] + \text{const.}  \; .
\end{split}
\label{eqn:tJ_bb}
\end{equation}
In this equation, the spin-1 operators 
$(\bolT_{r})^{a}=-i \varepsilon_{abc}b^{\dagger}_{b,r}b_{c,r}$ act only on 
occupied sites.   That is, triplet projection operators 
$P_{\text{triplet}}(r)$ on both sides of $\bolT_{r}$ are implied.   
From these, we can conclude that our ladder Hamiltonian 
is equivalent to an $S=1$ bosonic $t$-$J$-like model 
{\em on a chain} \cite{comment3};
the deviation from the self-dual models $B-D\neq 0$ introduces 
pair creation/annihilation processes.  
A special case of it ($C=-A<0$), where only the biquadratic 
interaction exists and the system has an enlarged 
SU(3)-symmetry, was investigated by 
Albertini \cite{Albertini-01}.       
As has been mentioned in section \ref{sec:intro}, 
this connection might be quite useful in 
realizing the unconventional phases in the system of ultra-cold 
atomic gases\cite{lewenstein}.   
%%%%%%%%%%%%%%%%%%%%%%%%%%%%%%%%%%%%%%%%%%%%%%%%%%%%%%%%%%%%%%%
\section{Continuum Limit}
%%%%%%%%%%%%%%%%%%%%%%%%%%%%%%%%%%%%%%%%%%%%%%%%%%
\label{sec:cont_limit}
%%%%%%%%%%%%%%%%%%%%%%%%%%%%%%%%%%%%%%%%%%%%%%%%%%%%%%%%%%%%%%%
\subsection{Field-theory description of SU(4) point}
\label{sec:SU4_FP}
%%%%%%%%%%%%%%%%%%%%%%%%%%%%%%%%%%%%%%%%%%%%%%%%%%%%%%%%%%%%%%%
In this section, we develop a low-energy approach to our problem 
starting from the SU(4)-symmetric point.  
Thanks to the exact Bethe-ansatz solution \cite{Sutherland-75}, 
we know that the SU(4) point ($A=B=C=D>0,E=F=G=0$) is gapless 
and the (conformally invariant) field theory describing this 
massless fixed point was obtained by several 
authors\cite{Affleck-88,Azaria-G-L-N-99,Itoi-Q-A-00}.  
It is given by the level-1 SU(4) Wess-Zumino-Witten (WZW) 
conformal field theory with central charge $c=3$ (for 
a review of WZW- and related models, 
see for instance Ref. \onlinecite{Gogolin-N-T-book}).  
For our purpose, it is more convenient to use an equivalent 
free-field description in terms of six real (Majorana) 
fermions\cite{Azaria-G-L-N-99}.   
The derivation is sketched briefly in Appendix \ref{sec:deriv_SU4}  
and the effective action corresponding to the fixed point is 
given by\cite{Azaria-G-L-N-99}   
\begin{equation}
{\cal H}_{\text{SO(6)}} = 
- \frac{ iv}{2} \sum_{a=1}^{3}
\left( \xi_{\text{R}}^a \partial_x \xi_\text{R}^a 
- \xi_\text{L}^a \partial_x \xi_\text{L}^a
+ \chi_\text{R}^a \partial_x \chi_\text{R}^a
- \chi_\text{L}^a \partial_x \chi_\text{L}^a \right) \; .
\label{eqn:ham}
\end{equation}
The above action describes six free massless Majorana fermions 
$(\xi^{a}_{\text{L,R}},\chi^{a}_{\text{L,R}})$ $(a=1,2,3)$ 
which are equivalent to six copies of critical 2D Ising models.  
In general, SO(6)-invariant marginally irrelevant interactions 
should be added to ${\cal H}_{\text{SO(6)}}$ 
in order to describe the low-energy physics of 
the SU(4) model (see Appendix \ref{sec:deriv_SU4}).    
In what follows, shorthand notations 
$\vec{\xi}_{\text{R,L}}=(\xi^{1}_{\text{R,L}},\xi^{2}_{\text{R,L}},
\xi^{3}_{\text{R,L}})$ etc. will be used. 

Using the quantum equivalence between the level-1 SU(4) WZW model 
and ${\cal H}_{\text{SO(6)}}$ (or, the level-1 SO(6) WZW model), 
we can express all the 15 generators of SU(4) 
in terms of the above six Majorana fermions.  
In general, we may expect that local operators have the following 
expansions:
\begin{equation}
{\cal O}_{r,\text{lattice}} \sim 
{\cal O} + \be^{i \pi r/2}{\cal N}_{\cal O}  
+ \be^{-i \pi r/2}{\cal N}^{\ast}_{\cal O} 
+ (-1)^{r}n_{\cal O}  \; .
\end{equation}  
For the SU(4)-generators, they read
\begin{equation}
X^{A}_{r} \sim X_{\text{R}}^{A}+X_{\text{L}}^{A} 
+ \be^{i\pi r/2} {\cal N}^{A} 
+ \be^{-i\pi r/2} {\cal N}^{A,\ast} 
+(-1)^{r} n^{A} \; .
\label{eqn:expansion_SU4}
\end{equation}
One can write down 
$X_{\text{R,L}}^{A}$ 
in terms of six Majorana fermions by using the 6$\times$6 
representation of the SU(4) generators (see Appendix
\ref{sec:SU4_generators} for the definition of $X^{A}$
and Eq. (\ref{su4gen})): 
\begin{subequations}
\begin{equation}
\begin{split}
& \bolS_{1,\text{L/R}} = -\frac{i}{2}\vec{\xi}_{\text{L/R}}\times
\vec{\xi}_{\text{L/R}} \, , \quad 
\bolS_{2,\text{L/R}} = -\frac{i}{2}\vec{\chi}_{\text{L/R}}\times
\vec{\chi}_{\text{L/R}}  \\
& G_{ab,\text{L/R}} = -i\, \xi_{\text{L/R}}^{a}\chi_{\text{L/R}}^{b} 
\quad (a,b=1,2,3)\; .
\end{split}
\label{eqn:SU4byMajorana1}
\end{equation}
The staggered part ($n^{A}$) is given similarly as 
\begin{equation}
\begin{split} 
& \bolS_{1} = iB\, \vec{\xi}_{\text{R}}\times\vec{\xi}_{\text{L}} \, , 
\quad 
\bolS_{2} = iB\, \vec{\chi}_{\text{R}}\times\vec{\chi}_{\text{L}} 
\\
& G_{ab} = iB\, (\xi_{\text{R}}^{a}\chi_{\text{L}}^{b}
-\chi_{\text{R}}^{b}\xi_{\text{L}}^{a}) \; ,
\end{split}
\label{eqn:SU4byMajorana2}
\end{equation}
where $B$ is a regularization-dependent constant.  
\end{subequations}
Therefore, both the uniform- and the staggered correlation of 
$X^{A}$ are written as products of two free fermion propagators 
and behave like $x^{-2}$.  
The second part, whose correlation decays as $x^{-3/2}$, 
is more complicated and given by a product 
of six order/disorder operators of the underlying Ising
models\cite{Azaria-G-L-N-99}.  
%%%%%%%%%%%%%%%%%%%%%%%%%%%%%%%%%%%%%%
%\begin{widetext}
%%%%%%%%%%%%%%%%%%%%%%%%%%%%%%%%%%%%%%%%%%%%%%%%%%%%%%%%%%%%%%
\subsection{Symmetry operations}
\label{sec:sym_op}
%%%%%%%%%%%%%%%%%%%%%%%%%%%%%%%%%%%%%%%%%%%%%%%%%%%%%%%%%%%%%%
Aside from the internal $\text{SO(6)}_{\text{L}}\times
\text{SO(6)}_{\text{R}}$ symmetry, various symmetry operations 
keep the fixed-point Hamiltonian (\ref{eqn:ham}) invariant.  
Among them, the followings will play important roles: 
%%%%%%%%%%%%%%%%%%%%%%%%%% 
\begin{itemize}
\item {\bf Time-reversal:} (${\cal T}$)
\begin{equation}
{\xi}^a_{\text{R,L}} \xrightarrow{{\cal T}}
- {\xi}^a_{\text{L,R}} \quad , \quad 
{\chi}^a_{\text{R,L}} \xrightarrow{{\cal T}}
{\chi}^a_{\text{L,R}} . 
\label{eqn:time_rev}
\end{equation}
A remark is in order here about ${\cal T}$.  
As is well-known, the time-reversal operation 
is anti-unitary and the complex-conjugation 
must be taken after the transformation (\ref{eqn:time_rev}) is applied.  
\item {\bf Translation by 1-site:} ($T_{\text{1-site}}$)        
\begin{align}
& {\xi}^a_{\text R} \xrightarrow{T_{\text{1-site}}}
- {\xi}^a_{\text R} \quad , \quad 
{\xi}^a_{\text L} \xrightarrow{T_{\text{1-site}}}
{\xi}^a_{\text L}
\nonumber \\
& {\chi}^a_{\text R} \xrightarrow{T_{\text{1-site}}}
- {\chi}^a_{\text R}
\quad , \quad 
{\chi}^a_{\text L} \xrightarrow{T_{\text{1-site}}}
{\chi}^a_{\text L},
\label{eqn:1_site_tr}
\end{align} 
In the field-theory language, $T_{\text{1-site}}$ is 
nothing but chiral symmetry generated by $\gamma^{5}$.          
\item {\bf Interchange of upper- and lower chains:} (${\cal P}_{12}\in 
\mathbb{Z}_{2}$)
\begin{equation}
{\xi}^a_{\text{R,L}} \xrightarrow{{\cal P}_{12}}
- {\chi}^a_{\text{R,L}}
\quad , \quad 
{\chi}^a_{\text{R,L}} \xrightarrow{{\cal P}_{12}}
{\xi}^a_{\text{R,L}} .
\label{eqn:P12}
\end{equation}
\item {\bf Site parity:} ($P_{\text{S}}$)
\begin{equation}
{\xi}^a_{\text{R,L}}(x) \xrightarrow{P_{\text{S}}}
{\xi}^a_{\text{L,R}}(-x)
\quad , \quad 
{\chi}^a_{\text{R,L}}(x) \xrightarrow{P_{\text{S}}}
{\chi}^a_{\text{L,R}}(-x)
\label{eqn:site_parity}
\end{equation} 
\item {\bf Link parity:} ($P_{\text{L}}$)
\begin{align}
{\xi}^a_{\text R}(x) \xrightarrow{P_{\text{L}}}
{\xi}^a_{\text L}(-x) \quad , \quad 
{\xi}^a_{\text L}(x) \xrightarrow{P_{\text{L}}}
-{\xi}^a_{\text R}(-x)  \notag \\
{\chi}^a_{\text R}(x) \xrightarrow{P_{\text{L}}}
{\chi}^a_{\text L}(-x) \quad , \quad 
{\chi}^a_{\text L}(x) \xrightarrow{P_{\text{L}}}
-{\chi}^a_{\text R}(-x)
\label{eqn:link_parity}
\end{align} 
\item {\bf Ising (or Kramers-Wannier) duality:} ($s_{1}$)
\begin{align}
& \xi^{a}_{\text{R,L}} \xrightarrow{s_{1}} \xi^{a}_{\text{R,L}}
\notag \\
& \chi^{a}_{\text{R}} \xrightarrow{s_{1}} -\chi^{a}_{\text{R}}
\quad , \quad 
\chi^{a}_{\text{L}} \xrightarrow{s_{1}} \chi^{a}_{\text{L}} 
\label{eqn:Kramers-Wannier}
\end{align}
%%%%%%%%%%%%%%%%%%%%%%%%%%%%%%%%%%%%
\end{itemize}
%%%%%%%%%%%%%%%%%%%%%%%%%%%%%%%%%%%%%%%%%%%%%%%%%%%%%%%%%%%%
\subsection{Spin-Chirality Transformation in the Continuum Limit}
%%%%%%%%%%%%%%%%%%%%%%%%%%%%%%%%%%%%%%%%%%%%%%%%%%%%%%%%%%%%
In this section, we look for the duality transformation 
for the Majorana fermions.  
To derive an expression of the `duality' transformation 
valid in the low-energy limit, it is convenient to use 
a generalized version (\ref{eqn:duality2}):
\[
  \calU(\theta) = \prod_{r\in\text{rung}}
\exp \left[ i\theta \left(\frac{3}{4}+ 
\bolS_{1,r}{\cdot}\bolS_{2,r} \right)\right] \; . 
\]
In terms of spin operators, ${\cal U}(\theta)$ is realized 
in a non-linear manner as we have seen in 
Eqs. (\ref{eqn:transf1},\ref{eqn:transf2}).   
Interestingly, 
it is realized in the continuum limit as 
\begin{equation}
{\cal U}(\theta)={\cal R}_{\text{R}}(\theta){\cal R}_{\text{L}}(\theta)
\label{eqn:SC_Majorana1}
\end{equation}
by using the following {\em spin-independent} SO(2) transformation 
${\cal R}_{\text{L,R}}(\theta_{\text{L,R}})$ 
for $(\xi^{a},\chi^{a})$\cite{Lecheminant-T-05}: 
\begin{equation}
\begin{split}
\tilde{\xi}_{\text{L,R}}^{a} &= 
\xi_{\text{L,R}}^{a}\cos \frac{\theta_{\text{L,R}}}{2} 
- \chi_{\text{L,R}}^{a}\sin \frac{\theta_{\text{L,R}}}{2} \; ,  \\
\tilde{\chi}_{\text{L,R}}^{a} &= 
\xi_{\text{L,R}}^{a}\sin \frac{\theta_{\text{L,R}}}{2} 
+ \chi_{\text{L,R}}^{a}\cos \frac{\theta_{\text{L,R}}}{2} \; . 
\end{split}
\label{eqn:SC_Majorana2}
\end{equation}
Strictly speaking, different notations should be used to denote 
the original ${\cal U}(\theta)$ defined by Eq. (\ref{eqn:duality2}) 
and its continuum version 
(\ref{eqn:SC_Majorana1},\ref{eqn:SC_Majorana2}).  
However, the meaning is obvious from the context 
and we shall use the same notation for the two transformations.  
 
In fact, the fixed-point Hamiltonian (\ref{eqn:ham}) 
is invariant under an even larger {\em chiral} 
(i.e. left-right independent) version 
$\text{SO(2)}_{\text{R}}\times\text{SO(2)}_{\text{L}}$ 
of the above SO(2) and this fact gives us a hint to find another 
set of order parameters.   
For readers who want to know more about the derivation 
of (\ref{eqn:SC_Majorana1}) and (\ref{eqn:SC_Majorana2}), we 
give it in the Appendix \ref{sec:derivation_duality}.  

The following special cases are worth mentioning:
\begin{itemize}
\item $\theta=\frac{\pi}{2}$: ${\cal U}(\pi/2)\equiv {\cal D}$ 
reduces to the original spin-chirality duality 
transformation\cite{Hikihara-M-H-03}, 
where the two fermions mix with equal weights.  
From Eqs. (\ref{eqn:duality_rule1}-\ref{eqn:duality_rule2}), 
it is clear that models with $E=0$ (no ${\cal T}$-odd term 
${\cal H}_{5}$) form a closed subset of the full Hamiltonian. 
\item $\theta=\pi$: This corresponds to the exchange of 
$\bolS_{1}$ and $\bolS_{2}$ (discrete $\mathbb{Z}_{2}$-symmetry 
${\cal P}_{12}$).  
In the case of Majorana fermions, it accompanies a sign change:
\[
 \tilde{\xi}_{\text{L/R}}^{a} = -\chi_{\text{L/R}}^{a} \quad , \quad 
\tilde{\chi}_{\text{L/R}}^{a} = \xi_{\text{L/R}}^{a} \; .
\]     
\item $\theta=2\pi$:  Although this is nothing but 
an identity operation {\em in the original spin-1/2-language}, 
Majorana fermions change their signs: 
\[
 \tilde{\xi}_{\text{L/R}}^{a} = -\xi_{\text{L/R}}^{a} \quad , \quad 
\tilde{\chi}_{\text{L/R}}^{a} = -\chi_{\text{L/R}}^{a} \; .
\]
This is already anticipated from what we have for $\theta=\pi$.  
That is, the $\mathbb{Z}_{2}$-exchange between $\bolS_{1}$ and 
$\bolS_{2}$ is {\em not} realized as a simple $\mathbb{Z}_{2}$ symmetry 
in terms of Majorana fermions \cite{commentunivconv}. 
Note that this sign inversion does not affect the physical operators 
since they are always written as fermion bilinears. 
\end{itemize}

%%%%%%%%%%%%%%%%%%%%%%%%%%%%%%%%%%%%%%%%%%%%%%%%%%%%%%%%%%%%%%
\subsection{Order Parameters as Duality Doublets}
%%%%%%%%%%%%%%%%%%%%%%%%%%%%%%%%%%%%%%%%%%%%%%%%%%%%%%%%%%%%%%
In section \ref{sec:Model_and_Sym},  
we have pointed out that two quantities
\[
 \bolS_{1}-\bolS_{2} \quad \text{and} \quad 
2(\bolS_{1}\times\bolS_{2})
\]
form a doublet under the spin-chirality rotation ${\cal U}(\theta)$.  
However, as our system 
is one-dimensional, the expectation values of these vector order 
parameters are identically zero: 
$\VEV{\bolS_{1}-\bolS_{2}}=
\VEV{2(\bolS_{1}\times\bolS_{2})}=\text{\bf 0}$.    
Instead, we adopt the following two rotationally invariant 
order parameters: 
\begin{subequations}
\begin{align}
& {\cal O}_{\text{SD}}^{\text{lattice}}= 
\bolS_{1,r}{\cdot}\bolS_{1,r+1}-\bolS_{2,r}{\cdot}\bolS_{2,r+1} 
\label{eqn:Osd} 
\\
\begin{split}
& {\cal O}_{\text{SC}}^{\text{lattice}}=
\left( \bolS_{1,r}+\bolS_{2,r}\right){\cdot}
\left(\bolS_{1,r+1}\times\bolS_{2,r+1}\right) \\
& \qquad \qquad \qquad 
+ \left(\bolS_{1,r}\times\bolS_{2,r}\right){\cdot}
\left( \bolS_{1,r+1}+\bolS_{2,r+1}\right) \; .  
\end{split}
\label{eqn:Osc}
\end{align}
\end{subequations}
In numerical
studies\cite{Lauchli-S-T-03}, 
it was shown that phases which are characterized by 
non-vanishing ${\cal O}_{\text{SD}}^{\text{lattice}}$ or 
${\cal O}_{\text{SC}}^{\text{lattice}}$ do exist 
in the phase diagram of the two-leg spin ladder with a ring-exchange interaction.
The appearance of the latter describes an exotic phase since 
a non-zero value of ${\cal O}_{\text{SC}}$ implies 
a non-magnetic order with ${\cal T}$-breaking.  

From Eqs. (\ref{eqn:O2_doublet}), (\ref{eqn:Osc}), and  
\[
 {\cal O}_{\text{SD}}^{\text{lattice}} = 
\frac{1}{2}\left(\bolS_{1}+\bolS_{2}\right)_{r}{\cdot}
\left(\bolS_{1}-\bolS_{2}\right)_{r+1}+
(r\leftrightarrow r+1) \; ,
\]
it follows that these two order parameters 
transform as an SO(2) doublet:
\begin{equation}
\begin{pmatrix} 
{\cal O}_{\text{SD}}^{\text{lattice}} \\ 
{\cal O}_{\text{SC}}^{\text{lattice}} 
\end{pmatrix} \mapsto 
\begin{pmatrix} \cos \theta & -\sin\theta \\
\sin\theta & \cos\theta 
\end{pmatrix}
\begin{pmatrix} 
{\cal O}_{\text{SD}}^{\text{lattice}} \\ {\cal O}_{\text{SC}}^{\text{lattice}} 
\end{pmatrix} \; .
\end{equation}
That is, as far as the spin-chirality SO(2) is concerned, 
${\cal O}_{\text{SD}}$ and ${\cal O}_{\text{SC}}$ 
behave like $(\bolS_{1}-\bolS_{2})$ and 
$2(\bolS_{1}\times \bolS_{2})$, respectively.  
In particular, by the duality ${\cal D}$, two order parameters 
${\cal O}_{\text{SD}}^{\text{lattice}}$ and 
${\cal O}_{\text{SC}}^{\text{lattice}}$ are 
interchanged\cite{Hikihara-M-H-03}.   

Now let us find the continuum expressions for the above order
parameters. 
As has been mentioned in section \ref{sec:SU4_FP}, 
any local operator on a lattice has an expansion of the following 
form:
\begin{equation}
 {\cal O}_{\text{lattice}} \sim 
{\cal O}(x) + \be^{i\pi x/2}{\cal N}(x)
+ \be^{-i\pi x/2}{\cal N}^{\ast}(x) 
+ (-1)^{x} {\cal O}^{\pi}(x)  \; .
\label{decompcont}
\end{equation}
By taking operator-product expansions (OPE), 
we obtain the expressions of the staggered 
parts (${\cal O}^{\pi}$) of ${\cal O}^{\text{lattice}}_{\text{SD}}$ 
and ${\cal O}^{\text{lattice}}_{\text{SC}}$: 
\begin{equation}
{\cal O}_{\text{SD}}^{\pi}
=i\left( \vec{\xi}_{\text{R}}{\cdot}\vec{\xi}_{\text{L}}
-\vec{\chi}_{\text{R}}{\cdot}\vec{\chi}_{\text{L}}\right) 
\quad \text{and}\quad 
{\cal O}_{\text{SC}}^{\pi}=i\left( \vec{\xi}_{\text{R}}{\cdot}
\vec{\chi}_{\text{L}}
+\vec{\chi}_{\text{R}}{\cdot}\vec{\xi}_{\text{L}}\right) \; .
\end{equation}
By using Eq. (\ref{eqn:SC_Majorana1}), it is straightforward to verify
\[
\begin{pmatrix} 
{\cal O}_{\text{SD}}^{\pi} \\ {\cal O}_{\text{SC}}^{\pi} 
\end{pmatrix} \mapsto 
\begin{pmatrix} \cos \theta & -\sin\theta \\
\sin\theta & \cos\theta 
\end{pmatrix}
\begin{pmatrix} 
{\cal O}_{\text{SD}}^{\pi} \\ {\cal O}_{\text{SC}}^{\pi} 
\end{pmatrix} \; .
\]

As has been mentioned before, the relations $B=D$ and $E=0$ define 
self-dual models, which are invariant under the {\em continuous} 
rotation ${\cal U}(\theta)\in \text{SO(2)}$.  
From the fact that the above two order 
parameters transform as an SO(2)-doublet, it readily follows that 
\begin{equation}
 \VEV{{\cal O}_{\text{SD}}^{\text{lattice}}} =
\VEV{{\cal O}_{\text{SC}}^{\text{lattice}}}=0 
\end{equation}
for generic models on the self-dual manifold
\cite{comment4}.
%%%%%%%%%%%%%%%%%%%%%%%%%%%%%%%%%%%%%%%%%%%%%%%%%%%%%%%%%%%%%%
\subsection{Second set of order parameters}
%%%%%%%%%%%%%%%%%%%%%%%%%%%%%%%%%%%%%%%%%%%%%%%%%%%%%%%%%%%%%%
It would be interesting to look for the possibility of other 
order parameters.  Let us restrict ourselves to those which are 
(i) spin singlet (that is, spin indices are contracted), 
(ii) Lorentz-invariant (i.e. Lorentz spin=0) and (iii) have 
a scaling dimension unity.  
Apparently, we have four such operators 
$\vec{\xi}_{\text{R}}{\cdot}\vec{\xi}_{\text{L}}$, 
$\vec{\chi}_{\text{R}}{\cdot}\vec{\chi}_{\text{L}}$, 
$\vec{\xi}_{\text{R}}{\cdot}\vec{\chi}_{\text{L}}$, and 
$\vec{\chi}_{\text{R}}{\cdot}\vec{\xi}_{\text{L}}$
made up of Majorana bilinears.   
We may recombine them into two scalars and two vectors 
under ${\cal U}(\theta)$: 
\begin{align}
& \text{scalar: } 
\begin{cases}
\vec{\xi}_{\text{R}}{\cdot}\vec{\xi}_{\text{L}}
+\vec{\chi}_{\text{R}}{\cdot}\vec{\chi}_{\text{L}} & \\
 \vec{\xi}_{\text{R}}{\cdot}\vec{\chi}_{\text{L}}
-\vec{\chi}_{\text{R}}{\cdot}\vec{\xi}_{\text{L}}
\end{cases}
\\
& \text{vector: }
\begin{cases}
\vec{\xi}_{\text{R}}{\cdot}\vec{\xi}_{\text{L}}-
\vec{\chi}_{\text{R}}{\cdot}\vec{\chi}_{\text{L}}
+i(\vec{\xi}_{\text{R}}{\cdot}\vec{\chi}_{\text{L}}+
\vec{\chi}_{\text{R}}{\cdot}\vec{\xi}_{\text{L}}) & \ldots \be^{i\theta} \\
\vec{\xi}_{\text{R}}{\cdot}\vec{\xi}_{\text{L}}-
\vec{\chi}_{\text{R}}{\cdot}\vec{\chi}_{\text{L}}
-i(\vec{\xi}_{\text{R}}{\cdot}\vec{\chi}_{\text{L}}+
\vec{\chi}_{\text{R}}{\cdot}\vec{\xi}_{\text{L}}) & \ldots \be^{-i\theta} 
\end{cases}
\end{align}
While the latter two have already appeared, the former are new.  
Therefore, it is suggested that we should add 
two more order parameters, which are scalars under the spin-chirality 
rotation ${\cal U}(\theta)$, to complete our analysis.  
Below, we shall use the following set of 
four order parameters\cite{Lecheminant-T-05}:
\begin{equation}
\begin{split}
& {\cal O}_{\text{SD}}^{\pi} = 
i\left( \vec{\xi}_{\text{R}}{\cdot}\vec{\xi}_{\text{L}}
-\vec{\chi}_{\text{R}}{\cdot}\vec{\chi}_{\text{L}}\right) \\
&{\cal O}_{\text{SC}}^{\pi} =
i\left( \vec{\xi}_{\text{R}}{\cdot}\vec{\chi}_{\text{L}}
+\vec{\chi}_{\text{R}}{\cdot}\vec{\xi}_{\text{L}}\right) \\
&{\cal O}_{\text{Q}}^{\pi} =  
i\left( \vec{\xi}_{\text{R}}{\cdot}\vec{\xi}_{\text{L}}
+\vec{\chi}_{\text{R}}{\cdot}\vec{\chi}_{\text{L}}\right)  \\
&{\cal O}_{\text{RQ}}^{\pi} =
i\left( \vec{\xi}_{\text{R}}{\cdot}\vec{\chi}_{\text{L}}
-\vec{\chi}_{\text{R}}{\cdot}\vec{\xi}_{\text{L}}
\right)  \; .
\end{split}
\label{eqn:orderparams}
\end{equation}

As we already know, the spin-chirality SO(2) 
transforms the first pair 
$({\cal O}_{\text{SD}},{\cal O}_{\text{SC}})$  
as a doublet, while keeping the second 
$({\cal O}_{\text{Q}},{\cal O}_{\text{RQ}})$ invariant:
%%%%%%%%%%%%%%%%%%%%%%%%%%%%%%
\begin{equation}
\begin{split}
{\cal U}(\theta): \quad 
&
\begin{pmatrix} 
{\cal O}^{\pi}_{\text{SD}} \\ {\cal O}^{\pi}_{\text{SC}} 
\end{pmatrix}
\mapsto 
\begin{pmatrix} 
\cos\theta & -\sin \theta \\ 
\sin\theta & \cos \theta 
\end{pmatrix} 
\begin{pmatrix} 
{\cal O}^{\pi}_{\text{SD}} \\ {\cal O}^{\pi}_{\text{SC}} 
\end{pmatrix}
\\
&
\begin{pmatrix} 
{\cal O}^{\pi}_{\text{Q}} \\ {\cal O}^{\pi}_{\text{RQ}} 
\end{pmatrix}
\mapsto 
\begin{pmatrix} 
{\cal O}^{\pi}_{\text{Q}} \\ {\cal O}^{\pi}_{\text{RQ}} 
\end{pmatrix} \; .
\end{split}
\end{equation}

A similar property holds for the second pair as well. 
To see this, let us introduce the following product\cite{Lecheminant-T-05}:
\begin{equation}
 \widetilde{\cal U}(\tilde{\theta}) \equiv 
s_{1}\, \calU(\tilde{\theta}) \, s_{1} 
= {\cal R}_{\text{R}}(\tilde{\theta})
{\cal R}_{\text{L}}(-\tilde{\theta})\; , 
\end{equation}
which is chiral (i.e. left-right asymmetric) 
and probably {\em non-local} in terms of the original lattice spins.  
Then, it is straightforward to show  
\begin{equation}
\begin{split}
\widetilde{\cal U}(\tilde{\theta}): \quad 
& 
\begin{pmatrix} 
{\cal O}^{\pi}_{\text{SD}} \\ {\cal O}^{\pi}_{\text{SC}} 
\end{pmatrix}
\mapsto 
\begin{pmatrix} 
{\cal O}^{\pi}_{\text{SD}} \\ {\cal O}^{\pi}_{\text{SC}} 
\end{pmatrix}
\\
&
\begin{pmatrix} 
{\cal O}^{\pi}_{\text{Q}} \\ {\cal O}^{\pi}_{\text{RQ}} 
\end{pmatrix}
\mapsto 
\begin{pmatrix} 
\cos\tilde{\theta} & -\sin\tilde{\theta} \\ 
\sin\tilde{\theta} & \cos\tilde{\theta} 
\end{pmatrix} 
\begin{pmatrix} 
{\cal O}^{\pi}_{\text{Q}} \\ {\cal O}^{\pi}_{\text{RQ}} 
\end{pmatrix} \; .
\end{split}
\end{equation}
%%%%%%%%%%%%%%%%%%%%%%%%%%%%%%
As in the case of ${\cal U}(\theta)$, we will use the notation 
$\widetilde{\cal D}\equiv\widetilde{\cal U}(\tilde{\theta}=\pi/2)$ 
to denote the second duality.  
The existence of the above two dualities (one is non-chiral and the other 
is chiral) will be useful in understanding the low-energy 
physics and the global phase structure.  
The transformation properties of these order parameters 
under the symmetry operations in section \ref{sec:sym_op} 
are summarized in Table \ref{OP_Sym}.  

However, this is not the end of the story.  
For the phases denoted by Q and RQ, it will turn out that 
the $q=\pi$ (i.e. period-two) components are not sufficient 
for the {\em full}  characterization of the phases.   
This point will be discussed 
in section \ref{sec:4-dominant-phases} in conjunction 
with the ground-state degeneracy.    

%Now that we have found four competing order parameters in our problem, 
%we try to understand what the order parameters 
%${\cal O}^{\pi}_{\text{Q}}$ and ${\cal O}^{\pi}_{\text{RQ}}$ mean on 
%the lattice.  
%%%%%%%%%%%%%%%%%%%%%%%%%%%%%%%%%%%%%%%
%The second one is related to a more conventional phase--%
%the {\em (staggered) rung dimer} phase:   
%\begin{equation}
%{\cal O}_{\text{RQ}}^{\text{lattice}} = \bolS_{1,r}{\cdot}\bolS_{2,r} 
%\; .
%\end{equation}
%If the uniform component of the above order parameter takes a finite 
%value, then the ground state is in the usual rung dimer phase.  
%What is relevant for our problem is the $q=\pi$ component of 
%${\cal O}_{\text{RQ}}^{\text{lattice}}$ and this is why we call 
%a phase characterized by 
%\begin{equation}
%{\cal O}_{\text{RQ}}^{\pi} =  
%i\left( \vec{\xi}_{\text{R}}{\cdot}\vec{\chi}_{\text{L}}
%- \vec{\chi}_{\text{R}}{\cdot}\vec{\xi}_{\text{L}}\right)  
%\end{equation}
%the `staggered' rung dimer phase.   
%(Note that the above expression can be readily obtained from 
%the $4k_{\text{F}}$-expression of $G_{ab}$
%given by Eq. (\ref{eqn:SU4byMajorana2})) 

%%%%%%%%%%%%%%%%%%%%% TABLE %%%%%%%%%%%%%%%%%%%%%%%%%%%%
\begin{table}[h]
\begin{center}
\begin{ruledtabular}
\begin{tabular}{c|cccccc|cc} 
& \multicolumn{6}{c}{symmetry} & 
\multicolumn{2}{c}{duality} 
%\multicolumn{2}{c|}{scalar chiral} 
\\
\cline{2-9}
\raisebox{1.5ex}[0pt]{order param.} & 
\makebox{${\cal T}$} & \makebox{$T_{\text{1-site}}$} & 
\makebox{${\cal P}_{12}$} & 
\makebox{$P_{\text{S}}$} & 
\makebox{$P_{\text{L}}$} &
\makebox{$s_{1}$} &
\makebox{${\cal D}$} &
\makebox{$\widetilde{\cal D}$} \\
\hline
${\cal O}_{\text{SD}}^{\pi}$ & $+1$ & $-1$ & $-1$ & $-1$ & $+1$ 
& ${\cal O}_{\text{Q}}^{\pi}$ 
& ${\cal O}_{\text{SC}}^{\pi}$ 
& inv. \\
\hline 
${\cal O}_{\text{SC}}^{\pi}$ & $-1$ & $-1$ & $-1$ & $-1$ & $+1$ 
& ${\cal O}_{\text{RQ}}^{\pi}$ 
& ${\cal O}_{\text{SD}}^{\pi}$ 
& inv. \\
\hline 
${\cal O}_{\text{Q}}^{\pi}$  & $+1$ & $-1$ & $+1$ & $-1$ & $+1$ 
& ${\cal O}_{\text{SD}}^{\pi}$ 
& inv. 
& ${\cal O}_{\text{RQ}}^{\pi}$ \\
\hline 
${\cal O}_{\text{RQ}}^{\pi}$ & $+1$ & $-1$ & $+1$ & $+1$ & $-1$ 
& ${\cal O}_{\text{SC}}^{\pi}$ 
& inv. 
& ${\cal O}_{\text{Q}}^{\pi}$ \\
\end{tabular}
\end{ruledtabular}
\caption{Four order parameters and discrete symmetries. 
Note that under $s_{1}$, ${\cal D}$, and $\widetilde{\cal D}$ 
the four order parameters transform onto each other.\label{OP_Sym}}
\end{center}
\end{table}
%%%%%%%%%%%%%%%%%%%%%%%%%%%%%%%%%%%%%%%%%%%%%%%%%%%%%%%%
%\end{widetext}
%%%%%%%%%%%%%%%%%%%%%%%%%%%%%%%%%%%%%%%%%%%%%%%%%%%%%%%%
%%%%%%%%%%%%%%%%%%%%%%%%%%%%%%%%%%%%%%%%%%%%%%%%%%%%%%%%%%%%%%
\subsection{Interactions in the continuum limit}
%%%%%%%%%%%%%%%%%%%%%%%%%%%%%%%%%%%%%%%%%%%%%%%%%%%%%%%%%%%%%%
Now that we have identified the order parameters, we proceed to 
constructing interactions in the low-energy effective action.  
Basically, we have three different contributions to the interactions. 
The first one comes from the uniform ($q=0$) terms 
in the continuum expressions of the spin operators 
(see Eq. (\ref{eqn:expansion_SU4})) while the other two from 
the contraction of the $\pm 2k_{\text{F}}$-terms 
or of two $4k_{\text{F}}$ terms ($k_{\text{F}}=\pi/4$).  

%%%%%%%%%%%%%%%%%%%%%%%%%%%%%%%%%%%%%%%%%%%%%%%%%%%%%%%%%%%%%%
%\subsubsection{uniform part}
%%%%%%%%%%%%%%%%%%%%%%%%%%%%%%%%%%%%%%%%%%%%%%%%%%%%%%%%%%%%%%
By using the Majorana expressions
(\ref{eqn:SU4byMajorana1},\ref{eqn:SU4byMajorana2}) 
for the spin operators, 
we can rewrite the non-oscillatory part of 
the seven interactions (\ref{eqn:int_1}-\ref{eqn:int_7}): 
\begin{subequations}
%%%%%%%%%%%%%
\begin{align}
%\begin{split}
{\cal V}^{0}_{1} &=
 (\vec{\xi}_{\text{R}}{\cdot}\vec{\xi}_{\text{L}})^{2}+
(\vec{\chi}_{\text{R}}{\cdot}\vec{\chi}_{\text{L}})^{2}+
(\vec{\xi}_{\text{R}}{\cdot}\vec{\chi}_{\text{L}})^{2}+
(\vec{\chi}_{\text{R}}{\cdot}\vec{\xi}_{\text{L}})^{2} 
 \\
%%%%%%%%%%%%%%%%%
{\cal V}^{0}_{2} &=
 (\vec{\xi}_{\text{R}}{\cdot}\vec{\xi}_{\text{L}})^{2}+
(\vec{\chi}_{\text{R}}{\cdot}\vec{\chi}_{\text{L}})^{2}-
(\vec{\xi}_{\text{R}}{\cdot}\vec{\chi}_{\text{L}})^{2}-
(\vec{\chi}_{\text{R}}{\cdot}\vec{\xi}_{\text{L}})^{2}  
%\end{split}
\\
%%%%%%%%%%%%%%%%%
{\cal V}^{0}_{3} &= 
 2\left[
(\vec{\xi}_{\text{R}}{\cdot}\vec{\xi}_{\text{L}})
(\vec{\chi}_{\text{R}}{\cdot}\vec{\chi}_{\text{L}}) -
(\vec{\xi}_{\text{R}}{\cdot}\vec{\chi}_{\text{L}})
(\vec{\chi}_{\text{R}}{\cdot}\vec{\xi}_{\text{L}})
\right]
\\
%%%%%%%%%%%%%%%%%
{\cal V}^{0}_{4} &= 2\left[
(\vec{\xi}_{\text{R}}{\cdot}\vec{\xi}_{\text{L}})
(\vec{\chi}_{\text{R}}{\cdot}\vec{\chi}_{\text{L}}) +
(\vec{\xi}_{\text{R}}{\cdot}\vec{\chi}_{\text{L}})
(\vec{\chi}_{\text{R}}{\cdot}\vec{\xi}_{\text{L}})
\right] 
%\end{split}
\\
%%%%%%%%%%%%%%%%%
{\cal V}^{0}_{5} &= 
 2(\vec{\xi}_{\text{R}}{\cdot}\vec{\xi}_{\text{L}}-
\vec{\chi}_{\text{R}}{\cdot}\vec{\chi}_{\text{L}})
(\vec{\xi}_{\text{R}}{\cdot}\vec{\chi}_{\text{L}}+
\vec{\chi}_{\text{R}}{\cdot}\vec{\xi}_{\text{L}})
\\
%%%%%%%%%%%%%%%%%
{\cal V}^{0}_{6} &= -\frac{i}{2}\left(
\vec{\xi}_{\text{R}}{\cdot}\vec{\chi}_{\text{R}}+
\vec{\xi}_{\text{L}}{\cdot}\vec{\chi}_{\text{L}}
\right) \\
%%%%%%%%%%%%%%%%%
{\cal V}^{0}_{7} &= -\frac{1}{2}
(\vec{\xi}_{\text{R}}{\cdot}\vec{\chi}_{\text{R}})
(\vec{\xi}_{\text{L}}{\cdot}\vec{\chi}_{\text{L}}) \; . 
\end{align}
\end{subequations}
%%%%%%%%%%%%%%%%%%%%%%%%%%%%%%%%%%%%%%%%%%%%%%%%%%%%%%%%%%%%
The superscripts `0' denote terms coming from the uniform ($q=0$) 
components of the (local) generators. 
Actually, products of two $n^A$-components
contribute the same (but with a permuted order) terms to 
the interactions ${\cal V}_{1},\ldots,{\cal V}_{7}$, 
which can be absorbed by the redefinition of the bare couplings. 
For this reason, we will suppress the superscripts `0' 
and consider the interaction
\begin{equation}
{\cal V} = \sum_{i=1}^{7}g_{i}{\cal V}_{i}
\end{equation}
in what follows.  
These expressions are consistent with the transformation properties 
(\ref{eqn:duality_rule1})--(\ref{eqn:duality_rule2}) under 
$\calU(\theta)$ and the discrete symmetries of the lattice
model (see Eqs. 
(\ref{eqn:time_rev})--(\ref{eqn:link_parity})).
Therefore, we may conclude that 
with an appropriate choice of bare couplings 
the above seven interactions 
will describe (weak) perturbations from the SU(4)-invariant model.  

It is also very suggestive to rewrite the main part of the interaction
as ($g_{6}=g_{7}=0$):
\begin{subequations}
\begin{equation}
\begin{split}
& 
{\cal V} = g_{1} {\cal V}_{1}+ g_{2}{\cal V}_{2}+ g_{3}{\cal V}_{3}
+g_{4}{\cal V}_{4} +g_{5}{\cal V}_{5}  \\
& \quad = \frac{1}{2}(g_{1}+g_{2}+g_{3}+g_{4})
\left(\vec{\xi}_{\text{R}}{\cdot}\vec{\xi}_{\text{L}}+
\vec{\chi}_{\text{R}}{\cdot}\vec{\chi}_{\text{L}}\right)^{2} \\
& \qquad +\frac{1}{2}(g_{1}-g_{2}+g_{3}-g_{4})
\left(\vec{\xi}_{\text{R}}{\cdot}\vec{\chi}_{\text{L}}-
\vec{\chi}_{\text{R}}{\cdot}\vec{\xi}_{\text{L}}\right)^{2}  \\
& \qquad +\frac{1}{2}(g_{1}+g_{2}-g_{3}-g_{4})
\left(\vec{\xi}_{\text{R}}{\cdot}\vec{\xi}_{\text{L}}-
\vec{\chi}_{\text{R}}{\cdot}\vec{\chi}_{\text{L}}\right)^{2} \\
& \qquad + \frac{1}{2}(g_{1}-g_{2}-g_{3}+g_{4})
\left(\vec{\xi}_{\text{R}}{\cdot}\vec{\chi}_{\text{L}}+
\vec{\chi}_{\text{R}}{\cdot}\vec{\xi}_{\text{L}}\right)^{2}
 + g_{5}\, {\cal H}_{5} ,
\end{split}
\label{eqn:interaction_1}
\end{equation}
which can be rewritten
into a more convenient form in terms of the order parameters:
\begin{equation}
\begin{split}
{\cal V}
& = -\frac{1}{2}(g_{1}+g_{2}+g_{3}+g_{4})
\left({\cal O}_{\text{Q}}^{\pi}\right)^{2} \\
& \qquad -\frac{1}{2}(g_{1}-g_{2}+g_{3}-g_{4})
\left({\cal O}_{\text{RQ}}^{\pi}\right)^{2} \\
& \qquad\quad  -\frac{1}{2}(g_{1}+g_{2}-g_{3}-g_{4})
\left({\cal O}_{\text{SD}}^{\pi}\right)^{2} \\
& \qquad\quad\quad - \frac{1}{2}(g_{1}-g_{2}-g_{3}+g_{4})
\left({\cal O}_{\text{SC}}^{\pi}\right)^{2} \\
& \qquad\quad\quad\quad
- g_{5}\, {\cal O}_{\text{SD}}^{\pi}{\cal O}_{\text{SC}}^{\pi}  \; .
\end{split}
\label{eqn:interaction_2}
\end{equation}
\end{subequations}
%%%%%%%%%%%%%%%%%%%%%%%%%%%%%%%%%%%%%%%%%%%%%%%%%%%%%%%%%%%%%%
From this, we may expect that one of the four competing quantum 
phases is selected according to the values of the seven couplings 
$g_{1}, \ldots, g_{7}$ in the low-energy limit.  
In order to investigate the low-energy behavior of the seven couplings, 
we shall carry out a RG analysis in the next section.   
%%%%%%%%%%%%%%%%%%%%%%%%%%%%%%%%%%%%%%%%%%%%%%%%%%%%%%%%%%%%%
\section{RG analysis and four competing orders}
%%%%%%%%%%%%%%%%%%%%%%%%%%%%%%%%%%%%%%%%%%%%%%%%%%%%%%%%%%%%%
\label{sec:RG-analysis}
%%%%%%%%%%%%%%%%%%%%%%%%%%%%%%%%%%%%%%%%%%%%%%%%%%%%%%%%%%%%%
\subsection{$\beta$-functions}
%%%%%%%%%%%%%%%%%%%%%%%%%%%%%%%%%%%%%%%%%%%%%%%%%%%%%%%%%%%%%
As has been shown in the last section, we have seven interactions 
around the SU(4) (or SO(6)) fixed point.  
Since we are interested in the {\em spontaneous} breaking of 
time-reversal symmetry, we will suppress the ${\cal H}_{5}$ 
interaction\cite{comment5}: $g_{5}=0$.
The sixth interaction ${\cal H}_{6}$ is a kind of 
{\em magnetic field} or {\em chemical potential} (see 
section \ref{sec:useful_analogy}) and may be incorporated 
after physics for $g_{6}=0$ is understood.   
For these reasons, we first consider 
the model with $E=F=0$: 
\begin{equation}
{\cal H}= {\cal H}_{\text{SO(6)}} 
+ {\cal V} \; .
\label{eqn:bare_action}
\end{equation}

At the one-loop level, the calculation of the RG $\beta$-function 
reduces to that of OPE defined below\cite{Cardy-book}:
\begin{equation}
 {\cal H}_{i}(z,\bar{z}){\cal H}_{j}(w,\bar{w}) 
\sim \frac{C_{ij}^{k}}{|z-w|^{2}}{\cal H}_{k}(w,\bar{w}) 
\quad (i,j=1,2,3,4,7) \; .
\end{equation}
The non-zero OPE coefficients listed in Appendix 
\ref{sec:non-zero_OPE} 
enable us to write down the RG $\beta$-function%
\cite{Lecheminant-T-05}: 
\begin{equation}
\begin{split}
\dot{g}_{1} &= g_{1}^{2}+g_{2}^{2}+5\,g_{3}^{2}+g_{4}^{2}  \\
\dot{g}_{2} &= 2\,g_{1}\,g_{2}+6\,g_{3}\,g_{4}+g_{4}\,g_{7} \qquad \qquad \\
\dot{g}_{3} &= 6\,g_{1}\,g_{3}+2\,g_{2}\,g_{4} \\ 
\dot{g}_{4} &= 2\,g_{1}\,g_{4}+6\,g_{2}\,g_{3}+g_{2}\,g_{7}  \\
\dot{g}_{7} &= -16(g_{1}\,g_{3}-g_{2}\,g_{4})   \; ,
\end{split}
\label{eqn:RGE}
\end{equation}
where dots denote the derivative with respect to the RG time: 
$\dot{g}=dg_{i}/(d\ln\! L)$.   

It is interesting to observe that this set of RG equations (RGE)
determines a gradient flow in a 5-dimensional space of coupling constants.  
In fact, if we make a change of variables: 
\begin{equation}
\begin{split}
& h_{1}=g_{1} \; , \; h_{2}=g_{2}\; , \; 
h_{3}=\sqrt{2}g_{3}+\frac{1}{\sqrt{128}}g_{7} \; , \; 
h_{4}=g_{4} \\
& h_{5}=\sqrt{\frac{5}{128}}g_{7} \; ,
\end{split}
\end{equation}
the RGE can be derived as 
\[
 \dot{h_{i}} = - \frac{\partial V}{\partial h_{i}}
\]
from a {\em single} RG potential:
\begin{multline}
V(h_{1},h_{2},h_{3},h_{4},h_{5}) = \\
-\frac{1}{3}h_{1}^{3}-h_{1}h_{2}^{2}- \frac{5}{2}h_{1}h_{3}^{2} 
-h_{1}h_{4}^{2}-\frac{1}{2}h_{1}h_{5}^{2} \\
+\sqrt{5}\, h_{1}h_{3}h_{5} -3\sqrt{2}\, h_{2}h_{3}h_{4}
-\sqrt{10}\, h_{2}h_{4}h_{5} \; .
\end{multline}
A similar property was pointed out in quasi one-dimensional 
electron systems\cite{Chen-C-L-C-M-04} to explain the simple 
structure of the phase diagrams.  
%%%%%%%%%%%%%%%%%%%%%%%%%%%%%%%%%%%%%%%%%%%%%%%%%%%%%%%%%%%%%%
\subsection{Symmetries of $\beta$-functions}
%%%%%%%%%%%%%%%%%%%%%%%%%%%%%%%%%%%%%%%%%%%%%%%%%%%%%%%%%%%%%%
\subsubsection{Sign-change- and permutation symmetries}
%%%%%%%%%%%%%%%%%%%%%%%%%%%%%%%%%%%%%%%%%%%%%%%%%%%%%%%%%%%%%%
The above set (\ref{eqn:RGE}) of $\beta$-functions is invariant 
under discrete symmetries (sign change and permutations).  
To be concrete, the RGE is invariant under 
the $s_{1}$ transformation (see Eq. (\ref{eqn:Kramers-Wannier})): 
\begin{equation}
(g_{1},g_{2},g_{3},g_{4},0,0,g_{7}) \xrightarrow{s_{1}}
(g_{1},g_{2},-g_{3},-g_{4},0,0,-g_{7})   \; .
\label{eqn:s1_RG}
\end{equation}

Of course, the spin-chirality duality ${\cal D}$ maps a set of 
couplings as 
\begin{equation}
(g_{1},g_{2},g_{3},g_{4},0,0,g_{7}) \xrightarrow{\cal D}
(g_{1},g_{4},g_{3},g_{2},0,0,g_{7}) 
\end{equation}
and keeps the self-dual manifold $g_{2}=g_{4}$ invariant.  
Combining this with $s_{1}$, 
we obtain for the second duality:
\begin{equation}
(g_{1},g_{2},g_{3},g_{4},0,0,g_{7}) \xrightarrow{\widetilde{\cal D}}
(g_{1},-g_{4},g_{3},-g_{2},0,0,g_{7}) \; . 
\end{equation} 
For the second duality $\widetilde{\cal D}$, 
the self-dual manifold is characterized by 
\begin{equation}
 g_{2} = -g_{4} \; .
\end{equation}
%%%%%%%%%%%%%%%%%%%%%%%%%%%%%%%%%%%%%%%%%%%%%%%%%%%%%%%%%%%%%%
\subsubsection{Self-dual manifolds}
%%%%%%%%%%%%%%%%%%%%%%%%%%%%%%%%%%%%%%%%%%%%%%%%%%%%%%%%%%%%%%
The existence of the two duality transformations ${\cal D}$ 
and $\tilde{\cal D}$ manifests itself in the invariance of 
the $\beta$-functions 
under the interchange: $g_{2} \leftrightarrow \pm g_{4}$ (note that 
the time-reversal breaking term $g_{5}$ is already suppressed).  
That is, the RG-flow is symmetric with respect to 
the four-dimensional self-dual manifolds defined by $g_{2}=\pm g_{4}$.  
Since it follows from the $\beta$-function (\ref{eqn:RGE}) that
\begin{equation*}
(\dot{g}_{2} \mp \dot{g}_{4}) 
= (2g_{1}\mp 6g_{3} \mp g_{7})(g_{2} \mp g_{4}) \, ,
\end{equation*}
systems which are self-dual initially will remain so even after
renormalization.  In fact, the original lattice model 
with $g_{2}=g_{4}$ has a continuous symmetry ${\cal U}(\theta)$--%  
the spin-chirality rotation by an arbitrary angle $\theta$--and  
the Mermin-Wagner-Coleman theorem guarantees the condition 
$g_{2}=g_{4}$ is preserved for all orders of perturbation.  
On this self-dual manifold, the $\beta$-function simplifies to 
\begin{align}
& \dot{g}_{1} = g_{1}^{2}+2g_{2}^{2}+5g_{3}^{2} \notag \\
& \dot{g}_{2} = g_{2}(2g_{1}+6g_{3}+g_{7}) \notag \\
& \dot{g}_{3} = 6g_{1}g_{3}+2g_{2}^{2} \notag \\
& \dot{g}_{7} =-16(g_{1}g_{3}-g_{2}^{2}) \; .
\end{align}
This reduced set of $\beta$-functions has a further invariant 
manifold characterized by SU(3) symmetry: 
\[
 g_{1} = g_{3} \; ,
\] 
where we have only three coupled equations:
\begin{align}
\dot{g}_{1} &= 6g_{1}^{2}+2g_{2}^{2} \notag \\
\dot{g}_{2} &= 8g_{1}g_{2} + g_{2}g_{7} \notag \\
\dot{g}_{7} &= -16(g_{1}^{2}-g_{2}^{2}) \; .
\end{align}
%%%%%%%%%%%%%%%%%%%%%%%%%%%%%%%%%%%%%%%%%%%%%%%%%%%%%%%%%%%%%%
\subsubsection{Other invariant manifolds}
\begin{figure}[h]
\begin{center}
\includegraphics[scale=0.6]{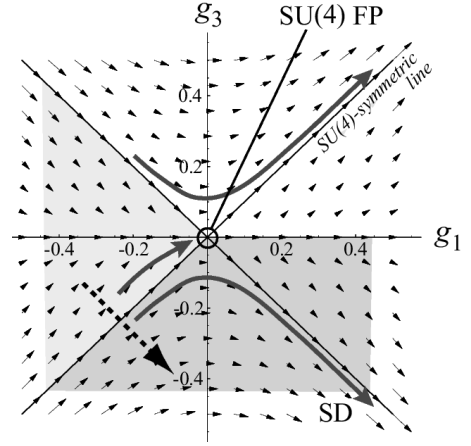}
\end{center}
\caption{RG flow for the SO- and dSO model. The origin (shown as 
SU(4) FP) corresponds to the level-1 SU(4) WZW model and a line 
$g_{1}=g_{3}$ to self-dual models (SU(4)-symmetric, in this case). 
A dashed line roughly corresponds to a path across the SU(4) Sutherland 
model (extended massless$\rightarrow$SU(4)$\rightarrow$staggered dimer (SD)) 
discussed in literature (e.g. Ref. \onlinecite{Azaria-G-L-N-99}).   
An asymptotic trajectory in the lower-right portion (highlighted by 
a thick arrow) is responsible for the spontaneous 
breakdown of translational- (for SO model) and time-reversal 
(for dSO model) symmetries.\label{fig:RG_SO_model}}
\end{figure}

%%%%%%%%%%%%%%%%%%%%%%%%%%%%%%%%%%%%%%%%%%%%%%%%%%%%%%%%%%%%%% 
Furthermore, by adding and subtracting equations in (\ref{eqn:RGE}), 
we obtain several invariant manifolds: \\
%%%%%%%%%%%%%%%%%%%%%%%%%%%%%%
\underline{\it Spin-Orbital (SO) manifold}:
\begin{equation}
g_{1}=g_{2}, \; g_{3}=g_{4}, \; g_{7}=0  \; .
\end{equation}
This means that as far as we consider the SO model (see 
Eq.(\ref{eqn:spin-orb-ham})), 
the rung-rung four-body interaction will {\em never} be generated 
radiatively on manifold.  In fact, the underlying 
SU(2)$\times$SU(2) ($\text{SO(3)}_{\xi}\times\text{SO(3)}_{\chi}$) 
symmetry guarantees that it persists even in {\em all} orders 
of perturbation expansion. 
On this manifold, only two couplings $g_{1}$ and $g_{3}$ suffice for 
describing the low-energy physics and 
the RG $\beta$-functions substantially simplify 
to the following two coupled equations:
\begin{equation}
 \dot{g}_{1}=2g_{1}^{2}+6g_{3}^{2} \quad , \quad 
\dot{g}_{3}=8g_{1}g_{3} \; .
\label{eqn:SO_RGE}
\end{equation}
When $g_{3}<0$ and $g_{3}<g_{1}$ (which may be interpreted as 
the SO-model with $J>K/4$. For a detailed discussion of 
the SO-model, see Ref. \onlinecite{Azaria-G-L-N-99}.),  
the system is attracted in the infrared limit 
by an asymptotic trajectory:
\[
 g_{1}^{\ast}(=g_{2}^{\ast})=-g_{3}^{\ast}(=-g_{4}^{\ast})  
\]
(see a region shown by darker gray in Fig. \ref{fig:RG_SO_model}) 
and breaks the translational symmetry 
spontaneously\cite{Azaria-G-L-N-99}.  
The exact ground state\cite{Kolezhuk-M-98} at $K=4J/3$ and 
the analysis for small $K$($\ll J$)\cite{Nersesyan-T-97} 
are consistent with the above result.  
The region $g_{1}<0, g_{3}< |g_{1}|$ (a portion of 
Fig. \ref{fig:RG_SO_model} marked by lighter gray) is the basin of 
the gapless $\text{SU(4)}_{1}$ fixed point.   

\noindent%
\underline{\it Dual Spin-Orbital (dSO) manifold}:
\begin{equation}
g_{1}=g_{4}, \; g_{2}=g_{3}, \; g_{7}=0
\end{equation}
This is the dual (${\cal D}$) partner of the above invariant manifold 
obtained by applying ${\cal D}$ to the SO-model and 
its existence is guaranteed by a hidden (dual) SU(2)$\times$SU(2) 
symmetry of the model.  
Now two couplings $g_{1}$ and $g_{2}$ describe the system and, again, 
$\beta$-functions reduce to the set of equations (\ref{eqn:SO_RGE}) 
with $g_{3}$ being replaced with $g_{2}$.   
For $g_{2}<0$ and $g_{2}<g_{1}$, the system flows toward  
another asymptotic trajectory:
\[
 g_{1}^{\ast}(=g_{4}^{\ast})=-g_{2}^{\ast}(=-g_{3}^{\ast}) 
\]
and, in this case, breaks time-reversal symmetry ${\cal T}$ 
(see the next subsection for the detail).  
The basin of $\text{SU(4)}_{1}$ fixed point of the SO manifold 
is now mapped onto $g_{1}<0,g_{2}< |g_{1}|$.   
In terms of lattice models, this corresponds to the model Hamiltonian 
\[
 {\cal H}_{\text{dSO}} = \frac{1}{2}J{\cal H}_{1}
+\frac{1}{8}K{\cal H}_{2} + \frac{1}{8}K{\cal H}_{3}+
\frac{1}{2}J{\cal H}_{4}  \; 
\]  
with $J < K$.   
Similarly, we have two more invariant manifolds 
\begin{equation}
\begin{split}
& g_{1}=-g_{2}\, , \; g_{3}=-g_{4}\, , \; g_{7}=0 \, ,\\
& g_{1}=-g_{4}\, , \; g_{2}=-g_{3}\, , \; g_{7}=0 \; ,
\end{split}
\end{equation}
on which the full RGE reduces to two coupled equations.  
Four RG-invariant lines (rays) defined by the intersections 
among these invariant manifolds (hyperplanes) will turn out 
to correspond to four dominant competing phases.   

Away from these invariant manifolds, the rung-rung four-body interaction 
$g_{7}$ will be generated in the course of renormalization even though it 
is absent initially.   
%%%%%%%%%%%%%%%%% FIGURE %%%%%%%%%%%%%%%%%%%%%%%%%%%%%%%%%%%%%%
%%%%%%%%%%%%%%%%%%%%%%%%%%%%%%%%%%%%%%%%%%%%%%%%%%%%%%%%%%%%%%
%
%
%%%%%%%%%%%%%%%%%%%%%%%%%%%%%%%%%%%%%%%%%%%%%%%%%%%%%%%%%%%%%%
\subsection{Asymptotic form of RG-flow}
%%%%%%%%%%%%%%%%%%%%%%%%%%%%%%%%%%%%%%%%%%%%%%%%%%%%%%%%%%%%%%
\label{sec:LBF-analysis}
%%%%%%%%%%%%%%%%%%%%%%%%%%%%%%%%%%%%%%%%%%%%%%%%%%%%%%%%%%%%%%
The set of RG $\beta$-functions (\ref{eqn:RGE}) 
is complicated and complete analysis of it is not so easy.  
We numerically integrated the equations and found that 
the RG flow exhibited a striking feature in the low-energy limit.   
To see this more closely, we apply an ansatz proposed 
by Lin, Balents, and Fisher\cite{Lin-B-F-98} in the study of 
a half-filled two-leg Hubbard ladder.  

Since only marginal interactions appear in the $\beta$-functions, 
a natural ansatz valid for the infrared asymptotics may be%
\cite{Lin-B-F-98} 
\begin{equation}
g_{i}(t) = \frac{G_{i}}{t_{0}-t} \qquad (i=1,2,3,4,7) \; , 
\end{equation}
where the constant $t_{0}$ marks the crossover point where 
the weak-coupling perturbation breaks down.  
Plugging these into the RG equations and requiring consistency,  
we have a set of non-linear equations which determines  
the IR-asymptotics.  We found various solutions 
and among them the followings are relevant for our analysis:  
%%%%%%%%%%%%%%%%%%%%%%%%%%%%%%%%%%%%%%%%%%%%%%%%%%%%%%%%%%
%\begin{widetext}
%%%%%%%%%%%%%%%%%%%%%%%%%%%%%%%%%%%%%%%%%%%%%%%%%%%%%%%%%%

\underline{\em Self-dual}:
\begin{subequations}
\begin{align} 
& \left(G_{1},G_{2},G_{3},G_{4},G_{7}\right)
=\left(\frac{1}{6},0,-\frac{1}{6},0,\frac{4}{9} \right) 
\qquad \text{(I)}  \label{eqn:SU3_ray1} \\
& \left(G_{1},G_{2},G_{3},G_{4},G_{7}\right)
=\left(\frac{1}{6},0,\frac{1}{6},0,-\frac{4}{9} \right) 
\qquad \text{(II)} \; . \label{eqn:SU3_ray2}
\end{align}
\end{subequations}
The meaning of these rays will become clear if we plug 
Eqs. (\ref{eqn:SU3_ray1},\ref{eqn:SU3_ray2}) into 
(\ref{eqn:interaction_2}):
\begin{subequations}
\begin{align}
\text{(I):}\quad & 
-\frac{1}{6}g_{\text{I}}^{\ast} \left(
\left({\cal O}^{\pi}_{\text{SD}}\right)^{2} 
+ \left({\cal O}^{\pi}_{\text{SC}}\right)^{2}
\right)
+\frac{4}{9}g_{\text{I}}^{\ast}{\cal H}_{7} \\
\text{(II):}\quad & 
-\frac{1}{6}g_{\text{II}}^{\ast} \left(
\left({\cal O}^{\pi}_{\text{Q}}\right)^{2}+ \left({\cal O}^{\pi}_{\text{RQ}}\right)^{2}
\right)
-\frac{4}{9}g_{\text{II}}^{\ast}{\cal H}_{7} \; .
\end{align}
\end{subequations}
Therefore, the above two rays (I) and (II) may be thought of as 
corresponding to SD-SC- and Q-RQ transitions, respectively.   

From the argument in section \ref{sec:Model_and_Sym}, it is obvious 
that the latter case-(II) has SU(3) symmetry,  
while the SU(3) symmetry of the former is hidden and appears 
only after the particle-hole transformation for the left movers 
is applied (see section \ref{sec:SU3_U1}).  
Also it is worth noting that they are related to each other 
by the Ising duality $s_{1}$.  
%%%%%%%%%%%%%%%%%%%%%%%%%%%%%%%%%%%%%%%%

\underline{\em SO(6)-symmetric rays}:
\begin{subequations}
\begin{alignat}{2}
& \left(G_{1},G_{2},G_{3},G_{4},G_{7} \right)=
\left(\frac{1}{8},\frac{1}{8},-\frac{1}{8},-\frac{1}{8},0 \right) 
& \qquad & \text{(SD)} 
\label{eqn:SO6_sym_ray1}
\\
& \left(G_{1},G_{2},G_{3},G_{4},G_{7} \right)=
\left(\frac{1}{8},-\frac{1}{8},-\frac{1}{8},\frac{1}{8},0 \right) 
& \qquad & \text{(SC)} 
\label{eqn:SO6_sym_ray2}
\\
& \left(G_{1},G_{2},G_{3},G_{4},G_{7}\right)
=\left(\frac{1}{8},\frac{1}{8},\frac{1}{8},\frac{1}{8},0 \right)
& \qquad & \text{(Q)} 
\label{eqn:SO6_sym_ray3}
\\
& \left(G_{1},G_{2},G_{3},G_{4},G_{7} \right)=
\left(\frac{1}{8},-\frac{1}{8},\frac{1}{8},-\frac{1}{8},0 \right) 
& \qquad & \text{(RQ)} 
\label{eqn:SO6_sym_ray4}
 \; .
\end{alignat}
\end{subequations}
Note that the latter two are invariant both under spin-chirality 
SO(2) and under SU(3), while the former two are not.  
It is important to note that the low-energy effective Hamiltonians 
for these rays assume the following simple forms:
\begin{equation}
{\cal H}_{\text{A}} = {\cal H}_{\text{SO(6)}} 
-g^{\ast}_{A}\, \left({\cal O}^{\pi}_{\text{A}}\right)^{2} \quad 
(A=\text{Q, RQ, SD, and SC}) \; .
\end{equation}  
Semiclassical argument based on Eq. (\ref{eqn:interaction_2}) 
tells us that these symmetric rays 
correspond to the four competing phases where one of the order 
parameters has a non-zero expectation value: 
$\VEV{{\cal O}^{\pi}_{A}}\neq 0$.   
In particular, from Table \ref{OP_Sym}, 
it follows that a spin-disordered ground state with broken ${\cal T}$ 
is realized along the ray `SC'.  
Furthermore, we can show that all these rays correspond to IR-massive 
flows along which translational symmetry is broken 
and the SO(6) symmetry is restored asymptotically (see next section).
%%%%%%%%%%%%%%%%%%%%%%%%%%%%%%%% 

\underline{\em Transitions among dominant phases}: 
\begin{subequations}
\begin{alignat}{2}
& G_{1}=G_{2}=\frac{1}{2} \; , \; G_{3}=G_{4}=0 \;, \; 
G_{7}=0 &\;\; & (\text{Q}\leftrightarrow\text{SD}) 
\label{eqn:ray-D-SD} \\
& G_{1}= - G_{2}=\frac{1}{2} \; , \; G_{3}=G_{4}=0 \;, \; 
G_{7}=0 & \;\; & (\text{RQ}\leftrightarrow\text{SC}) \\
& G_{1}= G_{4}=\frac{1}{2} \; , \; G_{2}=G_{3}=0 \; , \; G_{7}=0 
&\;\; &  
(\text{Q}\leftrightarrow\text{SC})  \\
& G_{1}= - G_{4}=\frac{1}{2} \; , \; G_{2}=G_{3}=0 \; , \; G_{7}=0 
&\;\; &  
(\text{RQ}\leftrightarrow\text{SD})
\label{eqn:ray-RD-SD}
\end{alignat}
\end{subequations}
None of these restores the original SO(6) symmetry.   
Actually, these rays correspond to transitions among 
the above four dominant phases.  

\underline{\em Case of four competing orders}:
\begin{equation}
G_{1}=1 \; , \; G_{2}=G_{3}=G_{4}=G_{7}=0 \; .
\label{eqn:unknown_ray}
\end{equation}
This ray describes non-trivial quantum criticality resulting from 
the competition among four different orders, which will be 
discussed in section \ref{sec:other_QCP}.   
%%%%%%%%%%%%%%%%%%%%%%%%%%%%%%%%%%%%%%%%%%%%%%%%%%%%%%%%%%

The point here is that in the low-energy limit our system 
is asymptotically characterized only by a {\em single} 
(diverging) coupling constant\cite{Lin-B-F-98,Chen-C-L-C-M-04} 
$\sim 1/(t_{0}-t)$.  

%%%%%%%%%%%%%%%%%%%%%%%%%%%%%%%%%%%%%%%%%%%%%%%%%%%%%%%%%
\subsection{Four dominant phases}
\label{sec:4-dominant-phases}
%%%%%%%%%%%%%%%%%%%%%%%%%%%%%%%%%%%%%%%%%%%%%%%%%%%%%%%%%
\subsubsection{SO(6)-restoration}
%%%%%%%%%%%%%%%%%%%%%%%%%%%%%%%%%%%%%%%%%%%%%%%%%%%%%%%%%
To illustrate how SO(6) symmetry, which is explicitly broken 
in the bare action (\ref{eqn:bare_action}), 
is restored along the four rays (i.e. `SD', `SC', `Q', and `RQ'),  
we take the two rays `Q' and `SD'.  From the continuum expressions 
(\ref{eqn:interaction_1}) and (\ref{eqn:interaction_2}),  
it is obvious that the model along the ray `Q' is explicitly 
$\text{SO(6)}$ invariant:
\begin{equation}
\begin{split}
{\cal H}_{\text{Q}} &= {\cal H}_{\text{SO(6)}} 
-g^{\ast} \left({\cal O}^{\pi}_{\text{Q}}\right)^{2}  \\
&= {\cal H}_{\text{SO(6)}} 
+ g^{\ast}(\vec{\xi}_{\text{R}}{\cdot}\vec{\xi}_{\text{L}}+
\vec{\chi}_{\text{R}}{\cdot}\vec{\chi}_{\text{L}})^{2} \; .
\end{split}
\label{eqn:SO6_Ham}
\end{equation}
This is nothing but the Hamiltonian of the SO(6) Gross-Neveu 
model \cite{gross}.  
This model is integrable. Its spectrum 
is known\cite{Zamolodchikov-Z-79,essler}  
to consist of the fundamental fermion with
mass $M$ together with a kink of mass $m = M/\sqrt{2}$.
The $\mathbb{Z}_{2}$-symmetry ${\cal O}^{\pi}_{\text{Q}}
\leftrightarrow -{\cal O}^{\pi}_{\text{Q}}$ is broken spontaneously
and, as a result, we have a finite expectation value 
$\langle {\cal O}^{\pi}_{\text{Q}}\rangle \neq 0$.  

The application of the Ising duality 
$s_{1}$ (Eq. (\ref{eqn:Kramers-Wannier})) to this Hamiltonian 
${\cal H}_{\text{Q}}$ takes us to another ray `SD' and 
the associated Hamiltonian ${\cal H}_{\text{SD}}$ since 
${\cal H}_{\text{SO(6)}}\mapsto {\cal H}_{\text{SO(6)}}$ 
and ${\cal O}_{\text{Q}}\mapsto {\cal O}_{\text{SD}}$.  
A similar argument applies to the other pair 
(${\cal H}_{\text{RQ}}$ and ${\cal H}_{\text{SC}}$) as well.  

\begin{figure}[h]
\begin{center}
\includegraphics[scale=0.5]{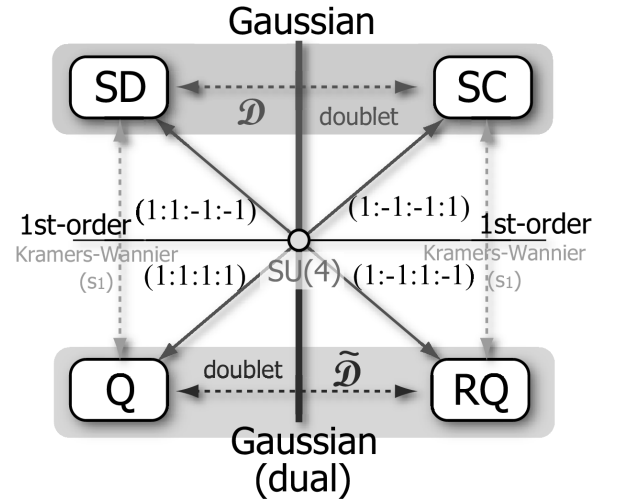}
\caption{Relationship among the four dominant phases. 
Ratio of the couplings $g_{1}$, $g_{2}$, $g_{3}$, and 
$g_{4}$ along each symmetric ray is shown.   
\label{fig:4phases}
}
\end{center}
\end{figure}

Moreover, the spin-chirality duality 
${\cal D}$ ($\widetilde{\cal D}$) maps ${\cal H}_{\text{SD}}$ 
(${\cal H}_{\text{Q}}$) onto 
${\cal H}_{\text{SC}}$ (${\cal H}_{\text{RQ}}$).    
These facts imply that all these four low-energy Hamiltonians 
${\cal H}_{\text{Q}}$, ${\cal H}_{\text{RQ}}$, 
${\cal H}_{\text{SD}}$, and ${\cal H}_{\text{SC}}$ can be transformed 
back to the SO(6)-invariant one (\ref{eqn:SO6_Ham}) 
by applying symmetry operations 
of $\text{O(6)}_{\text{L}}\times \text{O(6)}_{\text{R}}$.  
This is the reason why we named the asymptotic rays 
(\ref{eqn:SO6_sym_ray1}-\ref{eqn:SO6_sym_ray4}) 
`SO(6)-symmetric'.  
The relationship among these four phases is summarized in 
Fig. \ref{fig:4phases}.   
%%%%%%%% FIGS %%%%%%%%%%%%%%%%%%%%%%%%%%%%%%%%
%%%%%%%%%%%%%%%%%%%%%%%%%%%%%%%%%%%%%%%%%%%%%%%%%%%%%%%%%%%%%%%%%
\subsubsection{Ground-state degeneracy}
%%%%%%%%%%%%%%%%%%%%%%%%%%%%%%%%%%%%%%%%%%%%%%%%%%%%%%%%%%%%%%%%%
Now we know that four competing (gapped) phases exist 
around the SU(4)-point and that they are mapped onto  
each other by the two discrete transformations 
${\cal D}$ and $\widetilde{\cal D}$.   From this, 
we may conclude that all these phases have the same 
ground-state degeneracy (from the exact solution \cite{Kolezhuk-M-98} 
for the SD phase, it might seem natural that we should have 
the degeneracy 2 for the others as well.)  
But there is a pitfall here; the discrete transformation 
$\widetilde{\cal D}$  
that relates `SD' to `Q' is {\em not} a local one and there is no reason 
to believe that $\widetilde{\cal D}$ maps a phase onto another one 
with the same ground-state degeneracy.  
A typical example of this kind of {\em non-local} transformation 
is the so-called Kennedy-Tasaki transformation\cite{Kennedy-T-92}, 
which maps a unique (infinite-volume) ground state onto 
four-fold degenerate ones.   

To find the correct answer, we first note that 
Dirac fermions used to construct our SU(4) Hamiltonian 
(see Appendix \ref{sec:deriv_SU4}) are expressed as exponentials 
of boson operators and that the semiclassical bosonic 
ground states should be considered modulo the following 
gauge-equivalence:
\begin{equation}
\phi_{a,\sigma,\text{L/R}} \sim 
\phi_{a,\sigma,\text{L/R}} + \sqrt{\pi}\; N_{a,\sigma,\text{L/R}}
 \quad (N_{a,\sigma,\text{L/R}}\in \mathbb{Z} )\; ,
\label{eqn:gauge-redundancy}
\end{equation}
which keeps the bosonic expressions of the fermions 
(\ref{eqn:boson2fermion}) unchanged.  

We carefully counted the number of inequivalent ground states for all 
four cases `Q', `SD', `SC', and `RQ'.  
Although the calculation was slightly more complicated 
than that in Ref. \onlinecite{Lin-B-F-98}, the procedure was 
essentially the same (see Appendix \ref{sec:GS_degeneracy} 
for details) and the results are summarized as (see also Table \ref{tab:sym_4phases}): \\
\underline{\em `Q'-phase:} 4 ground states; $2k_{\text{F}}$-part 
of $\VEV{X_{r}^{A}X_{r+1}^{A}}$ is  
non-vanishing (a kind of quadrumerized phase) and period-4 
ground states appear\\
\underline{\em `SD'-phase:} 2 ground states; $2k_{\text{F}}$-part is 
vanishing (staggered phase) and the ground states have a period-2 \\
\underline{\em `SC'-phase:} 2 ground states; $2k_{\text{F}}$-part is 
vanishing \\
\underline{\em `RQ'-phase:} 4 ground states; $2k_{\text{F}}$-part is 
non-vanishing (period-4 density wave of the pseudo-spin 
$\VEV{\bolS_{1,r}{\cdot}\bolS_{2,r}}$).  

For some reasons, the period-4 ($q=\pi/2$) contribution in the 
order-parameter correlations vanishes for the SD- and SC phases, 
as is expected from the exact ground state\cite{Kolezhuk-M-98} 
for which we have $\VEV{{\cal O}_{r}{\cal O}_{r+1}}=(-1)^{r}3/4$ 
(when the amplitude of the order parameter is maximal).  
On the other hand, we have the $2k_{\text{F}}$-components 
in the Q- and RQ phases.  This implies that the $4k_{\text{F}}$-%
component is not sufficient to fully characterize the latter two phases.  

So far, the nature of the ground states appearing in the `Q'- or 
the `RQ' phase is not so clear.  This is mainly because 
the transformation $s_{1}$ relating ${\cal O}^{\pi}_{\text{SD}}$ to 
${\cal O}^{\pi}_{\text{Q}}$ is non-local and we do not know 
the microscopic (or, lattice) expressions for 
${\cal O}^{\pi}_{\text{Q,RQ}}$.    
By symmetry arguments, we can restrict the possible forms 
of lattice operators corresponding to ${\cal O}^{\pi}_{\text{Q}}$.  
We have five building blocks for the lattice order parameters 
all of which have (at least for the leading contributions) 
the same $4k_{\text{F}}$(or $q=\pi$)-term ${\cal O}^{\pi}_{\text{Q}}$ 
in the continuum.  
Exactly on the SU(4)-invariant line, these five will 
condense with the same amplitude to describe a phase with 
SU(4) quadrumerization (note that we have 4-fold degeneracy).  
Unfortunately, our continuum approach cannot tell anything 
about which of the five will become dominant when we move away 
from the SU(4) manifold.   

Similarly, since the $4k_{\text{F}}$-part of 
$\bolS_{1,r}{\cdot}\bolS_{2,r}$ is written as 
${\cal O}^{\pi}_{\text{RQ}}$, the RQ-phase may be thought of as 
a density-wave phase of the pseudo-spin ${\cal S}^{z}$. 
By using the analogy in section \ref{sec:spin-1-bose}, 
this is nothing but the bosonic charge-density wave state 
with period-4.   

%%%%%%%%%%%%%%%%%%%%% TABLE %%%%%%%%%%%%%%%%%%%%%%%%%%%%
\begin{table}[h]
\begin{center}
\begin{ruledtabular}
\begin{tabular}{cccccc} 
& \multicolumn{5}{c}{G.S. degeneracy and symmetry} \\
\cline{2-6}
\raisebox{1.5ex}[0pt]{phase} & 
\makebox[10mm]{degen.} &
\makebox[10mm]{${\cal T}$} & 
\makebox[10mm]{${\cal P}_{12}$} & 
\makebox[10mm]{$P_{\text{S}}$} & 
\makebox[10mm]{$P_{\text{L}}$} \\
\hline
SD & $2$ & $+1$ & $-1$ & $-1$ & $+1$ \\
\hline 
SC & $2$ & $-1$ & $-1$ & $-1$ & $+1$ \\
\hline 
Q  & $4$ & $+1$ & $+1$ & $-1$ & $+1$ \\
\hline 
RQ & $4$ & $+1$ & $+1$ & $+1$ & $-1$ \\
\end{tabular}
\end{ruledtabular}
\caption{Four dominant phases and the ground-state degeneracy. 
In all four phases, translation symmetry is broken.
\label{tab:sym_4phases}}
\end{center} 
\end{table}
%%%%%%%%%%%%%%%%%%%%%%%%%%%%%%%%%%%%%%%%%%%%%%%%%%%%%%%%%%%%%%%%
\section{Physics of self-dual models and phase transitions}
%%%%%%%%%%%%%%%%%%%%%%%%%%%%%%%%%%%%%%%%%%%%%%%%%%%%%%%%%%%%%%%%
\label{sec:physics_selfdual}
%%%%%%%%%%%%%%%%%%%%%%%%%%%%%%%%%%%%%%%%%%%%%%%%%%%%%%%%%%%%%%%%
Now that we have identified four dominant phases in our SO(6) problem, 
the next question to ask would be the quantum phase transitions 
among them. In this section, we discuss the nature of
the phase transition between SD and SC phases 
as well as the Q-RQ transition.
As described in section \ref{sec:LBF-analysis},
the transition should belong to the self-dual manifolds ($g_{2}=\pm g_{4}$)
and we first present a fully quantum description of the low-energy physics
along these manifolds.
%%%%%%%%%%%%%%%%%%%%%%%%%%%%%%%%%%%%%%%%%%%%%%%%%%%%%%%%%%%%%%
\subsection{SU(3)$\times$U(1) bosonization}
%%%%%%%%%%%%%%%%%%%%%%%%%%%%%%%%%%%%%%%%%%%%%%%%%%%%%%%%%%%%%%
\label{sec:SU3_U1}
%%%%%%%%%%%%%%%%%%%%%%%%%%%%%%%%%%%%%%%%%%%%%%%%%%%%%%%%%%%%%%
According to the analysis presented in section \ref{sec:LBF-analysis}, 
the one-loop RG  flow on the self-dual manifold, which is invariant under 
the spin-chirality rotation ${\cal U}(\theta)$ 
or $\widetilde{\cal U}(\theta)$,  
is attracted to one of the following two special rays: 
(I): $g_1 = - g_3 = g^{*}_{\text{I}}/6$, $g_2 = g_4 =0$, 
$g_7 = 4 g^{*}_{\text{I}}/9$ and 
(II): $g_1 = g_3 = g^{*}_{\text{II}}/6$, $g_2 = g_4 =0$, 
$g_7 = -4 g^{*}_{\text{II}}/9$ with $g_{\text{I,II}}^{*} >0$. 
Along these rays, the low-energy field theory 
reads as follows: 
%%%%%%%%%%%%%%%%%%%%%%%%%%%%%%%%%%%%%
\begin{widetext}
\begin{subequations}
\begin{align}
& {\cal H}_{\text{IR}}^{\text{(I)}} = 
{\cal H}_{\text{SO(6)}}
+ \frac{g^{*}_{\text{I}}}{6} \left[ \left( \vec{\xi}_\text{R} 
{\cdot} \vec{\xi}_\text{L} 
- \vec{\chi}_\text{R} {\cdot}\vec{\chi}_\text{L} \right)^2
+ 
\left( \vec{\xi}_\text{R}
{\cdot} \vec{\chi}_\text{L}
+ \vec{\chi}_\text{R} {\cdot} \vec{\xi}_\text{L} \right)^2 \right]
- \frac{2 g^{*}_{\text{I}}}{9}
\left(\vec{\xi}_\text{R} {\cdot} \vec{\chi}_\text{R}\right)
\left(\vec{\xi}_\text{L} {\cdot} \vec{\chi}_\text{L} \right)
\label{eqn:IRth1}
\\
& {\cal H}_{\text{IR}}^{\text{(II)}} = 
{\cal H}_{\text{SO(6)}}
+ \frac{g^{*}_{\text{II}}}{6} \left[ \left( \vec{\xi}_\text{R}
{\cdot} \vec{\xi}_\text{L}
+ \vec{\chi}_\text{R} {\cdot} \vec{\chi}_\text{L} \right)^2
+
\left( \vec{\xi}_\text{R}
{\cdot} \vec{\chi}_\text{L}
- \vec{\chi}_\text{R}{\cdot} \vec{\xi}_\text{L} \right)^2 \right]
+ \frac{2 g^{*}_{\text{II}}}{9}
\left(\vec{\xi}_\text{R}{\cdot} \vec{\chi}_\text{R}\right)
\left(\vec{\xi}_\text{L}{\cdot} \vec{\chi}_\text{L} \right) \; .
\label{eqn:IRth2}
\end{align}
\end{subequations}
\end{widetext}
%%%%%%%%%%%%%%%%%%%%%%%%%%%%%%%%%%%%%%%%%%%%%%%
It is interesting to observe that the two low-energy
field theories ${\cal H}_{\text{IR}}^{\text{(I)}}$ and 
${\cal H}_{\text{IR}}^{\text{(II)}}$ 
interchange themselves under the Ising duality symmetry $s_1$ 
(Eq.(\ref{eqn:Kramers-Wannier})).   

As has been shown in section \ref{sec:LBF-analysis}, 
the low-energy effective Hamiltonians 
(\ref{eqn:IRth1},\ref{eqn:IRth2}) describe 
the competition between two different orders 
(see Eqs.(\ref{eqn:SU3_ray1}), (\ref{eqn:SU3_ray2})): 
the staggered dimerized ${\cal O}^{\pi}_{\text{SD}}$ and the staggered
scalar chirality ${\cal O}^{\pi}_{\text{SC}}$ orders for the 
model ${\cal H}_{\text{IR}}^{\text{(I)}}$ (\ref{eqn:IRth1}),  
and the two quadrumerized phases ${\cal O}^{\pi}_{\text{Q}}$ 
and ${\cal O}^{\pi}_{\text{RQ}}$ for the model 
${\cal H}_{\text{IR}}^{\text{(II)}}$ (\ref{eqn:IRth2}). 
The main question concerns the nature of the quantum
phase transitions that result from the competition
between these orders (${\cal O}^{\pi}_{\text{SD}}\leftrightarrow 
{\cal O}^{\pi}_{\text{SC}}$ for the case I and 
${\cal O}^{\pi}_{\text{Q}}\leftrightarrow 
{\cal O}^{\pi}_{\text{RQ}}$ for the case II).  
In this section, we present a full 
quantum description to this issue.  

%%%%%%%%%%%%%%%%%%%%%%%%%%%%%%%%%%%%%%%%%%%%%%%%%%%%%%%%%%%%%%%%%
%\subsection{What's going on for self-dual models}
%%%%%%%%%%%%%%%%%%%%%%%%%%%%%%%%%%%%%%%%%%%%%%%%%%%%%%%%%%%%%%%%%
The key observation, which will be crucial in the following 
analysis, is that 
the low-energy theories (\ref{eqn:IRth1},\ref{eqn:IRth2})
display a hidden symmetry which is not SU(2)$\times$U(1)
but, in fact, a larger U(3) symmetry.  
In the following, we are going to discuss only  
the case of the first model (\ref{eqn:IRth1}), i.e. the nature 
of the phase transition between the staggered dimerized- 
and the scalar chirality phases.  
The physical properties of the second model (\ref{eqn:IRth2}) can then be 
derived readily by applying the Ising duality symmetry $s_{1}$ 
(\ref{eqn:Kramers-Wannier}).  

In the following analysis, it would be convenient to combine 
three pairs of Majorana fields $\xi^{a}_{\text{L/R}}$ and 
$\chi^{a}_{\text{L/R}}$ to form three Dirac fermions 
$\Psi_{a,\text{L/R}}$: 
\begin{equation}
\Psi_{a,\text{R}} = \frac{\xi_\text{R}^a + i \chi_\text{R}^a}{\sqrt{2}} 
\quad , \quad 
\Psi_{a,\text{L}} = \frac{\xi_\text{L}^a - i \chi_\text{L}^a}{\sqrt{2}} ,
\label{eqn:Diracfer}
\end{equation}
with $a=1,2,3$.  The reason for this left-right {\em asymmetric} definition 
is that SU(3)-symmetry is clear in this notation.  The symmetric 
notation will be useful in describing the second ray (II) 
(Eq. (\ref{eqn:SU3_ray2}) or the model (\ref{eqn:IRth2}))
\cite{comment6}.

From Eq. (\ref{eqn:Diracfer}), it is straightforward 
to express the order parameters (\ref{eqn:orderparams}) in terms 
of these Dirac fermions:
\begin{equation}
\begin{split}
& {\cal O}^{\pi}_\text{Q} = i \sum_{a=1}^{3}\left( \Psi_{a,\text{R}} 
\Psi_{a,\text{L}}
+ \Psi_{a,\text{R}}^{\dagger} \Psi_{a,\text{L}}^{\dagger} \right)
\\
& {\cal O}^{\pi}_\text{RQ} = \sum_{a=1}^{3}
\left( - \Psi_{a,\text{R}} \Psi_{a,\text{L}}
+ \Psi_{a,\text{R}}^{\dagger} \Psi_{a,\text{L}}^{\dagger} \right) 
 \\
& {\cal O}^{\pi}_\text{SD} = i \sum_{a=1}^{3}\left( \Psi_{a,\text{R}} 
\Psi_{a,\text{L}}^{\dagger}
+ \Psi_{a,\text{R}}^{\dagger} \Psi_{a,\text{L}} \right) 
\\
& {\cal O}^{\pi}_\text{SC} =  \sum_{a=1}^{3}\left( 
\Psi_{a,\text{R}} \Psi_{a,\text{L}}^{\dagger}
- \Psi_{a,\text{R}}^{\dagger} \Psi_{a,\text{L}} \right) \; .
\end{split}
\label{orderparadfer}
\end{equation}
In this notation, two order parameters ${\cal O}^{\pi}_{\text{Q}}$ and 
${\cal O}^{\pi}_{\text{RQ}}$, which are invariant under 
${\cal U}(\theta)$, transform like `Cooper pairs', 
while ${\cal O}^{\pi}_{\text{SD}}$ and ${\cal O}^{\pi}_{\text{SC}}$ 
look like the density operators.  
If we had adopted the left-right symmetric definition instead of 
Eq. (\ref{eqn:Diracfer}), 
${\cal O}^{\pi}_{\text{SD}}$ and ${\cal O}^{\pi}_{\text{SC}}$ would have 
behaved like the `Cooper pairs'.   
This analogy can also be revealed by mentioning 
the expression of the 
spin-chirality rotation ${\cal U}(\theta)$ (Eq. (\ref{eqn:SC_Majorana1}))
and the Ising duality symmetry $s_{1}$ 
(Eq. (\ref{eqn:Kramers-Wannier})) on these Dirac fermions:
\begin{subequations}
\begin{align}   
& {\Psi}_{a,\text{R}} \xrightarrow{{\cal U}(\theta)} 
\be^{+i\frac{\theta}{2}}\; {\Psi}_{a,\text{R}}  
\quad , \quad 
{\Psi}_{a,\text{L}} \xrightarrow{{\cal U}({\theta})}
\be^{-i\frac{\theta}{2}}\; {\Psi}_{a,\text{L}}
\label{eqn:Dirac_U}\\
& {\Psi}_{a,\text{R}} 
\xrightarrow{\widetilde{{\cal U}}(\widetilde{\theta})}
\be^{-i\frac{\tilde{\theta}}{2}} \; {\Psi}_{a,\text{R}}  
\quad , \quad 
{\Psi}_{a,\text{L}} 
\xrightarrow{\widetilde{{\cal U}}(\widetilde{\theta})}
\be^{-i\frac{\tilde{\theta}}{2}} \; {\Psi}_{a,\text{L}} 
\label{eqn:Dirac_Utilde} \\
& {\Psi}_{a,\text{R}} \xrightarrow{s_1}
{\Psi}_{a,\text{R}}^{\dagger}
\quad , \quad 
{\Psi}_{a,\text{L}} \xrightarrow{s_1}
{\Psi}_{a,\text{L}}   \; .
\label{eqn:Dirac_s1}
\end{align}
\end{subequations}
The appearance of the left-right asymmetric expressions 
for ${\cal U}(\theta)$ is just 
an artifact of our definition (\ref{eqn:Diracfer}); 
if the symmetric definition had been used, 
Eqs. (\ref{eqn:Dirac_U}) and (\ref{eqn:Dirac_Utilde}) 
would have been interchanged.  
One thus observes that the spin-chirality rotation acts
as a pseudo-charge U(1) symmetry on these fermions.

The next step is to express the low-energy field theory 
corresponding to model (\ref{eqn:IRth1}) in such a way that 
U(3)-symmetry of our problem is manifest.  
To this end, it is useful to introduce the chiral SU(3)$_1$ 
currents built from the three Dirac fermions:
\begin{equation}
{\cal J}^A_{\text{R/L}} = \sum_{a,b} \Psi_{a,\text{R/L}}^{\dagger} 
T^A_{ab}\, \Psi_{b,\text{R/L}}, 
\label{su3curr}
\end{equation}
where the 3$\times$3-matrices $T^A, A=1,..,8$ are 
the generators of SU(3) in the fundamental representation {\bf 3}, 
which are normalized so that ${\rm Tr}( T^A T^B) = \delta^{AB}/2$. 
Using the identity: 
\begin{equation}
\sum_A T^A_{ab}\, T^A_{ c d}  = \frac{1}{2} \left(
\delta_{a d} \delta_{b c} - \frac{1}{3} 
\delta_{a b} \delta_{c d} \right),
\end{equation}
and the definition (\ref{eqn:Diracfer}), one obtains 
the following identity: 
\begin{multline}
\sum_A {\cal J}^A_{\text{R}} {\cal J}^A_{\text{L}} = 
\frac{1}{8} \left[ \left( \bolxi_\text{R}
{\cdot} \bolxi_\text{L}
- \bolchi_\text{R}{\cdot}\bolchi_\text{L} \right)^2
+
\left( \bolxi_\text{R}{\cdot} \bolchi_\text{L}
+ \bolchi_\text{R} {\cdot}\bolxi_\text{L} \right)^2 \right] \\
- \frac{1}{6}
\left(\bolxi_\text{R}{\cdot} \bolchi_\text{R} \right)
\left(\bolxi_\text{L}{\cdot} \bolchi_\text{L} \right) .       
\label{iden}
\end{multline}
We thus deduce that the low-energy field theory (\ref{eqn:IRth1})
associated to the ray (I) exhibits an exact U(3)-symmetry:
\begin{multline}
{\cal H}_{\text{IR}}^{\text{(I)}} =
- iv \sum_{a=1}^{3}
\left( \Psi_{a,\text{R}}^{\dagger} \partial_x \Psi_{a,\text{R}}
- \Psi_{a,\text{L}}^{\dagger} \partial_x \Psi_{a,\text{L}}
 \right) \\
+ \frac{4 g^{*}}{3} \sum_A {\cal J}^A_{\text{R}} {\cal J}^A_{\text{L}}
\; . 
\label{eqn:u3ham}
\end{multline}
Noting the well-known fact\cite{Gogolin-N-T-book} that the three 
massless Dirac fermions (the first term) can be bosonized 
in terms of a single scalar field and the above $\text{SU(3)}_{1}$ 
currents, we can further recast the above Hamiltonian as
\begin{equation}
\begin{split}
{\cal H}_{\text{IR}}^{\text{(I)}} &=
\frac{v}{2} \left[\left(\partial_x \varphi\right)^2 
+ \left(\partial_x \vartheta\right)^2 \right] \\
& + \frac{\pi v}{2} \sum_{A=1}^{8} \left({\cal J}^A_{\text{R}} 
{\cal J}^A_{\text{R}}
+ {\cal J}^A_{\text{L}} {\cal J}^A_{\text{L}} \right)
+ \frac{4 g^{*}}{3} \sum_{A=1}^{8} {\cal J}^A_{\text{R}} {\cal J}^A_{\text{L}}
\; ,
\end{split}
\label{eqn:u3hamsep}
\end{equation}                                    
where $\varphi$ and $\vartheta$ are the U(1) free bosonic field and 
its dual, respectively.  
There is a `{\em spin-charge separation}' between 
the `charge' sector described by the Tomonaga-Luttinger (TL) model 
(first term in Eq. (\ref{eqn:u3hamsep}))
and the SU(3) non-Abelian `spin'  sector described by the second- 
and third term--% 
the SU(3) Gross-Neveu (GN) model.   
The appearance of SU(3) is not so surprising; 
as has been mentioned in section \ref{sec:Model_and_Sym}, 
we can embed SU(3)-symmetry into the triplet (spin-1) sector of 
the original ladder models.  Therefore, we may conclude that 
the SU(3)-sector describes the dynamics of the (real) spin 
degrees of freedom.  The physical meaning of the remaining 
`charge' sector will be clarified below by using a slightly 
different bosonization scheme.  

As is known from the exact solution\cite{Andrei-L-80}, 
for $g^{*} >0$ a spectral gap (spin gap) 
is formed in the `spin' (SU(3)) sector and 
the low-energy physics is dictated by massive SU(3) spinons 
and antispinons which transform like 
the fundamental representations {\bf 3} and 
$\bar{\text{\bf 3}}$, respectively. 
The low-energy field theory (\ref{eqn:u3hamsep}) displays 
nevertheless a $c=1$ quantum criticality due to 
the decoupled `charge' degrees of freedom which are described
by a massless free boson field (TL model) at the free-fermion point.   
Therefore, the quantum phase transition between the 
staggered dimerized phase and the scalar chirality 
phase generically continuous and belongs to the $c=1$ universality class.   

%%%%%%%%%%%%%%%% Abelain %%%%%%%%%%%%%%%%%%%%%%%%
To clarify the role of the `charge' U(1) ($\varphi$) 
and the `spin' SU(3)-sector, 
we apply the Abelian bosonization to the Hamiltonian 
(\ref{eqn:u3ham}).   
As in Appendix \ref{sec:dictionary}, three bosonic fields are  
introduced to bosonize the three Dirac fermions as follows:
\begin{equation}
\begin{split}
&\Psi_{a,\text{R}} = \frac{{\tilde \kappa}_{a}}{\sqrt{2\pi a_0}}
\exp\left(i \sqrt{4\pi} \phi_{a,\text{R}}\right)  \\
&\Psi_{a,\text{L}} = \frac{{\tilde \kappa}_{a}}{\sqrt{2\pi a_0}}
\exp\left(-i \sqrt{4\pi} \phi_{a,\text{L}}\right)  \; .
\end{split}
\label{eqn:bosofer}
\end{equation}
Since we have adopted the left-right asymmetric definition 
in the above equations to bosonize our three Dirac fermions, 
the spin-chirality rotation simply reads (see Eq.(\ref{eqn:Dirac_U}))
\begin{equation}
 \phi_{a,\text{L/R}} \mapsto \phi_{a,\text{L/R}}
+\frac{\theta}{4\sqrt{\pi}} \quad (a=1,2,3) \; .
\end{equation}
                           
The next step is to switch to a new basis 
where the U(1)- and the SU(3) degrees 
of freedom are separated from each other:
\begin{equation}
\begin{split}
& \varphi = \frac{1}{\sqrt{3}}\left( \phi_1 + 
\phi_2 + \phi_3\right)  \\
& \varphi_\text{s} = \frac{1}{\sqrt{2}}\left( \phi_1 - 
\phi_2 \right)  \\
& \varphi_\text{f} = \frac{1}{\sqrt{6}}\left( \phi_1 + 
\phi_2 - 2 \phi_3\right)  \; .
\end{split}
\label{eqn:su3basis}
\end{equation}
The two bosonic fields $\varphi_\text{s}$ and $\varphi_\text{f}$ 
are compactified fields with special radii $R_{\text{s,f}}$ 
so as to capture the underlying SU(3) symmetry of the problem:
\begin{equation}
\varphi_{\text{s,f}} \sim \varphi_{\text{s,f}} + 2\pi R_{\text{s,f}},
\label{compact}
\end{equation}
with $R_\text{s} = 1/\sqrt{2 \pi}$ and $R_\text{f} = \sqrt{3/2\pi}$.  
The spin-chirality transformation ${\cal U}(\theta)$ affects only 
the `charge' sector:
%%%%%%%%%%%%%%%%%%%%%
\begin{subequations}
\begin{equation}
\varphi \mapsto \varphi + \sqrt{\frac{3}{4\pi}}\, \theta 
\quad , \quad 
\vartheta \mapsto \vartheta \; ,
\end{equation}
while the dual spin-chirality $\widetilde{\cal U}(\tilde{\theta})$ 
changes only the dual field $\vartheta$:
\begin{equation}
\varphi \mapsto \varphi 
\quad , \quad 
\vartheta \mapsto \vartheta + \sqrt{\frac{3}{4\pi}}\, 
\tilde{\theta} \; .
\end{equation}
\end{subequations}
%%%%%%%%%%%%%%%%%%%%%
Now the physical meaning of the `charge' fields $\varphi$ and 
$\vartheta$ is clear; 
the field $\varphi$ (respectively $\vartheta$) describes 
the angular (or, phase) fluctuation of 
the doublet $({\cal O}^{\pi}_{\text{SD}},{\cal O}^{\pi}_{\text{SC}})$ 
(respectively $({\cal O}^{\pi}_{\text{Q}},{\cal O}^{\pi}_{\text{RQ}})$).  
In the bosonization approach, two fields $\varphi$ and 
$\vartheta$ are conjugate (or, dual) to each other.  
Therefore, two doublets $({\cal O}^{\pi}_{\text{SD}},{\cal O}^{\pi}_{\text{SC}})$ 
and $({\cal O}^{\pi}_{\text{Q}},{\cal O}^{\pi}_{\text{RQ}})$ are mutually 
dual objects which are analogous to the Cooper pairs and 
the density-wave operators in strongly-correlated electron 
systems.  

Using this Abelian bosonization, it is then possible
to write down the SU(3) current-current interaction 
in Eq. (\ref{eqn:u3ham}) in terms of the bosonic fields:
\begin{equation}
\begin{split}
\sum_A {\cal J}^A_R {\cal J}^A_L & \\
=&
\frac{1}{2\pi} \left(\partial_x \varphi_{\text{s,R}} 
\; \partial_x \varphi_{\text{s,L}} + 
\partial_x \varphi_{\text{f,R}} \;\partial_x \varphi_{\text{f,L}} \right)
\\
&  \quad 
- \frac{1}{2 \pi^2 a_0^2} \cos\left(\sqrt{2\pi} \;\varphi_{\text{s}}
\right) \cos\left(\sqrt{6\pi} \;\varphi_{\text{f}}\right)  \\
& \qquad
- \frac{1}{4 \pi^2 a_0^2} \cos\left(\sqrt{8\pi} \;\varphi_\text{s}\right)
 \\ 
=& 
\frac{1}{2\pi} \left(\partial_x \varphi_\text{s,R} 
\; \partial_x \varphi_\text{s,L} + 
\partial_x \varphi_\text{f,R} \;\partial_x \varphi_\text{f,L} \right) \\
& \qquad \qquad 
- \frac{1}{8}\left(\left({\cal O}^{\pi}_{\text{SD}}\right)^{2}+ 
\left({\cal O}^{\pi}_{\text{SC}}\right)^{2}
\right) .
\label{eqn:ccbose}
\end{split}
\end{equation}

A straightforward semiclassical analysis of the potential
part of Eq. (\ref{eqn:ccbose}) together with the identification (\ref{compact})
show that the ground state in the SU(3) ``spin'' sector
is three-fold degenerate with expectation values:
\begin{equation}
\begin{split}
\langle \varphi_\text{s} \rangle &= 0 \, , \; 
\langle \varphi_\text{f} \rangle = 0  \\
\langle \varphi_\text{s} \rangle &= \sqrt{\frac{\pi}{2}}\, , \; 
\langle \varphi_\text{f} \rangle = \sqrt{\frac{\pi}{6}} \\
\langle \varphi_\text{s} \rangle &= 0 \, , \; 
\langle \varphi_\text{f} \rangle = 2\sqrt{\frac{\pi}{6}} .
\end{split}
\label{eqn:vevs}
\end{equation}
Now the role of the current-current interaction 
$\sum {\cal J}^{A}_{\text{R}}{\cal J}^{A}_{\text{L}}$ is 
clear.  For the semiclassical vacuum configurations  
$\langle \varphi_{\text{s}}\rangle$ and $\langle \varphi_{\text{f}}\rangle$, 
the modulus of the order-parameter doublet 
$({\cal O}^{\pi}_{\text{SD}},{\cal O}^{\pi}_{\text{SC}})$ is given by 
\[
 \langle \left({\cal O}^{\pi}_{\text{SD}}\right)^{2}+
\left({\cal O}^{\pi}_{\text{SC}}\right)^{2} \rangle 
= \frac{6}{\pi^{2}a_{0}^{2}} \; .
\] 
That is, fluctuations of the doublet in the {\em radial} 
direction is suppressed by the marginally relevant interaction  
$\sum_{A}{\cal J}_{\text{R}}^{A}{\cal J}_{\text{L}}^{A}$.    
The only remaining massless fluctuations in the azimuthal 
direction are described 
by the TL (or, Gaussian) model (the first term in Eq. (\ref{eqn:u3hamsep})).  

It is then interesting to express the order parameters
of Eq. (\ref{orderparadfer}) in terms of these bosonic fields:
\begin{widetext}
%%%%%%%%%%%%%%%%%%%%%%%%%%%
\begin{subequations}
\begin{align}
& {\cal O}^{\pi}_\text{Q} = - \frac{1}{\pi a_0} 
\left[ 2 \cos\left(\sqrt{2\pi} \;\vartheta_\text{s}\right)
\cos\left(\sqrt{4\pi/3} \;\vartheta + 
\sqrt{2\pi/3} \;\vartheta_\text{f} \right)
+ 
\cos\left(\sqrt{4\pi/3} \;\vartheta -
\sqrt{8\pi/3} \;\vartheta_\text{f} \right) \right] 
\label{eqn:orderpara_bose1}  \\
& {\cal O}^{\pi}_\text{SD} = \frac{1}{\pi a_0}
\left[ 2 \cos\left(\sqrt{2\pi} \;\varphi_\text{s}\right)
\cos\left(\sqrt{4\pi/3} \;\varphi +
\sqrt{2\pi/3} \;\varphi_\text{f} \right)
+
\cos\left(\sqrt{4\pi/3} \;\varphi -
\sqrt{8\pi/3} \;\varphi_f \right) \right]  
\label{eqn:orderpara_bose2}  \\  
& {\cal O}^{\pi}_\text{SC} = \frac{1}{\pi a_0}
\left[ 2 \cos\left(\sqrt{2\pi} \;\varphi_\text{s}\right)
\sin\left(\sqrt{4\pi/3} \;\varphi +
\sqrt{2\pi/3} \;\varphi_\text{f} \right)
+
\sin\left(\sqrt{4\pi/3} \;\varphi -
\sqrt{8\pi/3} \;\varphi_\text{f} \right) \right] 
\label{eqn:orderpara_bose3}  \\
& {\cal O}^{\pi}_\text{RQ} = - \frac{1}{\pi a_0}
\left[ 2 \cos\left(\sqrt{2\pi} \;\vartheta_\text{s}\right)
\sin\left(\sqrt{4\pi/3} \;\vartheta +
\sqrt{2\pi/3} \;\vartheta_\text{f} \right)
+
\sin\left(\sqrt{4\pi/3} \;\vartheta -
\sqrt{8\pi/3} \;\vartheta_\text{f} \right) \right].
\label{eqn:orderpara_bose4}
\end{align}
\end{subequations}
%%%%%%%%%%%%%%%%%%%%%%%%%%%%%%%%
\end{widetext}
It is straightforward to observe that 
for the type-(I) self-dual model, i.e. 
for the model described by the field theory (\ref{eqn:IRth1}),
one has $\langle {\cal O}^{\pi}_\text{D} \rangle = 
\langle {\cal O}^{\pi}_\text{SD} \rangle =
\langle {\cal O}^{\pi}_\text{SC} \rangle =
\langle {\cal O}^{\pi}_\text{RD} \rangle =0$ due 
to the quantum-criticality  of the 
``charge'' degrees field in Eq. (\ref{eqn:u3hamsep}).
From Eqs. (\ref{eqn:orderpara_bose1}-\ref{eqn:orderpara_bose4}),
we also deduce that the first doublet ${\cal O}^{\pi}_{\text{SD}}$ 
and ${\cal O}^{\pi}_{\text{SC}}$ 
has correlation functions decaying as $x^{-2/3}$, 
i.e. has quasi long-range coherence, 
whereas the second one ${\cal O}^{\pi}_{\text{Q}}$ and 
${\cal O}^{\pi}_{\text{RQ}}$ 
is exponentially decaying due to strong quantum fluctuations.
The situation is completely reversed for the second (type-(II)) 
self-dual model.  
%%%%%%%%%%%%%%%%%%%%%%%%%%%%%%%%%%%%%%%%%%%%%%%%%%%%%%%%%%%%
\subsection{Small deviation from selfdual models}
\label{sec:effect_of_symbreaking}
%%%%%%%%%%%%%%%%%%%%%%%%%%%%%%%%%%%%%%%%%%%%%%%%%%%%%%%%%%%%
Now let us discuss the effect of small deviation 
from the self-dual model (\ref{eqn:IRth1}). 
The deviation may be incorporated by adding the following  
symmetry-breaking perturbation
to the Hamiltonian (\ref{eqn:u3hamsep}): 
\begin{equation}
\begin{split}
{\cal H}_\text{SB} &= \epsilon \left({\cal H}_2 
- {\cal H}_4 \right)  \\
&= \epsilon \left[
\left(\vec{\xi}_\text{R} {\cdot}\vec{\xi}_\text{L} 
- \vec{\chi}_\text{R} {\cdot}\vec{\chi}_\text{L}\right)^2
- \left(\vec{\xi}_\text{R} {\cdot} \vec{\chi}_\text{L} 
+ \vec{\chi}_\text{R} {\cdot}\vec{\xi}_\text{L}\right)^2
\right] .
\label{symmb}
\end{split}
\end{equation}
We can also express this symmetry-breaking perturbation
in terms of the three Dirac fermions using Eq. (\ref{eqn:Diracfer}):
\begin{equation}
{\cal H}_\text{SB} = 2 \epsilon \sum_{a,b} 
\left[ \Psi_{a,\text{R}} \Psi_{a,\text{L}}^{\dagger} 
\Psi_{b,\text{R}} \Psi_{b,\text{L}}^{\dagger} + 
\Psi_{a,\text{R}}^{\dagger} \Psi_{a,\text{L}}
\Psi_{b,\text{R}}^{\dagger} \Psi_{b,\text{L}} \right].
\label{diracpertu}
\end{equation}
The effect of this term can be elucidated 
by means of the Abelian bosonization 
(\ref{eqn:bosofer}) of the Dirac fermions. 
Moving to the basis (\ref{eqn:su3basis}),
we obtain the bosonized form of the 
symmetry-breaking term (\ref{symmb}):
\begin{equation}
\begin{split}
{\cal H}_\text{SB} =& 
\frac{-2 \epsilon}{\pi^2 a_0^2}
\biggl[ \cos\left(\sqrt{2\pi} \;\varphi_{\text{s}}\right)
\cos\left(2\sqrt{4\pi/3} \;\varphi -
\sqrt{2\pi/3} \;\varphi_{\text{f}} \right)  \\
& + \cos\left(2\sqrt{4\pi/3} \;\varphi +
\sqrt{8\pi/3} \;\varphi_{\text{f}} \right) \biggr]  \; .
\end{split}
\label{bospert}
\end{equation}
The stability of the critical line described 
by the model (\ref{eqn:u3hamsep}) with respect
to the small perturbation (\ref{bospert}) with 
$|\epsilon| \ll 1$ can be investigated by
a naive semiclassical analysis:  
for $|\epsilon| \ll 1$, the bosonic fields $\varphi_{\text{s,f}}$ 
are still frozen to one of the ground-state configurations
$(\VEV{\varphi_{\text{s}}}, \VEV{\varphi_{\text{f}}})$ 
and the non-abelian ``spin'' degrees of freedom are still gapfull.

\begin{figure}[h]
%%%%%%%%%%%%%%%%%%%%%%%%%%%%%%%%%%%%%%%%%%%%%%%%%%%%%%
%\begin{widetext}
\begin{center}
\includegraphics[scale=0.5]{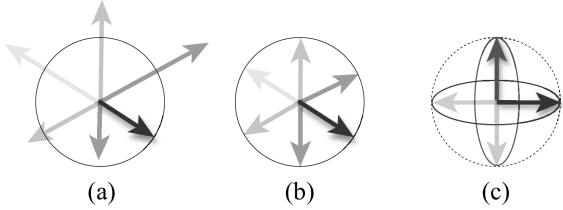}
\caption{Schematic picture illustrating how the doublet of 
order parameters is fixed: (a) At the SU(4) point, the doublet 
fluctuate both in the radial- and angular directions. 
(b) For the self-dual models, the modulus of it is fixed by 
the interactions in the SU(3)-sector, whereas fluctuations in 
the azimuthal direction are still gapless. 
(c) The remaining angular fluctuation gets locked by a small 
deviation from the self-dual models.\label{fig:doublet_fluc}}
\end{center}
%%%%%%%%%%%%%%%%%%%%%%%%%%%%%%%%%%%%%%%%%%%%%%%%%%%%%%%%
%\end{widetext}
%%%%%%%%%%%%%%%%%%%%%%%%%%%%%%%%%%%%%%%%%%%%%%%%%%%%%%%%
\end{figure}

The only difference from the previous case is that 
here we have couplings between $\varphi$ and $\varphi_{\text{s,f}}$ 
and  they may shift the (semiclassical) values 
$(\VEV{\varphi_{\text{s}}},\VEV{\varphi_{\text{f}}})$ 
from those obtained for $B=D$ (Eq. (\ref{eqn:vevs})).  
However, as far as the value of $\varepsilon$ is small enough, 
we may expect that $\varphi_{\text{s,f}}$ still are pinned  
at the same values.  
Assuming that $\varphi_{\text{s,f}}$ are locked to, 
for instance, the first set in (\ref{eqn:vevs}), 
we deduce that the U(1) ``charge'' degrees of freedom acquire
now an extra term:
\begin{equation}
{\cal H}_c \simeq
\frac{v}{2} \left[\left(\partial_x \varphi\right)^2
+ \left(\partial_x \vartheta\right)^2 \right]
 - \frac{4 \epsilon}{\pi^2 a_0^2} \cos \left( 
 2\sqrt{4 \pi/3}\ \varphi \right) .
\label{eqn:SG_model}
\end{equation}
The low-energy field theory for the ``charge'' degrees of freedom
takes thus the form of a quantum sine-Gordon model.
The interaction has scaling dimension $\Delta = 4/3 < 2$
so that the perturbation is relevant and the 
``charge'' degrees of freedom acquire a gap.
For $\epsilon > 0$ i.e. $g_{2}>g_{4}$ ($\epsilon < 0$ i.e. 
$g_{2}<g_{4}$), 
the U(1) bosonic field is locked on: 
$\langle \varphi \rangle = 0$ or $\langle \varphi \rangle = \sqrt{3\pi}/2$ 
($\langle \varphi \rangle = \sqrt{3 \pi}/4$ or 
$\langle \varphi \rangle = 3\sqrt{3 \pi}/4$).
Using the identifications 
(\ref{eqn:orderpara_bose1}--\ref{eqn:orderpara_bose4}),
we then deduce:
$\langle {\cal O}^{\pi}_{\text{SD}} \rangle \ne 0$,
and $\langle {\cal O}^{\pi}_{\text{Q}} \rangle =
\langle {\cal O}^{\pi}_{\text{SC}} \rangle =
\langle {\cal O}^{\pi}_{\text{RQ}} \rangle =0$ for $\epsilon > 0$,
i.e. for $g_2 > g_4$, so that one enters the staggered dimerized 
phase SD. 
In contrast, when $\epsilon < 0$, i.e. $g_2 < g_4$,
we have:
$\langle {\cal O}^{\pi}_{\text{SC}} \rangle \ne 0$,
and $\langle {\cal O}^{\pi}_{\text{Q}} \rangle =
\langle {\cal O}^{\pi}_{\text{SD}} \rangle =
\langle {\cal O}^{\pi}_{\text{RQ}} \rangle =0$ and the 
scalar chirality (SC) phase is stabilized by the small
symmetry-breaking term.  
A similar result can be obtained by considering
the second ground state 
$\langle \varphi_{\text{s}} \rangle = \sqrt{\pi/2}, \;
\langle \varphi_{\text{f}} \rangle = \sqrt{\pi/6}$.
In Fig. \ref{fig:doublet_fluc}, we illustrate how the fluctuation  
of the order parameter doublet 
is frozen  as the symmetry is lowered.  

Finally, we remark that a similar approach can  
be applicable also to the competition between 
the quadrumerized phase (Q) and the RQ phase 
described by the low-energy field 
theory (\ref{eqn:IRth2}).   Repeating the same steps as before, 
we see that the theory obtained by making the replacement 
$\varphi \leftrightarrow \vartheta$ in Eq. (\ref{eqn:SG_model}) 
dictates the competition. 
%%%%%%%%%%%%%%%%%%%%%%%%%%%%%%%%%%%%%%%%%%%%%%%%%%%%%%
%%%%%%%%%%%%%%%%%%%%%%%%%%%%%%%%%%%%%%%%%%%%%%%%%%%%%%
%%%%%%%%%%%%%%%%%%%%%%%%%%%%%%%%%%%%%%%%%%%%%%%%%%%%%%%%%
\subsection{On the possibility of umklapp} 
%%%%%%%%%%%%%%%%%%%%%%%%%%%%%%%%%%%%%%%%%%%%%%%%%%%%%%%%%
In the previous subsection, we have seen that 
on the self-dual manifold the enlarged U(1)-symmetry 
forbids the cosine interactions with explicit $\varphi$-dependence 
from appearing in the low-energy Hamiltonian.   
However, we are on a lattice and 
any interactions compatible with both the U(1) symmetry and 
the {\em discrete} lattice symmetries are allowed in principle. 
Now let us discuss briefly about the possibility of a gap 
generation by such umklapp interactions.   

To this end, it is necessary to know how the discrete 1-site 
translation is realized in terms of our bosons.   
Apparently, the 1-site translation concerns only the {\em charge} 
boson $\varphi$ and its dual $\vartheta$.   
In order for the expression of the translation 
(see Eq. (\ref{eqn:1_site_tr}))
\[
 (\vec{\xi}_{\text{L}},\vec{\chi}_{\text{L}},\vec{\xi}_{\text{R}},
\vec{\chi}_{\text{R}}) \mapsto 
(\vec{\xi}_{\text{L}},\vec{\chi}_{\text{L}},-\vec{\xi}_{\text{R}},
-\vec{\chi}_{\text{R}}) 
\] 
written in terms of the Majorana sextet to be translated correctly 
to a bosonic expression, we should take 
\begin{equation}
\varphi \mapsto \varphi + \sqrt{\frac{3\pi}{4}} \quad , \quad 
\vartheta \mapsto \vartheta - \sqrt{\frac{3\pi}{4}} \; .
\end{equation}
From Eqs. (\ref{eqn:orderpara_bose1}--\ref{eqn:orderpara_bose4}),
it is easy to check that it indeed changes the sign of four order parameters 
${\cal O}^{\pi}_{A}$ (note our order parameters are staggered ones). 
Using this, we can obtain a selection rule for the umklapp. 

In what follows, we shall look 
for terms which (i) does not contain $\varphi$-field and are 
(ii) translationally invariant.     
To this end, we note that the charge bosons $\varphi$ and $\vartheta$ 
enter into the expressions of (chiral) Dirac fermions 
$\Psi_{a,\text{L/R}}$ like 
\begin{align*}
\Psi_{a, \text{R}} \sim \be^{i\sqrt{\pi}\left(\varphi_{a}-\vartheta_{a}\right)}
\sim \be^{i\sqrt{\frac{\pi}{3}}(\varphi-\vartheta)+\cdots} \\
\Psi_{a,\text{L}} \sim \be^{-i\sqrt{\pi}\left(\varphi_{a}+\vartheta_{a}\right)}
\sim \be^{-i\sqrt{\frac{\pi}{3}}(\varphi+\vartheta)+\cdots}  \; ,
\end{align*}
where the ellipses denote the contributions from $\varphi_{\text{s}}$ 
($\vartheta_{\text{s}}$) and $\varphi_{\text{f}}$ 
($\vartheta_{\text{f}}$).  
In particular, `Cooper pairs' are U(1)-invariant and 
will take the following form:
\[
 \Psi_{a,\text{R}}\Psi_{b,\text{L}} \sim \be^{-i \sqrt{
\frac{4\pi}{3}} \vartheta +\cdots} \; .
\]
Since an operator 
\[
  \left(\Psi_{a,\text{R}}\Psi_{b,\text{L}}\right)^{N} 
\sim \be^{-i \sqrt{\frac{4\pi}{3}}N \vartheta+\cdots}
\]
acquires a phase $N\pi$ after the 1-site translation, 
the integer $N$ should be even.  This gives us the first constraint. 

However, this is not the whole story.   
In general, interactions constructed this way will contain 
not only $\varphi_{\text{s}}$ and $\varphi_{\text{f}}$  but also 
their duals $\vartheta_{\text{s,f}}$ which are not pinned by 
the interaction $\sum {\cal J}^{A}_{\text{R}}{\cal J}^{A}_{\text{L}}$;  
if these interactions include the dual fields they will be suppressed by 
strong quantum fluctuations 
(as was pointed out by Schulz\cite{Schulz-86}  
in the context of spin chains).   
For this reason, we have to look for 
combinations with (i) $N=$even and (ii) {\em no} explicit 
$\vartheta_{\text{s}}$ ($\vartheta_{\text{f}}$) dependence 
(these interactions will be generated by perturbations 
even if they do not exist in the bare Hamiltonian).    
By a direct enumeration, we checked 
no such interactions appear up to $N=4$.   For $N=6$, we have 
several combinations {\em e.g.} 
\begin{equation}
\begin{split}
& \left(\Psi_{\text{1,R}}\Psi_{1,\text{L}}\right)
\left(\Psi_{2,\text{R}}\Psi_{2,\text{L}}\right)
\left(\Psi_{2,\text{R}}\Psi_{1,\text{L}}\right)
\\
& \left(\Psi_{2,\text{R}}\Psi_{3,\text{L}}\right)
\left(\Psi_{3,\text{R}}\Psi_{2,\text{L}}\right)
\left(\Psi_{3,\text{R}}\Psi_{3,\text{L}}\right) 
\end{split}
\end{equation}
(note that due to the Fermi statistics these expressions should be 
understood as short-distance expansions).   
In general, terms containing both 
$\Psi_{a,\text{R}}\Psi_{b,\text{L}}$ and 
$\Psi^{\dagger}_{c,\text{R}}\Psi^{\dagger}_{d,\text{L}}$ 
are also allowed.  However, up to $N=6$ we did not find such combinations 
satisfying our requirements.  (In principle, we can consider 
interactions made up of $N/2$ pieces of 
 $\Psi_{a,\text{R}}\Psi_{b,\text{L}}$ 
and the same number of $\Psi^{\dagger}_{a,\text{R}}
\Psi^{\dagger}_{b,\text{L}}$.  But they do not contain $\vartheta$-field 
and are not umklapp.)      
    
Therefore, possible umklapp interactions with the lowest dimensions 
will be of the following form: 
\begin{equation}
\cos \left( 6\sqrt{\frac{4\pi}{3}}\vartheta + \text{const.} \right) 
\; .
\label{umklappsdual}
\end{equation}
Around the SU(4) point, it has scaling dimensions larger than 
two and is thus irrelevant; strong renormalization of the Luttinger 
$K$ is needed for the umklapp (\ref{umklappsdual}) to stabilize a gapped phase 
on the self-dual manifold.  
%%%%%%%%%%%%%%%%%%%%%%%%%%%%%%%%%%%%%%%%%%%%%%%%%%%%%%%%%%%%%%
\section{Other quantum criticalities}
%%%%%%%%%%%%%%%%%%%%%%%%%%%%%%%%%%%%%%%%%%%%%%%%%%%%%%%%%%%%%%
\label{sec:other_QCP}
%%%%%%%%%%%%%%%%%%%%%%%%%%%%%%%%%%%%%%%%%%%%%%%%%%%%%%%%%%%%%%
Now let us discuss the properties of the low-energy effective 
Hamiltonian on the asymptotic rays (\ref{eqn:ray-D-SD})--%
(\ref{eqn:ray-RD-SD}) and (\ref{eqn:unknown_ray}) 
which correspond to the transitions among the four dominant phases.   
Combination of the Abelian- and non-Abelian bosonization 
enables us to obtain non-perturbative solutions to the problem. 
Although it looks more complicated to treat the last one 
(\ref{eqn:unknown_ray}), technically it is slightly simpler 
and we will begin with the case of four competing orders 
(\ref{eqn:unknown_ray}).  
%%%%%%%%%%%%%%%%%%%%%%%%%%%%%%%%%%%%%%%%%%%%%%%%%%%%%%%%%%%%%%
\subsection{Case of four competing orders}
%%%%%%%%%%%%%%%%%%%%%%%%%%%%%%%%%%%%%%%%%%%%%%%%%%%%%%%%%%%%%%
On the ray (\ref{eqn:unknown_ray}), the effective action reads as follows:
\begin{equation}
\begin{split}
& {\cal H}_{\text{Q-SD-SC-RQ}} =
{\cal H}_{\text{SO(6)}}- \frac{g^{*}}{2} \left( \left({\cal O}^{\pi}_{\text{SD}}\right)^2 + 
\left({\cal O}^{\pi}_{\text{SC}}\right)^2 \right. \\
& \left. \quad  \qquad \qquad + \left({\cal O}^{\pi}_{\text{Q}}\right)^2 
+ \left({\cal O}^{\pi}_{\text{RQ}}\right)^2 \right) \\
=& - \frac{ iv}{2} \sum_{a=1}^{3}
\left( \xi_{\text{R}}^a \partial_x \xi_{\text{R}}^a 
- \xi_{\text{L}}^a \partial_x \xi_{\text{L}}^a
+ \chi_{\text{R}}^a \partial_x \chi_{\text{R}}^a
- \chi_{\text{L}}^a \partial_x \chi_{\text{L}}^a \right) \\
& \quad + g^{*} \biggl(
\left(
{\vec \xi}_{\text{R}} \cdot {\vec \xi}_{\text{L}} \right)^2
+ \left( {\vec \chi}_{\text{R}} \cdot {\vec \chi}_{\text{L}} 
\right)^2 \\
& \quad \qquad \qquad
+ \left( {\vec \xi}_{\text{R}} \cdot {\vec \chi}_{\text{L}} 
\right)^2
+ \left( {\vec \chi}_{\text{R}} \cdot {\vec \xi}_{\text{L}} 
\right)^2 \biggr) \; ,
\end{split}
\label{eqn:Ham_unknown_ray}
\end{equation}
with $g^{*} > 0$.

As has been mentioned in section \ref{sec:LBF-analysis}, 
the model (\ref{eqn:Ham_unknown_ray}) describes 
the competition between the four different orders of our problem:
the staggered dimerized (SD), 
scalar chiral (SC), 
and the two quadrumerized (i.e. period-4) orders (Q and RQ).   

We apply the SU(3)$\times$U(1) bosonization scheme to our Hamiltonian 
(\ref{eqn:Ham_unknown_ray}).  After some algebra, we obtain  
the following simple Hamiltonian:
\begin{equation}
\begin{split}
& {\cal H}_{\text{Q-SD-SC-RQ}}  = 
\frac{v}{2} \left[ \left(\partial_x \varphi\right)^2 + 
\left(\partial_x \vartheta\right)^2 \right] \\
& + 
\frac{v}{2}\left[ \left(\partial_x {\vec \phi}\right)^2 + 
\left(\partial_x {\vec \theta}\right)^2 \right] \\
& - \frac{g^{*}}{\pi^2} \, \sum_{i=1}^{3} \left[
\cos\left(\sqrt{8\pi} \; {\vec \alpha}_i \cdot {\vec \phi} \right) 
+ \cos\left(\sqrt{8\pi} \; {\vec \alpha}_i \cdot {\vec \theta} \right) 
\right] \; ,
\label{eqn:gensg}
\end{split}
\end{equation}
where $\vec \phi = (\varphi_{\text{s}}, \varphi_{\text{f}})$, 
$\vec \theta = (\vartheta_{\text{s}}, \vartheta_{\text{f}})$, 
and ${\vec \alpha}_i$ are the positive 
roots of the SU(3) algebra: ${\vec \alpha}_1 = (1/2, \sqrt{3}/2)$, 
${\vec \alpha}_2 = (1/2, -\sqrt{3}/2)$, and   
${\vec \alpha}_3 = (1,0)$.  
Eq. (\ref{eqn:gensg}) describes a Lie-algebraic generalization\cite{Boyanovsky-H-91,sierra} 
of the self-dual sine-Gordon models considered 
in Ref. \onlinecite{Lecheminant-G-N-02}.  

From Eq. (\ref{eqn:gensg}), one immediately observes that 
Hamiltonian (\ref{eqn:Ham_unknown_ray}) 
displays U(1) quantum critical behavior 
due to the non-interacting bosonic field $\varphi$.  
The interacting part of (\ref{eqn:Ham_unknown_ray}) 
takes the form of an SU(3) self-dual sine-Gordon model 
with a marginal interaction.  
From the self-duality symmetry $\vec \phi \leftrightarrow \vec \theta$ 
in the SU(3)-part, 
we may naively expect additional critical degrees of freedom 
resulting from this symmetry.  
However, this is not the case. To see this, we first note that 
the interaction part of Eq. (\ref{eqn:Ham_unknown_ray}) 
is written compactly as 
\begin{equation}
\begin{split}
{\cal H}_{\rm int} &= - \frac{g^{*}}{2} \left( {\vec {\xi}}_{\text{R}}
\times {\vec {\xi}}_{\text{R}}
+ {\vec {\chi}}_{\text{R}} \times {\vec {\chi}}_{\text{R}} 
\right) {\cdot}  
\left( {\vec {\xi}}_{\text{L}} \times {\vec {\xi}}_{\text{L}} 
+ {\vec {\chi}}_{\text{L}} \times {\vec {\chi}}_{\text{L}} \right) \\
& = 2g^{\ast}\, \vec{\cal I}_{\text{R}}{\cdot}\vec{\cal I}_{\text{L}} ,
\end{split}
\label{eqn:newint}
\end{equation}
where we have introduced the level-2 SO(3) currents 
${\vec {\cal I}}_{R,L}$ defined by: 
\begin{equation}
{\vec{\cal I}}_{\text{R,L}} \equiv - \frac{i}{2} \left( 
{\vec {\xi}}_{\text{R,L}} 
\times {\vec {\xi}}_{\text{R,L}}
+ {\vec{\chi}}_{\text{R,L}} \times {\vec {\chi}}_{\text{R,L}} \right) \; .
\label{eqn:cursu24}
\end{equation}
It is interesting to note that $\vec{{\cal I}}_{\text{L,R}}$ 
are simply written as 
\begin{equation}
 {\cal I}^{a}_{\text{R,L}}
=\vec{\Psi}^{\dagger}_{\text{R,L}}({\cal S}^{a})
\vec{\Psi}_{\text{R,L}} \; ,
\end{equation}
where $({\cal S}^{a})_{bc}=-i\varepsilon_{abc}$ ($a=x,y,z$) 
is a 3$\times$3 representation of spin 
operators which form an SO(3) subgroup of SU(3).  

By using these SO(3)-currents, we can then rewrite  
the initial Hamiltonian (\ref{eqn:Ham_unknown_ray}),
which governs the competition of the four orders, 
in a current-current form: 
\begin{equation}
\begin{split}
{\cal H}_{\text{Q-SD-SC-RQ}} 
=& \frac{v}{2} \left( \left(\partial_x \varphi\right)^2 +
\left(\partial_x \vartheta\right)^2 \right)  \\
& + \frac{\pi v}{3} \left( {\vec {\cal I}}_{\text{R}}^2 
+ {\vec {\cal I}}_{\text{L}}^2 \right) 
+ 2 g^{*} \; \;{\vec {\cal I}}_{\text{R}} {\cdot} 
{\vec {\cal I}}_{\text{L}} \; .
\end{split}
\label{eqn:finalham_SO3}
\end{equation}
The second part of this Hamiltonian is that of the level-2 
SO(3) WZW model perturbed by an SO(3)-invariant current-current 
interaction and is integrable\cite{Andrei-D-84}.   
Now the relation to the previous model (Eq. (\ref{eqn:u3hamsep})) 
is clear; 
if one adds current interactions coming from the remaining five 
SU(3)-currents, the $\vec{\theta}$-dependent part of (\ref{eqn:gensg}) 
is canceled and the form (\ref{eqn:ccbose}) is recovered.  

For $g^{*} >0$, which is relevant to our problem, 
the second part has a spectral gap and 
a non-trivial structure of massive spinon.
We thus deduce that the initial Hamiltonian (\ref{eqn:Ham_unknown_ray}) 
exhibits only a U(1) quantum criticality 
due to the free boson $\varphi$. 
The transition that results from the competition between the four
orders is thus of a U(1) Gaussian type.
%%%%%%%%%%%%%%%%%%%%%%%%%%%%%%%%%%%%%%%%%%%%%%%%%%%%%%%%%%%%%%
\subsection{Transitions among dominant phases}
%%%%%%%%%%%%%%%%%%%%%%%%%%%%%%%%%%%%%%%%%%%%%%%%%%%%%%%%%%%%%%
The transitions between two dominant phases within the same (spin-chirality) 
doublet (i.e. SD-SC and Q-RQ) have been already discussed 
in the previous section in conjunction with the low-energy 
physics of the selfdual models.  Here we concentrate on 
the transitions between different doublets (e.g. Q-SD).  

Let us consider, for instance, the Q-SD transition 
(Eq. \ref{eqn:ray-D-SD}) and begin with rewriting the interaction term 
$({\cal O}^{\pi}_{\text{Q}})^{2}+ ({\cal O}^{\pi}_{\text{SD}})^{2}$.  
Plugging the bosonized expressions (\ref{eqn:orderpara_bose1}) and 
(\ref{eqn:orderpara_bose2}) into the above, we obtain 
\begin{equation}
\begin{split}
& \left({\cal O}^{\pi}_{\text{Q}}\right)^{2} +
\left({\cal O}^{\pi}_{\text{SD}}\right)^{2}  \\
& =-\frac{1}{\pi} \left[
\left(\partial_{x}\vec{\varphi} \right)^{2}
+\left(\partial_{x}\vec{\vartheta}\right)^{2}\right]
\\
& \qquad + \frac{1}{\pi^{2} a_0^2} \sum_{r=1}^{6}\left[
\cos(\sqrt{8\pi}\, \vec{\beta}_{r}{\cdot}
\vec{\varphi}) +
 \cos(\sqrt{8\pi}\, \vec{\beta}_{r}{\cdot}
\vec{\vartheta}) 
\right] \; ,
\end{split}
\end{equation}
where 
$\vec{\varphi}=(\varphi,\varphi_{\text{s}},\varphi_{f})$, 
$\vec{\vartheta}=(\vartheta,\vartheta_{\text{s}},
\vartheta_{f})$,  
and the last summation here is taken over all six positive roots of 
SU(4): $\vec \beta_{r}=(1/2,\pm\sqrt{3}/2,0)$,
$(1/2,\pm 1/(2\sqrt{3}),\mp \sqrt{2/3})$, 
$(0,1/\sqrt{3},\sqrt{2/3})$, and $(1,0,0)$ normalized so that 
$|\beta_{r}|=1$.   
Therefore, the effective Hamiltonian corresponding to the ray 
(\ref{eqn:ray-D-SD}) reads
\begin{equation}
\begin{split}
{\cal H}_{\text{Q-SD}} & = \frac{v^{\prime}}{2} \left[
\left(\partial_{x}\vec{\varphi} \right)^{2}
+\left(\partial_{x}\vec{\vartheta}\right)^{2}\right]
\\
& \quad - \frac{g^{\ast}}{\pi^{2} a_0^2} \sum_{r=1}^{6}\left[
\cos(\sqrt{8\pi}\, \vec{\beta}_{r}{\cdot}
\vec{\varphi}) +
 \cos(\sqrt{8\pi}\, \vec{\beta}_{r}{\cdot}
\vec{\vartheta}) 
\right] \; ,
\end{split}
\label{eqn:SUN_SSG}
\end{equation}
where we have rescaled the velocity $v$ so that 
$(\partial_{x}\vec{\varphi})^{2}$ terms coming from 
interactions may be absorbed.  If we replace 
SU(3) ($\vec{\alpha}$) by SU(4) ($\vec{\beta}$) and 
$\vec{\phi}$ by $\vec{\varphi}$ in the previous section, 
this is nothing but the second part of (\ref{eqn:gensg}).  
Therefore, the argument here goes similarly to that in the previous 
section; 
the effective Hamiltonian (\ref{eqn:SUN_SSG}) 
can be written here in terms of SO(4)-currents 
${\cal J}_{\text{R,L}}^{ij}$ $(1\leq i < j \leq 4)$ 
as\cite{Lecheminant-T-06}
\begin{equation}
\begin{split}
{\cal H}_{\text{Q-SD}} =&
\frac{\pi v^{\prime}}{4}\sum_{i<j}\bigl[
({\cal J}_{\text{R}}^{ij})^{2}  
+ ({\cal J}_{\text{L}}^{ij})^{2}  
\bigr] \\
& \qquad + 2g^{\ast} \,\, 
\sum_{i<j}{\cal J}_{\text{R}}^{ij}{\cal J}_{\text{L}}^{ij}
 \; . 
\end{split}
\end{equation}
That is, we have obtained the level-2 SO(4) WZW model 
(with central charge $c=3$) 
perturbed by an SO(4)-invariant marginal current-current interaction which
is an integrable field theory \cite{Andrei-D-84}.  
In contrast to the previous case, we have no critical degrees of 
freedom here and we expect that the transition between Q-phase 
and SD-phase is of first order.   
A similar result holds for RQ-SC transition as well.  
%%%%%%%%%%%%%%%%%%%%%%%%%%%%%%%%%%%%%%%%%%%%%%%%%%%%%%%%%%%%%%
\section{Global structure of the phase diagram}
%%%%%%%%%%%%%%%%%%%%%%%%%%%%%%%%%%%%%%%%%%%%%%%%%%%%%%%%%%%%%%
\label{sec:global-str}

In this section, we shall investigate the main effects of the interactions
(${\cal H}_{5,6}$) in Eq. (\ref{eqn:family}) that we have so far neglected in our field-theoretical approach.
In addition, we shall also use variational and strong-coupling analyses to
figure out the global phase diagram of model (\ref{eqn:family}).

%%%%%%%%%%%%%%%%%%%%%%%%%%%%%%%%%%%%%%%%%%%%%%%%%%%%%%%%%%%%%%
\subsection{Effects of ${\cal H}_{5}$}
%%%%%%%%%%%%%%%%%%%%%%%%%%%%%%%%%%%%%%%%%%%%%%%%%%%%%%%%%%%%%%
So far, we have neglected the ${\cal T}$-breaking three-spin 
interaction ${\cal H}_{5}$ of Eq. (\ref{eqn:int_5}).  In this section, we shall give a hand-waving 
argument for its main effect.   
In section \ref{sec:physics_selfdual}, it has been argued that 
the ${\cal T}$-breaking and the appearance of the SD/SC phases 
can be understood as a pinning of the vector doublet 
$({\cal O}^{\pi}_{\text{SD}},{\cal O}^{\pi}_{\text{SC}})$ 
in the spin-chirality plane (see Fig. \ref{fig:doublet_fluc}).  
Now the role of ${\cal H}_{5}$ can be discussed in a similar 
manner by means of a semiclassical argument applied to 
the bosonized Hamiltonian.  

According to Eq. (\ref{eqn:interaction_2}),  
the ${\cal T}$-breaking interaction ${\cal H}_{5}$ 
changes the form of the effective potential as 
\begin{equation}
\begin{split}
& -\frac{g^{\ast}}{6}\left[\left({\cal O}^{\pi}_{\text{SD}}\right)^{2}
+\left({\cal O}^{\pi}_{\text{SC}}\right)^{2}
\right] \\ 
& \qquad \rightarrow 
-\frac{g^{\ast}}{6} \left[\left({\cal O}^{\pi}_{\text{SD}}\right)^{2}
+\left({\cal O}^{\pi}_{\text{SC}}\right)^{2}
\right] -g_{5}{\cal O}^{\pi}_{\text{SD}}{\cal O}^{\pi}_{\text{SC}}
\; .
\end{split}
\end{equation}
In the presence of ${\cal H}_{5}$, the spin-chirality plane 
is no longer isotropic and the principal axes of the ellipsoid 
(see Fig. \ref{fig:doublet_fluc}) are tilted by $\pi/4$ 
(the sign of $g_{5}$ determines the direction 
of the longer axis.).  Then, we may expect that the two order 
parameters ${\cal O}^{\pi}_{\text{SD}}$ and ${\cal O}^{\pi}_{\text{SC}}$ 
simultaneously take non-zero values.  

To make the above argument more quantitative, we add the $g_{5}$-term 
to the interaction (\ref{eqn:ccbose}).  
The interaction takes its minima when 
$\VEV{\varphi_{\text{s}}}$ and $\VEV{\varphi_{\text{f}}}$ are given by 
Eq. (\ref{eqn:vevs}).   Then the expectation values of the $\varphi$-field 
are determined by finding the minima of trigonometric functions. 
For $\VEV{\varphi_{\text{s}}}=\VEV{\varphi_{\text{f}}}=0$, 
$\VEV{\varphi}$ is determined by minimizing 
$-9 g_{5}/2 \sin(\sqrt{16\pi/3}\varphi)$.  
The calculation goes similarly for the other cases as well.  

This analysis 
tells us that the doublet 
$({\cal O}^{\pi}_{\text{SD}},{\cal O}^{\pi}_{\text{SC}})$ 
takes one of the following values when ${\cal T}$-breaking interaction $g_{5}$ 
is present:
\begin{equation}
({\cal O}^{\pi}_{\text{SD}},{\cal O}^{\pi}_{\text{SC}})=
\begin{cases}
\left(\pm \frac{3}{\sqrt{2}\pi a_{0}},
\pm \frac{3}{\sqrt{2}\pi a_{0}}\right) & 
\text{ for }g_{5}>0 \\
\left(\pm \frac{3}{\sqrt{2}\pi a_{0}},
\mp \frac{3}{\sqrt{2}\pi a_{0}}\right) & 
\text{ for }g_{5}<0 \; .
\end{cases}
\end{equation}
The above argument can be easily generalized to the case of 
non-self-dual models.  
%%%%%%%%%%%%%%%%%%%%%%%%%%%%%%%%%%%%%%%%%%%%%%%%%%%%%%%%%%%%%%
\subsection{Effects of ${\cal H}_{6}$}
%%%%%%%%%%%%%%%%%%%%%%%%%%%%%%%%%%%%%%%%%%%%%%%%%%%%%%%%%%%%%%
Now let us discuss the effect of the ${\cal H}_{6}$ interaction 
(Eq. (\ref{eqn:int_6}))
which has been neglected in the preceding analysis.  
As has been mentioned in section \ref{sec:Model_and_Sym}, 
our model (spin) Hamiltonian can be rewritten in terms of 
spin-1 (hardcore) boson $b_{r,a}$ $(a=x,y,z)$ as:
\begin{multline}
{\cal H}= A{\cal H}_{1}+B{\cal H}_{2}+C{\cal H}_{3}+
D{\cal H}_{4}+E{\cal H}_{5}+F{\cal H}_{6}+G{\cal H}_{7} \\
= (B+D) \sum_{r,a}
\left(b_{r,a}^{\dagger}b_{r+1,a}+b_{r+1,a}^{\dagger}b_{r,a}\right) \\
+(B-D)\sum_{r,a}
\left(b_{r,a}^{\dagger}b^{\dagger}_{r+1,a}+b_{r+1,a}b_{r,a}\right)  \\
 + E \sum_{r}\varepsilon_{abc}\left(
b_{r,a}^{\dagger}b_{r+1,b}^{\dagger}b_{r,c}
+b_{r,c}^{\dagger}b_{r,a}b_{r+1,b} + (r\leftrightarrow r+1)
\right) \\
+  \sum_{r}\left[
(A+C)\bolT_{r}{\cdot}\bolT_{r+1}+2C \left(\bolT_{r}{\cdot}
\bolT_{r+1}\right)^{2}\right]  \\
+ G \sum_{r}n_{r}^{\text{B}}n_{r+1}^{\text{B}}
+ \left(-4C + F -\frac{3}{2}G \right)\sum_{r}n_{r}^{\text{B}}
 \; ,
\label{eqn:spin1_tJ}
\end{multline}
where the projection onto occupied states is implied for 
the fourth term (the triplet-triplet interaction).  
From this, it can be easily read off that 
the coupling $F$ of ${\cal H}_{6}$ affects the chemical potential 
of the hardcore boson.  Therefore, for sufficiently large 
values of $|F|$, the system becomes either a carrierless 
insulator (i.e. $n_{r}^{\text{B}}=0$ for all sites $r$) or 
a fully occupied state (i.e. $n_{r}^{\text{B}}=1$ for all $r$).  
Apparently, the ground state of the carrierless insulator 
is trivial and is, in terms of the original spins, 
given by a tensor product of local (rung) singlets. 
The fate of the spin wavefunction in the latter case depends 
strongly on the two couplings $A$ and $C$ which dictate 
the spin-spin interaction between hardcore bosons.   
Here, we will mainly focus on the case 
$A+C>2|C|$ which corresponds to the Haldane phase\cite{Affleck-K-L-T-88}.  

When $|F|$ is decreased, we have quantum phase transitions 
to a conducting state.  Let us first consider a transition from 
the carrierless insulator.  Then, the transition is equivalent 
to the superfluid(SF)-onset transition of spin-1 bosons.  
For the moment, let us neglect the interactions among 
different species of particles (including the hardcore repulsion). 
As is well-known\cite{Sachdev-S-S-94}, 
the bosonic two-body interactions are relevant at the $z=2$ 
SF-onset transition and the system flows toward the strong-coupling 
fixed point.   In the presence of U(1)-breaking anisotropy ($B\neq D$), 
the particle number is no longer conserved and 
the fixed point is replaced by that of the $z=1$ critical 2D Ising 
model\cite{Damle-S-96,Totsuka-98}.  Since we have three copies of 
such Ising models, we may expect that the fixed point here is 
described by the level-2 SU(2) WZW model which is 
equivalent to three massless Majorana fermions with $c=3/2$.  
Now let us estimate the impact of the neglected boson-boson 
interactions.  Since they are given by O(3)-invariant 
(i.e. spin-symmetric) products of four Fermi fields,  
the only possible one should be the marginal interaction of the form 
$\bolJ_{\text{R}}{\cdot}\bolJ_{\text{L}}$ ($\bolJ_{\text{L,R}}$ 
are the level-2 SU(2)-currents of the WZW model).  
Therefore, we may conclude that the SF-onset transition from 
the rung-singlet phase ({\em carrierless insulator}) to 
the staggered dimer- (SD) or the scalar-chirality (SC) phase 
belongs to the level-2 SU(2) WZW criticality; 
the SD- or SC order appears as a consequence of the SF ordering 
in the presence of the U(1)-breaking anisotropy.   
This is consistent with the results of weak-coupling 
approach\cite{Nersesyan-T-97,Muller-V-M-02,gritsev} 
and numerical analysis\cite{Hijii-Q-N-03}.    
Since the number of the bosonic particles is conserved, 
the transition at $B=D$ (self-dual) is exceptional and 
belongs essentially to 
the Japaridze-Nersesyan-Pokrovsky-Talapov 
(JNPT) universality class\cite{Japaridze-N-P-T-79} 
(note that we have three copies of the JNPT criticality here).  

\begin{figure}[h]
\begin{center}
\includegraphics[scale=0.3]{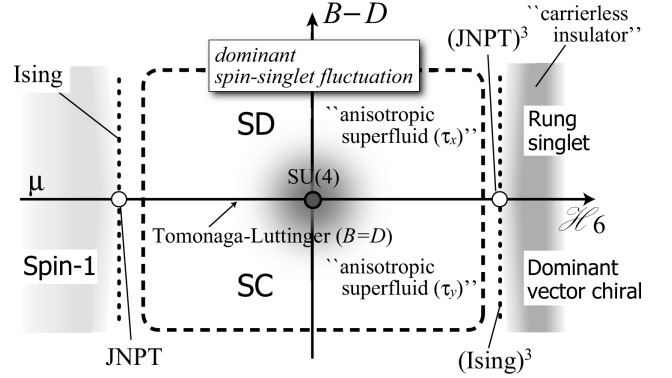}
\end{center}
\caption{A schematic picture illustrating the effects of 
${\cal H}_{6}$ and the anisotropy $\varepsilon=B-D$ 
in the spin-chirality plane 
on the global structure of the phase diagram of our model.  
According to the sign of the anisotropy, we have 
two different orderings; an anisotropic SF in the $x$-direction 
(${\cal T}$-even) for $\varepsilon>0$ and one in the $y$-direction 
(${\cal T}$-odd) for $\varepsilon<0$.  
The duality symmetry ${\cal D}$ interchanges the region with 
$\varepsilon >0$ and that with $\varepsilon <0$.    
On both sides of the region enclosed by a dashed line 
({\em dominant singlet fluctuations}), 
we find conventional phases (spin-1 or rung-singlet), which can 
be interpreted as insulating ones. 
\label{fig:global_phase_diag}}
\end{figure}

A similar argument applies to the transition from 
the fully-occupied state as well.    
The elementary particle here is a rung singlet moving in 
the sea of rung triplets (note that we have assumed that 
we are in the spin-gapped phase and the magnon excitations occur 
at the energy scale of the `Haldane' gap).  Then the dynamics of 
these excited singlets may be modeled by the interacting 
spinless fermion.  The sign of the fermion-fermion interaction 
will be positive (negative) when $G$ is much larger (smaller) 
than $A$ and $C$.  When $A$ and $C$ are not so large and 
the anisotropy is absent ($B=D$), the $F$-driven transition 
is of the JNPT universality class again.  
The $xy$-anisotropy ($B\neq D$) alters the transition to that of 
the Ising-type.   
We show a schematic phase diagram illustrating the effects 
of ${\cal H}_{5}$ and ${\cal H}_{6}$ in Fig.\ref{fig:global_phase_diag}.
%%%%%%%%%%%%%%%%%%%%%%%%%%%%%%%%%%%%%%%%%%%%%%%%%%%%%%%%%%%%
%%%%%%%%%%%%%%%%%%%%%%%%%%%%%%%%%%%%%%%%%%%%%%%%%%%%%%%%%%%%%%
%%%%%%%%%%%%%%%%%%%%%%%%%%%%%%%%%%%%%%%%%%%%%%%%%%%%%%%%%%%%%%
\subsection{XXZ-analogy and pseudo-spin description}
\label{sec:XXZ_analogy}
%%%%%%%%%%%%%%%%%%%%%%%%%%%%%%%%%%%%%%%%%%%%%%%%%%%%%%%%%%%%%%
In order to demonstrate the relevance of the pseudo-spin picture 
(see Section II D 1), 
we present two approaches which are complementary to 
that developed in the previous sections.  In both approaches, 
the key is how to identify the pseudo-spin degrees of freedom 
on the lattice.   
%%%%%%%%%%%%%%%%%%%%%%%%%%%%%%%%%%%%%%%%%%%%%%%%%%%%%%%%%%%%%%
\subsubsection{Variational approach}
%%%%%%%%%%%%%%%%%%%%%%%%%%%%%%%%%%%%%%%%%%%%%%%%%%%%%%%%%%%%%%
Let us begin with the self-dual models.   
In the above, we have seen how the U(1) Gaussian model emerges 
as the low-energy effective field theory after the spin sector 
is gapped.  The spin and the (pseudo)charge decouple 
from each other at low energies and 
the physics of the spin sector may be described 
by the following bilinear-biquadratic interaction 
(see section \ref{sec:Model_and_Sym} and 
Appendix \ref{sec:U1_SU3} for more details)
\[
A{\cal H}_{1}+C{\cal H}_{3}\sim \sum\left[(A+C)\bolT_{r}
{\cdot}\bolT_{r+1}+2C (\bolT_{r}{\cdot}\bolT_{r+1})^{2}\right]
\; .
\]
For this reason, we may expect that the Haldane-gap physics {\em a la} 
Affleck-Kennedy-Lieb-Tasaki\cite{Affleck-K-L-T-88} dominates 
in a reasonably large region of the phase diagram.   
Of course, we have the additional pseudo-charge degrees of 
freedom (i.e. motion of triplet rungs in the rung-singlet background) 
here and the stability of the Haldane state is not so obvious.  
However, from a simple argument\cite{Zhang-A-89} we know that 
the valence-bond-solid state proposed in 
Ref. \onlinecite{Affleck-K-L-T-88} is stable against the motion 
of singlet rungs (or, holes in the spin-1 $t$-$J$ model given 
by Eqs. (\ref{eqn:tJ_hop})-(\ref{eqn:tJ_bb}) or 
Eq. (\ref{eqn:spin1_tJ})).   
In fact, numerical simulations carried out for the spin-1 
bosonic $t$-$J$ model\cite{Nishiyama-S-96} 
(the set of parameters $B=D=t/2$, $A=J$, $C=0$ was used there) 
show the existence of a finite spin gap and 
the Luttinger-liquid behavior in the pseudo-charge sector, 
which is consistent with our conclusion in section \ref{sec:SU3_U1}.  

Now that we know that the spin sector is {\em generically} gapped 
and decoupled from the pseudo-charge sector, the next question 
would be what kind of degrees of freedom determines the global 
structure of the phase diagram.   
From the bosonization analysis presented above, 
a natural guess would be the pseudo-spin ${\cal S}=1/2$ 
(or `charge' in the $t$-$J$ language).  
As was described in section \ref{sec:useful_analogy}, 
${\cal H}_{2,4}$ correspond to the pseudo-spin-flipping 
processes and ${\cal H}_{6}$ and ${\cal H}_{7}$ to 
magnetic field and the ${\cal S}^{z}{\cal S}^{z}$-interaction, 
respectively.   
The remaining parts concern dynamics in the (true)spin sector, 
which is separated from the low-energy sector by a finite spin gap.  

\begin{figure}[h]
\begin{center}
\includegraphics[scale=0.6]{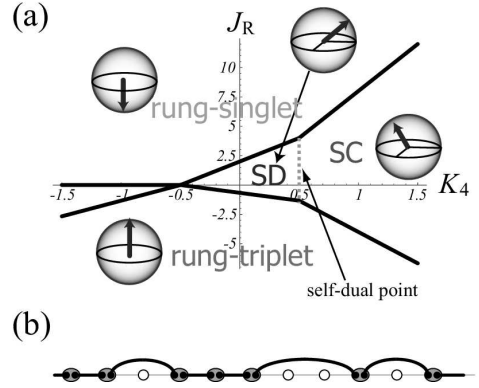}
\end{center}
\caption{(a) Variational phase diagram obtained by the ansatz 
(\ref{eqn:variational_wf}). 
Shown is the `pseudo-spin' direction $(\phi,\theta)$ for each phase: 
conventional rung-singlet- and rung-triplet (Haldane or VBS) phases 
are {\em spin-polarized states} 
while the SD- and SC phases may be viewed as {\em canted states} 
with different XY-projections. 
(b) Typical spin-singlet configuration contained in our 
variational wave function (\ref{eqn:variational_wf}). 
Ovals and open circles respectively denote triplet- (`occupied sites') 
and singlet (`unoccupied sites') rungs. 
The spin-1 bosons totally form a spin-singlet valence-bond 
state. 
Due to motion and pair creation/annihilation of triplet bosons, 
the wave function (\ref{eqn:variational_wf}) 
describes a strongly fluctuating state. 
\label{fig:var_phase_diag}}
\end{figure}

To develop a variational theory for this kind of 
spin liquid, we first construct a {\em coherent state} of 
the pseudo-spin ${\cal S}$ by combining a local singlet 
$|\text{s}\rangle_{r}$ and a triplet $|\text{\bf t}\rangle_{r}$:
\[
 |\boldsymbol{\Omega}\rangle^{\prime} 
= \bigotimes_{r\in \text{rung}}
\left[ \be^{-i\frac{\phi(r)}{2}}\cos\frac{\theta(r)}{2}
\left(\be^{i\phi(r)}\tan\frac{\theta(r)}{2}|\text{s}\rangle_{r} 
+ |\text{\bf t}\rangle_{r} \right)\right] \; ,
\]
where $\phi(r)$ and $\theta(r)$ are respectively azimuthal- and polar 
angles of the spin vector $\boldsymbol{\Omega}_{r}$ at rung $r$.  
Apparently, this is unsatisfactory because we still have 
local spin degrees of freedom represented by 
$|\text{\bf t}\rangle_{r}$.  To kill the spin degrees of freedom 
and construct a spin-gapped wave function out of 
the above coherent states, the most natural way 
would be to replace triplet states $|\text{\bf t}\rangle_{r}$ 
on $r$-th rung by the following 
2$\times$2-matrix\cite{Klumper-S-Z-92,Totsuka-S-95}:
\[
 \boldsymbol{g}_{r} \equiv \frac{1}{\sqrt{3}}
\begin{pmatrix} |\text{t}_{0}\rangle_{r} & 
-\sqrt{2}|\text{t}_{1}\rangle_{r} \\
+\sqrt{2}|\text{t}_{-1}\rangle_{r} & -|\text{t}_{0}\rangle_{r} 
\end{pmatrix}
\] 
($|\text{t}_{a}\rangle$ is the rung-triplet state with $\text{T}^{z}=a$) 
and use the modified ansatz:
\begin{equation}
|\boldsymbol{\Omega}\rangle 
= \bigotimes_{r\in \text{rung}}\left(
\sqrt{3} \, \be^{i\phi(r)}\tan\frac{\theta(r)}{2}|\text{s}\rangle_{r} 
\text{\bf 1} 
+ \boldsymbol{g}_{r} \right)  \; . 
\label{eqn:variational_wf}
\end{equation}
Note that this is essentially the same as the building block 
of the matrix-product ground state adopted 
in Ref. \onlinecite{Kolezhuk-M-98}.   
Taking the trace of the above matrix product, we can obtain 
the desired (unnormalized) spin-singlet wave function 
with a finite correlation length.  
This wave function may be thought of as the valence-bond-solid 
(VBS) state\cite{Affleck-K-L-T-88} randomly `diluted' 
by vacancies (see Fig.\ref{fig:var_phase_diag}(b)).   

To illustrate the usefulness of our pseudo-spin picture, 
we map out the variational phase diagram of the two-leg 
spin ladder with four-spin cyclic exchange 
${\cal H}_{\text{ladder}+\text{4-spin}}$ 
(see Eq. (\ref{eqn:4-spin-ladder})).  
As a variational ansatz, we choose 
\begin{equation}
\left(\phi(r),\theta(r)\right) = 
\begin{cases}
\left(\phi_{1},\theta_{1}\right) & \text{ for $r$=even}   \\
\left(\phi_{2},\theta_{2}\right) & \text{ for $r$=odd} \; .
\end{cases}
\end{equation}
Thanks to the special form of $|\Omega\rangle$, 
we can reduce the computation of the ground-state expectation 
values $\langle\Omega|{\cal H}_{\text{ladder+4-spin}}|\Omega\rangle$ 
to that of 4$\times$4 matrices\cite{Klumper-S-Z-92,%
Totsuka-S-95} and we obtain the variational energy 
$E_{\text{var}}(\phi_{1},\theta_{1},\phi_{2},\theta_{2})$ 
to minimize.  Since the explicit form of 
$E_{\text{var}}(\phi_{1},\theta_{1},\phi_{2},\theta_{2})$ is 
unimportant, we only show the resulting phase diagram 
in Fig. \ref{fig:var_phase_diag}.   
%%%%%%%%%%%%%%%%%%%%%%%%%%%%%%%%%%%%%%%%%%%%%%%%%%%%%%%%%%%%%%   
%%%%%%%%%%%%%%%%%%%%%%%%%%%%%%%%%%%%%%%%%%%%%%%%%%%%%%%%%%%%%%
It should be compared with the phase diagram of the same model 
obtained by large-scale numerical simulations\cite{Lauchli-S-T-03} (DMRG).  
The path searched in Ref. \onlinecite{Lauchli-S-T-03} 
corresponds to the line $J_{\text{R}}=1$  
in Fig. \ref{fig:var_phase_diag}.   Three dominant phases 
(rung-singlet, SD, and SC) found there appear 
in our variational phase diagram as well, 
while the phase denoted by `dominant vector chirality'
\cite{Hikihara-M-H-03,Lauchli-S-T-03}, 
whose nature is not clear currently, is missing in ours.  
%%%%%%%%%%%%%%%%%%%%%%%%%%%%%%%%%%%%%%%%%%%%%%%%%%%%%%%%%%%%%

Similarly, the ferromagnetic phase is beyond the scope of 
our simple variational calculation since our variational 
wave function (\ref{eqn:variational_wf}) 
contains only spin-singlet phases.     
In our picture, both rung-triplet- and rung-singlet phases 
correspond to {\em (pseudo)spin-polarized states} as is shown in 
Fig. \ref{fig:var_phase_diag}.   The spin-chirality transformation 
${\cal U}(\theta)$ rotates the pseudo-spin vector along 
the $z$-axis.  When the external magnetic field (${\cal H}_{6}$) 
is not very strong, {\em spin-canted states} (SD and SC) 
are favored by the $xy$-coupling ${\cal H}_{2,4}$; 
in the SD (SC) phase, the $xy$-projection of the pseudo-spin 
lies in the $x$($y$)-direction.   
%%%%%%%%%%%%%%%%%%%%%%%%%%%%%%%%%%%%%%%%%%%%%%%%%%%%%%%%%%%%%%
\subsubsection{Strong-coupling approach}
\begin{figure}[h]
\begin{center}
\includegraphics[scale=0.4]{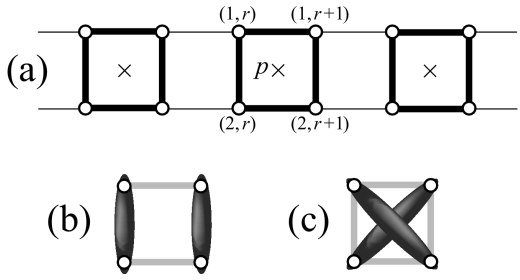}
\end{center}
\caption{(a): A two-leg ladder as a weakly-coupled 
plaquettes. Strongly-coupled plaquettes, on which 
pseudo-spin 1/2s and coarse-grained spin-1s are defined, 
are shown by thick lines.  
All interactions (both two-spin and four-spin) connecting 
these plaquettes are multiplied by $\lambda (\ll 1)$.  
(b) and (c): Two triplet states which constitute 
a (nearly) degenerate ground-state manifold in the vicinity 
of the SU(4) point.  Two spin-1/2s contained in each oval 
form a spin triplet.  In both (b) and (c), two 
triplets (ovals) are anti-symmetrized to form a total spin triplet 
on a plaquette.  
\label{fig:plaquette}}
\end{figure}

%%%%%%%%%%%%%%%%%%%%%%%%%%%%%%%%%%%%%%%%%%%%%%%%%%%%%%%%%%%%%%
Another way to see the role of the pseudo-spin degrees of freedom 
would be a strong-coupling expansion.  
A natural starting point might be isolated rungs (as in the usual 
ladder systems).  Unfortunately, however, all the phases that we have 
found in the field-theory analysis break the periodicity of 
the original Hamiltonian and the limit of isolated rungs is 
not quite helpful (the phases accessible from the limit would be 
more or less trivial).  Instead, we divide the whole lattice 
into plaquettes (see Fig. \ref{fig:plaquette}(a)) 
and introduce a new coupling 
$\lambda$ which controls the coupling between neighboring 
plaquettes.  On these plaquettes, a kind of {\em coarse-grained} 
degrees of freedom will be defined by which the low-energy sector 
of our ladder system may be described.  

To explore the vicinity of the SU(4) point, 
we adopt the following Hamiltonian:
\begin{equation}
\begin{split}
{\cal H} &= \sum_{r=\text{rung}}J_{r}\bigl\{
h_{1}(r)+(1+\delta_{2})h_{2}(r)+(1+\delta_{3})h_{3}(r) \\
& \qquad +(1+\delta_{4})h_{4}(r)
\bigr\}
+\delta_{6} \sum_{r}\bolS_{1,r}{\cdot}\bolS_{2,r}
\; ,
\end{split}
\end{equation}
where the couplings $J_{r}$ alternate like 
\begin{equation}
J_{r} = 1 \quad (\text{for $r$ even}) \quad , \quad 
J_{r} = \lambda (\ll 1) \quad (\text{for $r$ odd}) 
\end{equation} 
and $h_{i}(r)$ denote local Hamiltonians obtained by 
restricting ${\cal H}_{i}$ to two rungs $r$ and $r+1$.  
%%%%%%%%%%%%%%%%%%%%%%%%%%%%%%%%%%%%%%%%%%%%%%%%%%%%%%%%%%%%%%%%
%%%%%%%%%%%%%%%%%%%%%%%%%%%%%%%%%%%%%%%%%%%%%%%%%%%%%%%%%%%%%%

Let us begin with the case of $\lambda=0$, i.e. 
the limit of isolated plaquettes.  
On a single plaquette, we have 16 states which can be decomposed into 
the following multiplets:
\begin{equation}
({\bf 0})^{2}\oplus({\bf 1})^{3}\oplus({\bf 2}) \; .
\end{equation}
Around the SU(4) point, two of the three triplets 
(see Fig. \ref{fig:plaquette}(b,c)) form 
nearly degenerate ground states (exactly at the SU(4) point, 
they are completely degenerate).  
The explicit forms of these states are given as 
\begin{subequations}
%%%%%%%%%
\begin{equation}
\bigl| \,
\begin{pspicture}[0.35](0,0)(0.5,0.5)
\psline[doubleline=true](0,0.0)(0.5,0.5)
\psline[doubleline=true](0,0.5)(0.5,0.0)
\psline[linestyle=dashed](0,0.0)(0.5,0.0)
\psline[linestyle=dashed](0.0,0.5)(0.5,0.5)
\psline[linestyle=dashed](0,0)(0.0,0.5)
\psline[linestyle=dashed](0.5,0.0)(0.5,0.5)
\end{pspicture} 
\;
\bigr\rangle 
= \frac{1}{\sqrt{2}}\left(
%%%%%%%%%%%%%%%%%
\bigl| \, 
\begin{pspicture}[0.35](0,0)(0.5,0.5)
\psline{->}(0,0.5)(0,0)
\psline[doubleline=true](0.5,0.5)(0.5,0)
\end{pspicture} 
\;
\bigr\rangle 
-
 \bigl| \, 
\begin{pspicture}[0.35](0,0)(0.5,0.5)
\psline[doubleline=true](0,0)(0,0.5)
\psline{->}(0.5,0.5)(0.5,0)
\end{pspicture} 
\;
\bigr\rangle 
%%%%%%%%%%%%%%%%%
\right)
= \frac{1}{\sqrt{2}}\left(
%%%%%%%%%%%%%%%%%
 \bigl| \, 
\begin{pspicture}[0.35](0,0)(0.5,0.5)
\psline{->}(0,0.5)(0.5,0.5)
\psline[doubleline=true](0.0,0.0)(0.5,0)
\end{pspicture} 
\;
\bigr\rangle 
-
 \bigl| \, 
\begin{pspicture}[0.35](0,0)(0.5,0.5)
\psline[doubleline=true](0,0.5)(0.5,0.5)
\psline{->}(0.0,0.0)(0.5,0)
\end{pspicture} 
\;
\bigr\rangle 
%%%%%%%%%%%%%%%%%
\right)
\end{equation}
%%%%%%%%%%%%%%%%%%%%%%%%
\begin{equation}
\bigl| \,
\begin{pspicture}[0.35](0,0)(0.5,0.5)
\psline[doubleline=true](0.0,0.0)(0.0,0.5)
\psline[doubleline=true](0.5,0.0)(0.5,0.5)
\psline[linestyle=dashed](0,0.0)(0.5,0.0)
\psline[linestyle=dashed](0.0,0.5)(0.5,0.5)
\psline[linestyle=dashed](0,0)(0.0,0.5)
\psline[linestyle=dashed](0.5,0.0)(0.5,0.5)
\end{pspicture} 
\;
\bigr\rangle 
= \frac{1}{\sqrt{2}}\left(
%%%%%%%%%%%%%%%%%
 \bigl| \, 
\begin{pspicture}[0.35](0,0)(0.5,0.5)
\psline[doubleline=true](0,0.5)(0,0)
\psline[doubleline=true](0.5,0.5)(0.5,0)
\end{pspicture} 
\;
\bigr\rangle 
-
 \bigl| \, 
\begin{pspicture}[0.35](0,0)(0.5,0.5)
\psline[doubleline=true](0,0)(0,0.5)
\psline[doubleline=true](0.5,0.5)(0.5,0)
\end{pspicture} 
\;
\bigr\rangle 
%%%%%%%%%%%%%%%%%
\right)
= \frac{1}{\sqrt{2}}\left(
%%%%%%%%%%%%%%%%%
 \bigl| \, 
\begin{pspicture}[0.35](0,0)(0.5,0.5)
\psline{->}(0,0.5)(0.5,0.5)
\psline[doubleline=true](0.0,0.0)(0.5,0)
\end{pspicture} 
\;
\bigr\rangle 
+
 \bigl| \, 
\begin{pspicture}[0.35](0,0)(0.5,0.5)
\psline[doubleline=true](0,0.5)(0.5,0.5)
\psline{->}(0.0,0.0)(0.5,0)
\end{pspicture} 
\;
\bigr\rangle 
%%%%%%%%%%%%%%%%%
\right) \; ,
\end{equation}
\end{subequations}
%%%%%%%%%%%%%%%%%%%%%%%%
where arrows and double lines in the above 
respectively denote spin singlets and triplets (ovals 
in Fig. \ref{fig:plaquette}).  

It should be remarked that we shall construct an effective theory 
in terms of new degrees of freedom (spin-1 and species of 
triplets) defined on these plaquettes.  
In this plaquette picture, our ladder is already dimerized and  
the two staggered phases (SD and SC) should be understood 
as {\em uniform} ones.  Correspondingly, we shall introduce 
order parameters defined on each plaquette:  
\begin{equation}
\begin{split}
& {\cal O}_{\text{SD}}(p) = 
\bolS_{1,r}{\cdot}\bolS_{1,r+1} 
-\bolS_{2,r}{\cdot}\bolS_{2,r+1}  \\
& {\cal O}_{\text{SC}}(p)  
= \left(\bolS_{1,r}\!\times\!\bolS_{2,r}\right){\cdot}
\left(\bolS_{1,r+1}+\bolS_{2,r+1}\right) + (r \leftrightarrow r+1) \; ,
\end{split}
\end{equation}
where $p$ labels the plaquette formed by 
four points $(1,r)$, $(2,r)$, $(2,r+1)$, and $(1,r+1)$ 
(see Fig. \ref{fig:plaquette}).  

The point here is that the eigenstates 
of ${\cal O}_{\text{SD}}$ and ${\cal O}_{\text{SC}}$ can be 
constructed out of the two triplets shown in 
Fig. \ref{fig:plaquette}:
\begin{equation}
\begin{split}
& |\pm 1 \rangle_{\text{SD}} =
\frac{1}{\sqrt{2}} \left(
%%%%%%%%%%%%%%%%%
 \bigl| \,
\begin{pspicture}[0.35](0,0)(0.5,0.5)
\psline[doubleline=true](0.0,0.0)(0.0,0.5)
\psline[doubleline=true](0.5,0.0)(0.5,0.5)
\psline[linestyle=dashed](0,0.0)(0.5,0.0)
\psline[linestyle=dashed](0.0,0.5)(0.5,0.5)
\psline[linestyle=dashed](0,0)(0.0,0.5)
\psline[linestyle=dashed](0.5,0.0)(0.5,0.5)
\end{pspicture} 
\;
\bigr\rangle 
\mp 
 \bigl| \,
\begin{pspicture}[0.35](0,0)(0.5,0.5)
\psline[doubleline=true](0,0.5)(0.5,0.0)
\psline[doubleline=true](0,0.0)(0.5,0.5)
\psline[linestyle=dashed](0,0.0)(0.5,0.0)
\psline[linestyle=dashed](0.0,0.5)(0.5,0.5)
\psline[linestyle=dashed](0,0)(0.0,0.5)
\psline[linestyle=dashed](0.5,0.0)(0.5,0.5)
\end{pspicture} 
\;
\bigr\rangle 
%%%%%%%%%%%%%%%%%
\right) 
\\
& |\pm 1 \rangle_{\text{SC}} =
\frac{1}{\sqrt{2}} \left(
%%%%%%%%%%%%%%%%%
 \bigl| \,
\begin{pspicture}[0.35](0,0)(0.5,0.5)
\psline[doubleline=true](0.0,0.0)(0.0,0.5)
\psline[doubleline=true](0.5,0.0)(0.5,0.5)
\psline[linestyle=dashed](0,0.0)(0.5,0.0)
\psline[linestyle=dashed](0.0,0.5)(0.5,0.5)
\psline[linestyle=dashed](0,0)(0.0,0.5)
\psline[linestyle=dashed](0.5,0.0)(0.5,0.5)
\end{pspicture} 
\;
\bigr\rangle 
\mp i \,
 \bigl| \,
\begin{pspicture}[0.35](0,0)(0.5,0.5)
\psline[doubleline=true](0,0.5)(0.5,0.0)
\psline[doubleline=true](0,0.0)(0.5,0.5)
\psline[linestyle=dashed](0,0.0)(0.5,0.0)
\psline[linestyle=dashed](0.0,0.5)(0.5,0.5)
\psline[linestyle=dashed](0,0)(0.0,0.5)
\psline[linestyle=dashed](0.5,0.0)(0.5,0.5)
\end{pspicture} 
\;
\bigr\rangle 
%%%%%%%%%%%%%%%%%
\right) \; ,
\end{split}
\label{eqn:eigenstates_OP}
\end{equation}
where $\pm 1$ denote the eigenvalues of 
${\cal O}_{\text{SD,SC}}(p)$.   

The next step is to identify the states having 
the pseudo-spin ($\boldsymbol{\tau}$) up and down.  
To this end, we consider the coherent state of a spin-1/2:
\[
 | \boldsymbol{\Omega}\rangle = 
\cos \frac{\theta}{2}|\uparrow \rangle +
\be^{+i \phi} \sin \frac{\theta}{2}|\downarrow \rangle \; .
\] 
If we identify the following eigenstates of 
$\tau^{x}$ and $\tau^{y}$ 
\begin{equation}
\begin{split}
& |\tau^{x}=\pm 1/2 \rangle = \frac{1}{\sqrt{2}}\left(
|\uparrow \rangle \pm |\downarrow \rangle
\right) 
\\
& |\tau^{y}=\pm 1/2 \rangle = \frac{1}{\sqrt{2}}\left(
|\uparrow \rangle \pm i\, |\downarrow \rangle
\right) 
\end{split}
\end{equation}
with the eigenstates (\ref{eqn:eigenstates_OP}) of the order parameters 
(${\cal O}_{\text{SD}}$ and ${\cal O}_{\text{SC}}$), 
it is suggested that we should take
\begin{subequations}
\begin{equation}
| \uparrow \rangle  \leftrightarrow 
 \bigl| \,
\begin{pspicture}[0.35](0,0)(0.5,0.5)
\psline[doubleline=true](0.0,0.0)(0.0,0.5)
\psline[doubleline=true](0.5,0.0)(0.5,0.5)
\psline[linestyle=dashed](0,0.0)(0.5,0.0)
\psline[linestyle=dashed](0.0,0.5)(0.5,0.5)
\psline[linestyle=dashed](0,0)(0.0,0.5)
\psline[linestyle=dashed](0.5,0.0)(0.5,0.5)
\end{pspicture} 
\;
\bigr\rangle 
\quad , \quad 
| \downarrow \rangle  \leftrightarrow 
- \bigl| \,
\begin{pspicture}[0.35](0,0)(0.5,0.5)
\psline[doubleline=true](0,0.5)(0.5,0.0)
\psline[doubleline=true](0,0.0)(0.5,0.5)
\psline[linestyle=dashed](0,0.0)(0.5,0.0)
\psline[linestyle=dashed](0.0,0.5)(0.5,0.5)
\psline[linestyle=dashed](0,0)(0.0,0.5)
\psline[linestyle=dashed](0.5,0.0)(0.5,0.5)
\end{pspicture} 
\;
\bigr\rangle  
\end{equation}
and 
\begin{equation}
\begin{split}
& {\cal O}_{\text{SD}}(p) = 2\tau_{p}^{x} \quad , \quad 
{\cal O}_{\text{SC}}(p) = 2\tau_{p}^{y}  \\
&  \tau_{p}^{z} = 
\bolS_{1,r}{\cdot}\bolS_{2,r}+\bolS_{1,r+1}{\cdot}\bolS_{2,r+1} \; .
\end{split} 
\label{eqn:pseudospin-orderparam}
\end{equation}
\end{subequations}
Obviously, $\tau^{z}_{p}$ is a generalization of ${\cal S}^{z}$ 
discussed in section \ref{sec:useful_analogy} to a plaquette system.  

Now that we have identified the pseudo-spin degrees of freedom, 
it is straightforward to carry out a perturbation expansion 
for (nearly) degenerate ground-state manifold made up of 
$
 \bigl| \,
\begin{pspicture}[0.35](0,0)(0.3,0.3)
\psline[doubleline=true](0,0.4)(0.4,0.0)
\psline[doubleline=true](0,0.0)(0.4,0.4)
\psline[linestyle=dashed](0,0.0)(0.4,0.0)
\psline[linestyle=dashed](0.0,0.4)(0.4,0.4)
\psline[linestyle=dashed](0,0)(0.0,0.4)
\psline[linestyle=dashed](0.4,0.0)(0.4,0.4)
\end{pspicture} 
\;
\bigr\rangle 
$ and $ \bigl| \,
\begin{pspicture}[0.35](0,0)(0.4,0.4)
\psline[doubleline=true](0.0,0.0)(0.0,0.4)
\psline[doubleline=true](0.4,0.0)(0.4,0.4)
\psline[linestyle=dashed](0,0.0)(0.4,0.0)
\psline[linestyle=dashed](0.0,0.4)(0.4,0.4)
\psline[linestyle=dashed](0,0)(0.0,0.4)
\psline[linestyle=dashed](0.4,0.0)(0.4,0.4)
\end{pspicture} 
\;
\bigr\rangle$.  
Within this manifold, operators for spin, $p$-nematic, 
and $n$-nematic read
\begin{subequations}
\begin{align} 
& \bolS_{1,r}+\bolS_{2,r}=\bolS_{1,r+1}+\bolS_{2,r+1}=
\frac{1}{2}\text{\bf 1}\otimes\bolT_{p} \\
& \bolS_{1,r}-\bolS_{2,r}=\bolS_{1,r+1}-\bolS_{2,r+1}=
\tau^{x}\otimes\bolT_{p} \\
& 2(\bolS_{1,r}\!\times\!\bolS_{2,r})
=2(\bolS_{1,r+1}\!\times\!\bolS_{2,r+1})=
\tau^{y}\otimes\bolT_{p} \\
\begin{split}
& Q^{\alpha\beta}_{r}\equiv 
(S^{\alpha}_{1,r}S^{\beta}_{2,r}+S^{\beta}_{1,r}S^{\alpha}_{2,r}) \\
& \phantom{T^{\alpha\beta}_{r}}
= - \tau^{z}\otimes
\left\{
\frac{1}{2}(\text{T}_{p}^{\alpha}\text{T}_{p}^{\beta}
+\text{T}_{p}^{\beta}\text{T}_{p}^{\alpha})
-\delta^{\alpha\beta}
\right\} = Q_{r+1}^{\alpha\beta}
\; ,
\end{split}
\end{align}
\end{subequations}
where $\bolT_{p}$ denotes a spin-1 operator {\em on the plaquette} $p$.  
Plugging these equations into the original Hamiltonian, 
we obtain the (1st-order) effective Hamiltonian 
which describes the physics around the SU(4) point:
\begin{align}
\begin{split}
& {\cal H}_{\text{eff}} = \\
& \quad \lambda \sum_{p=\text{plaq}}
\left\{
(1+\delta_{2})\tau^{x}_{p}\tau^{x}_{p+1}
+(1+\delta_{4})\tau^{y}_{p}\tau^{y}_{p+1}+\frac{1}{4}
\right\} \bolT_{p}{\cdot}\bolT_{p+1} \\
& \quad 
+\lambda(1+\delta_{3})\sum_{p}
\left(\tau^{z}_{p}\tau^{z}_{p+1}\right)
\left\{
\bolT_{p}{\cdot}\bolT_{p+1}
+2\left(\bolT_{p}{\cdot}\bolT_{p+1}\right)^{2}-2
\right\} \\
& \quad 
+\sum_{p}\left(\delta_{2}-\delta_{3}+\delta_{4}+\delta_{6}\right)
\tau^{z}_{p}   
\end{split}
\notag \\
\begin{split}
& = \lambda \sum_{p=\text{plaq}}
\left\{
(1+\delta_{2})\tau^{x}_{p}\tau^{x}_{p+1}
+(1+\delta_{4})\tau^{y}_{p}\tau^{y}_{p+1}+\frac{1}{4}
\right\} \bolT_{p}{\cdot}\bolT_{p+1} \\
& \quad 
+ 2 \lambda(1+\delta_{3})\sum_{p}
\left(\tau^{z}_{p}\tau^{z}_{p+1}\right)
\left\{
\sum_{\alpha,\beta}\widetilde{Q}^{\alpha\beta}_{p}
\widetilde{Q}^{\alpha\beta}_{p+1} -1
\right\} \\
& \quad +\sum_{p}\left(\delta_{2}-\delta_{3}+\delta_{4}+\delta_{6}\right)
\tau^{z}_{p}   \; ,
\end{split}
\label{eqn:plaquette_Heff1}
\end{align}
where we have introduced the coarse-grained $n$-type nematic operators 
$\widetilde{Q}_{p}^{\alpha\beta}
\equiv(\text{T}_{p}^{\alpha}\text{T}_{p}^{\beta}
+\text{T}_{p}^{\beta}\text{T}_{p}^{\alpha})/2$ on plaquettes.  
This is similar to the well-known Kugel-Khomskii effective 
Hamiltonian\cite{Kugel-K-82} for the orbital-degenerate systems; here 
the role of two degenerate orbitals is played 
by two types of triplets 
$ \bigl| \,
\begin{pspicture}[0.35](0,0)(0.3,0.3)
\psline[doubleline=true](0,0.4)(0.4,0.0)
\psline[doubleline=true](0,0.0)(0.4,0.4)
\psline[linestyle=dashed](0,0.0)(0.4,0.0)
\psline[linestyle=dashed](0.0,0.4)(0.4,0.4)
\psline[linestyle=dashed](0,0)(0.0,0.4)
\psline[linestyle=dashed](0.4,0.0)(0.4,0.4)
\end{pspicture} 
\;
\bigr\rangle$ 
and 
$\bigl| \,
\begin{pspicture}[0.35](0,0)(0.4,0.4)
\psline[doubleline=true](0.0,0.0)(0.0,0.4)
\psline[doubleline=true](0.4,0.0)(0.4,0.4)
\psline[linestyle=dashed](0,0.0)(0.4,0.0)
\psline[linestyle=dashed](0.0,0.4)(0.4,0.4)
\psline[linestyle=dashed](0,0)(0.0,0.4)
\psline[linestyle=dashed](0.4,0.0)(0.4,0.4)
\end{pspicture} 
\;
\bigr\rangle$.   
It should be noted that the $xy$-components of the pseudo spin 
(AF $\tau^{x}$ and $p$-nematic $\tau^{y}$) couple to 
the dipolar part ($\bolT_{p}{\cdot}\bolT_{p+1}$) 
of the magnetic Hamiltonian 
while the $z$-component is associated with the ($n$-type) nematic 
part ($\widetilde{Q}^{\alpha\beta}\widetilde{Q}^{\alpha\beta}$).   

In order to further simplify the effective Hamiltonian, we again 
assume that the spin sector is well described by the VBS wave 
function\cite{Affleck-K-L-T-88}. 
Then the dipolar- and the nematic part of the effective Hamiltonian 
may be replaced by the following expectation values:
\begin{equation}
\VEV{\bolT_{p}{\cdot}\bolT_{p+1}}_{\text{VBS}} 
= -\frac{4}{3} \quad , \quad \biggl\langle 
\sum_{\alpha,\beta}\widetilde{Q}^{\alpha\beta}_{p}
\widetilde{Q}^{\alpha\beta}_{p+1}
\biggr\rangle_{\text{VBS}}
= +\frac{4}{3} 
\end{equation}
and the above effective Hamiltonian (\ref{eqn:plaquette_Heff1}) 
reduces to that for the pseudo-spins:
\begin{equation}
\begin{split}
{\cal H}_{\text{eff}} 
=& -\frac{4}{3}\lambda \sum_{p=\text{plaq}}
\left\{
(1+\delta_{2})\tau^{x}_{p}\tau^{x}_{p+1}
+(1+\delta_{4})\tau^{y}_{p}\tau^{y}_{p+1}+\frac{1}{4}
\right\}  \\
& \quad 
+ \frac{2}{3}\lambda(1+\delta_{3})\sum_{p}
\tau^{z}_{p}\tau^{z}_{p+1} \\
& \qquad 
+\sum_{p}\left(\delta_{2}-\delta_{3}+\delta_{4}+\delta_{6}\right)
\tau^{z}_{p}  \; . 
\end{split}
\label{eqn:plaquette_Heff2}
\end{equation}
This is nothing but the $S=1/2$ XYZ Hamiltonian in 
a magnetic field\cite{giamarchixyz}.  
For small $\delta_{3}$ the system is $xy$-like (easy plane) and 
{\em ferromagnetic} ordering will occur mainly in the $xy$-plane;   
when $\delta_{2}>\delta_{4}$ ($\delta_{2}<\delta_{4}$) 
the pseudo spins align in the $x$($y$)-direction 
and the SD (SC) phase will be stabilized 
(see Eq. (\ref{eqn:pseudospin-orderparam})).  
For $\delta_{2}=\delta_{4}$ (the self-dual models), 
on the other hand, the system reduces to the well-known 
XXZ Hamiltonian and the low-energy sector is generically 
described by the TL model (\ref{eqn:TL_ham}).  
These are exactly what we have found by a field-theoretical analysis 
in section \ref{sec:physics_selfdual}.   

For appropriate choices of $\delta_{2}$, $\delta_{3}$, and 
$\delta_{4}$, the $zz$-part will dominate and staggered ordering 
along the $z$-axis will occur (note that the sign of 
$\tau^{z}\tau^{z}$ is antiferromagnetic).   
Since this ground state has a period two in the `plaquette' picture, 
it has, when translated back to the original ladder, a period 4.  
%%%%%%%%%%%%%%%%%%%%%%%%%%%%%%%%%%%%%%%%%%%%%%%%%%%%%%%%%%%%%%
\section{Concluding remarks}
%%%%%%%%%%%%%%%%%%%%%%%%%%%%%%%%%%%%%%%%%%%%%%%%%%%%%%%%%%%%%%
In the present paper, we have developed a unifying approach to
the problem of unconventional phases appearing in generalized 
two-leg spin ladders. Although the low-energy properties of 
the two-leg spin ladder with only two-spin (or, ordinary exchange) 
interactions are fairly well-known, not much is known 
for the case with four-spin interactions. 

To treat the problem from a wider viewpoint, we have constructed, 
out of $S=1/2$ spins and their bilinears, a general lattice model 
Hamiltonian with SU(2) (rotational) invariance 
and symmetry under the exchange of two constituent chains.  
We have adopted the so-called SU(4)-symmetric model as the starting 
point to elucidate the interplay between two-spin- and
four-spin interactions.  
In particular, the SU(4) symmetry is the maximal continuous 
symmetry that one can have in general two-leg $S=1/2$ spin ladders.
The greatest advantage of this extended symmetry approach 
is that it can unify the unconventional phases 
(e.g. SC phase) of the two-leg spin ladder with 4-spin exchange interactions.  

In the present paper, we have shown that four unconventional
phases (Q, SD, SC, and RQ) of the generalized two-leg 
spin ladder are unified at the SU(4) multicritical point.
These phases, which emerge in regions characterized by
substantial strength of the four-spin interactions, is quite  
difficult to describe by means of a perturbative approach 
starting from the limit of two decoupled chains.   
Fortunately, however, we have a well-controlled starting point%
--the SU(4) point--in our problem.  
In the field-theory language, the low-energy properties of 
the SU(4) point are described by the level-1 SO(6) WZW model 
which is equivalent to six copies of the 2D Ising model 
at its critical point.  
The continuum expressions of the interactions can then be written 
in terms of this basis and we have revealed the existence of 
a duality symmetry ${\cal D}$, the so-called 
spin-chirality duality\cite{Hikihara-M-H-03}, 
and a new hidden one $\widetilde{\cal D}$ which is an emergent symmetry. 
The one-loop RG flow exhibits quite a simple structure
(four dominant phases and symmetric rays toward them) in spite of 
the complexity in the RG equations.  
The duality transformation ${\cal D}$ and its dual 
($\widetilde{\cal D}$) map the dominant phases onto each other 
(see Fig. \ref{fig:4phases}).  
The existence of these duality symmetries enables us to investigate
the nature of the quantum phase transitions between these competing
orders which occur along the self-dual manifolds.
The behavior on the self-dual manifold $B=D$
is intriguing; the spin sector dies away by opening a gap
and in turn the {\em pseudo-spin} degrees of freedom 
come into play in the low-energy physics.  
Their gapless spin-singlet fluctuations are described 
by the $c=1$ TL model.   
In fact, what controls the competition and quantum phase transitions
between the unconventional spin-singlet phases (SD and SC)
is this pseudo-spin degrees of freedom and we demonstrated
this picture by a simple variational calculation in
section \ref{sec:XXZ_analogy}.  
The fact that the pseudo-spin sector displays an U(1) quantum critical
behavior helps us understand the global phase 
structure in the same manner as we do in the $S=1/2$ XXZ model.
Note that the above U(1) criticality is {\em not} an emergent
one, whereas the other one within the manifold $g_{2}=-g_{4}$ is emergent
in the sense of Ref. \onlinecite{Senthil-V-B-S-F-04}.

Another interesting point concerning the physics 
on the self-dual manifold is the {\em emergent} SU(3) symmetry.  
Although this is not so obvious 
in the lattice analysis, our field-theory analysis tells us that
instead of the original SO(3) an enlarged SU(3) symmetry appears 
in the low-energy limit. 
As a byproduct, we established an interesting description
of our problem in terms of the spin-1 (hardcore)
bosons.  
In this respect, we hope that our results obtained for spin systems will
be of some help in understanding the phases of
the spin 1 boson systems\cite{Imambekov-L-D-03}.   
%%%%%%%%%%%%%%%%%%%%%%%%%%%%%%%%%%%%%%%%%%%%%%%%%%%%%%%%%
% Acknowledgement
%%%%%%%%%%%%%%%%%%%%%%%%%%%%%%%%%%%%%%%%%%%%%%%%%%%%%%%%%
\begin{acknowledgements}
The authors would like also to thank P.~Azaria,  E.~Boulat, A.~L{\"a}uchli, 
H-H.~Lin, T.~Momoi, and A. A. ~Nersesyan for very useful discussions.  
The author (K.T.) was supported by Grant-in-Aid 
No.16740222 from Ministry of Education, Culture, Sports, 
Science, and Technology of Japan.  He is grateful to 
LPTM at Cergy-Pontoise university and the condensed-matter-theory 
laboratory of RIKEN for hospitality extended to him during his stay.  
The author (P.L.) is grateful to 
the Abdus Salam ICTP for hospitality where this work was completed.
\end{acknowledgements}

%%%%%%%%%%%%%%%%%%%%%%%%%%%%%%%%%%%%%%%%%%%%%%%%%%%%%%%%%%%%%%
\appendix
%%%%%%%%%%%%%%%%%%%%%%%%%%%%%%%%%%%%%%%%%%%%%%%%%%%%%%%%%%%%%%
\section{Summary of SU(4)}
%%%%%%%%%%%%%%%%%%%%%%%%%%%%%%%%%
\label{sec:SU4_generators}
%%%%%%%%%%%%%%%%%%%%%%%%%%%%%%%%%%%%%%%%%%%%%%%%%%%%%%%%%%%%%%
\subsection{Generators}
In section \ref{sec:Model_and_Sym}, we introduced 15 SU(4) 
generators $X^{i}$ in the form of 4$\times$4 matrices.   
Below, we give the explicit forms of them.  
Note that we use the singlet-triplet basis defined by
\begin{equation}
\begin{split}
& |\text{s}\rangle = \frac{1}{\sqrt{2}}
\left(|\uparrow\downarrow\rangle-
|\downarrow\uparrow\rangle\right) \, , \; 
|\text{t}_{x}\rangle=\frac{-1}{\sqrt{2}}
\left(|\uparrow\uparrow\rangle-
|\downarrow\downarrow\rangle\right)  \\
& |\text{t}_{y}\rangle=\frac{i}{\sqrt{2}}
\left(|\uparrow\uparrow\rangle +
|\downarrow\downarrow\rangle\right)  \, , \; 
|\text{t}_{z}\rangle =  \frac{1}{\sqrt{2}}
\left(|\uparrow\downarrow\rangle + 
|\downarrow\uparrow\rangle\right) \; .
\end{split}
\end{equation}   
%%%%%%%%%%%%%%%%%%%%%%%%%%%%%%%%%%%%%%%%%%%%%%%%%%%
%\begin{widetext}
%%%%%%%%%%%%%%%%%%%
%%%%%%%%%%%%%%%%%%%%%%%%%%%%%%%%%%%
\begin{subequations}
%%%%%%%%%%%%%%%%%%%%%%%%%%%%%%%%%%%
\begin{align}
\begin{split}
& S^{1}_{1}=X^{1} =
\begin{pmatrix}
0&\frac{1}{2}&0&0 \\ 
\frac{1}{2}&0&0&0 \\
0&0&0&\frac{-i}{2} \\
0&0&\frac{i}{2}&0
\end{pmatrix}
\quad 
S^{2}_{1}=X^{2} =
\begin{pmatrix}
0&0&\frac{1}{2}&0\\
0&0&0&\frac{i}{2}\\
\frac{1}{2}&0&0&0\\
0&\frac{-i}{2}&0&0
\end{pmatrix}
\\
& S^{3}_{1}=X^{3} =
\begin{pmatrix}
0&0&0&\frac{1}{2}\\
0&0&\frac{-i}{2}&0\\
0&\frac{i}{2}&0&0\\
\frac{1}{2}&0&0&0
\end{pmatrix}
\end{split}
\end{align}
\begin{align}
\begin{split}
& S^{1}_{2}=X^{4} =
-
\begin{pmatrix}
0&\frac{1}{2}&0&0\\
\frac{1}{2}&0&0&0\\
0&0&0&\frac{i}{2}\\
0&0&\frac{-i}{2}&0
\end{pmatrix}
\,
S^{2}_{2}=X^{5} =
-
\begin{pmatrix}
0&0&\frac{1}{2}&0\\
0&0&0&\frac{-i}{2}\\
\frac{1}{2}&0&0&0\\
0&\frac{i}{2}&0&0
\end{pmatrix}
\\
& S^{3}_{2}=X^{6} =
-
\begin{pmatrix}
0&0&0&\frac{1}{2}\\
0&0&\frac{i}{2}&0\\
0&\frac{-i}{2}&0&0\\
\frac{1}{2}&0&0&0
\end{pmatrix} 
\end{split}
\end{align}
%%%%%%%%%%%%%%%%%%%%%%%%%%%%%%%%%%%
The remaining 9 generators are essentially given by tensor product 
of two Pauli matrices:
\[
G_{ab} = \frac{1}{2}\, \sigma^{a}\! \otimes \sigma^{b} = 
2S_{1}^{a}\, S_{2}^{b} \qquad (a,b=1,2,3) \; .
\]
%%%%%%%%%%%%%%%%%%%%%%%%%%%%%%%%%%%%%%%%%%%%%%%%%%%%
\begin{widetext}
%%%%%%%%%%%%%%%%%%%%%%%%%%%%%%%%%%%%%%%%%%%%%%%%%%%%
\begin{align}
& G_{11}=X^{7} =
\begin{pmatrix}
-\frac{1}{2}&0&0&0\\
0&-\frac{1}{2}&0&0\\
0&0&\frac{1}{2}&0\\
0&0&0&\frac{1}{2}
\end{pmatrix} 
\quad 
G_{22}=X^{8}=
\begin{pmatrix}
-\frac{1}{2}&0&0&0\\
0&\frac{1}{2}&0&0\\
0&0&-\frac{1}{2}&0\\
0&0&0&\frac{1}{2}
\end{pmatrix}
\quad 
G_{33}=X^{9}=
\begin{pmatrix}
-\frac{1}{2}&0&0&0\\
0&\frac{1}{2}&0&0\\
0&0&\frac{1}{2}&0\\
0&0&0&-\frac{1}{2}
\end{pmatrix}
\\
& G_{12}=X^{10} =
\begin{pmatrix}
0&0&0&\frac{i}{2}\\
0&0&-\frac{1}{2}&0\\
0&-\frac{1}{2}&0&0\\
\frac{-i}{2}&0&0&0
\end{pmatrix} 
\quad 
G_{13}=X^{11}=
\begin{pmatrix}
0&0&\frac{-i}{2}&0\\
0&0&0&-\frac{1}{2}\\
\frac{i}{2}&0&0&0\\
0&-\frac{1}{2}&0&0
\end{pmatrix}
\quad 
G_{21}=X^{12}=
\begin{pmatrix}
0&0&0&\frac{-i}{2}\\
0&0&-\frac{1}{2}&0\\
0&-\frac{1}{2}&0&0\\
\frac{i}{2}&0&0&0
\end{pmatrix}
\\
& G_{23}=X^{13} =
\begin{pmatrix}
0&\frac{i}{2}&0&0\\
\frac{-i}{2}&0&0&0\\
0&0&0&-\frac{1}{2}\\
0&0&-\frac{1}{2}&0
\end{pmatrix} 
\quad 
G_{31}=X^{14}=
\begin{pmatrix}
0&0&\frac{i}{2}&0\\
0&0&0&-\frac{1}{2}\\
\frac{-i}{2}&0&0&0\\
0&-\frac{1}{2}&0&0
\end{pmatrix}
\quad 
G_{32}=X^{15}=
\begin{pmatrix}
0&\frac{-i}{2}&0&0\\
\frac{i}{2}&0&0&0\\
0&0&0&-\frac{1}{2}\\
0&0&-\frac{1}{2}&0
\end{pmatrix}
\end{align}
%%%%%%%%%%%%%%%%%%
\end{widetext}
%%%%%%%%%%%%%%%%%%%%%%%%%%%%%%%%%%%%%%%%%
\end{subequations}
%%%%%%%%%%%%%%%%%%%%%%%%%%%%%%%%%%%%%%%%%%%%%%%%%%%%%%%%%%%%%
%%%%%%%%%%%%%%%%%%%%%%%%%%%%%%%%%%%%%%%%%%%%%%%%%%%%%%%%%%%%%%%%
\subsection{SU(3)$\times$U(1)}
%%%%%%%%%%%%%%%%%%%%%%%%%%%%%%%%%%%%%%%%%%%%%%%%%%%%%%%%%%%%%%%%
\label{sec:U1_SU3}
%%%%%%%%%%%%%%%%%%%%%%%%%%%%%%%%%%%%%%%%%%%%%%%%%%%%%%%%%%%%%%%%
The spin-chirality transformation introduces a natural 
categorization of the above generators in terms of 
a subgroup SU(3)$\times$U(1).  
The easiest way of understanding the appearance of 
SU(3) would be to decompose the whole 
Hilbert space into a singlet and a triplet (of course, the singlet 
corresponds to U(1)-factor); the SU(3)-part acts only to the triplet 
subspace.  

With this identification, the SU(3) Gell-Mann matrices are 
given as follows:
%%%%%%%%%%%%%%%%%%%%%%%%%%%%%%%%%%%%%%%%%%%%%%%%%%%%
\begin{widetext}
%%%%%%%%%%%%%%%%%%%%%%%%%%%%%%%%%%%%%%%%%%%%%%%%%%%%
\begin{align*}
& 
G^{1}=-(G_{12}+G_{21})=
\begin{pmatrix}
0 & 0 & 0 & 0 \cr 0 & 0 & 1 & 0 \cr 0 & 1 & 0 & 0 \cr 0 & 0 & 0 & 0 
\end{pmatrix} \quad
G^{2}=S_{1}^{3}+S_{2}^{3}=
\begin{pmatrix}
0 & 0 & 0 & 0 \cr 0 & 0 & -i  & 0 \cr 0 & i
    & 0 & 0 \cr 0 & 0 & 0 & 0
\end{pmatrix}  \\ 
&
G^{3}=-G_{11}+G_{22}=
\begin{pmatrix}
0 & 0 & 0 & 0 \cr 0 & 1 & 0 & 0 \cr 0 & 0 & -1 & 0 \cr 0 & 0 & 0 & 0
\end{pmatrix}  \quad
G^{4}= -(G_{13}+G_{31})=
\begin{pmatrix}
0 & 0 & 0 & 0 \cr 0 & 0 & 0 & 1 \cr 0 & 0 & 0 & 0 \cr 0 & 1 & 0 & 0
\end{pmatrix} \\
&
G^{5}= -\left(S_{1}^{2}+S_{2}^{2}\right)=
\begin{pmatrix}
0 & 0 & 0 & 0 \cr 0 & 0 & 0 & -i \cr 0 & 0 & 0 & 0 \cr 0 & i
    & 0 & 0 
\end{pmatrix} \quad
G^{6}= -(G_{23}+G_{32})= 
\begin{pmatrix}
0 & 0 & 0 & 0 \cr 0 & 0 & 0 & 0 \cr 0 & 0 & 0 & 1 \cr 0 & 0 & 1 & 0
\end{pmatrix} \\
&
G^{7}= S_{1}^{1}+S_{2}^{1}=
\begin{pmatrix}
0 & 0 & 0 & 0 \cr 0 & 0 & 0 & 0 \cr 0 & 0 & 0 & -i  \cr 0 & 0 & i
    & 0 
\end{pmatrix} \quad 
G^{8}= -\frac{1}{\sqrt{3}}\left(
G_{11}+G_{22}-2G_{33}
\right) = 
\begin{pmatrix}
0 & 0 & 0 & 0 \cr 0 & \frac{1}{{\sqrt{3}}} & 0 & 0 \cr 0 & 0 & \frac{1}
   {{\sqrt{3}}} & 0 \cr 0 & 0 & 0 & \frac{-2}{{\sqrt{3}}}
\end{pmatrix}  
\;  .
\end{align*}
%%%%%%%%%%%%%%%%%%%%%%%%%%%%%%%%%%%%%%%%%%%%%%%%%%%%%%
\end{widetext}
%%%%%%%%%%%%%%%%%%%%%%%%%%%%%%%%%%%%%%%%%%%%%%%%%%%%%%
They are supplemented by the generator of (spin-chirality) U(1):
\[
G_{\text{U(1)}} = -\sqrt{\frac{2}{3}}\left(
G_{11}+G_{22}+G_{33} \right)   \; ,
\]
which satisfies 
\[
 [\, G_{\text{U(1)}}\, , \, G^{a}\, ]=0 \quad (a=1,\ldots,8)\; .
\]
These 9 operators generate a subgroup SU(3)$\times$U(1).   

It is a straightforward task to rewrite our building blocks 
Eqs. (\ref{eqn:int_1}-\ref{eqn:int_7})
in terms of the above generators.
In particular, one has:
\begin{multline}
A {\cal H}_{1} + C{\cal H}_{3} \\
= \sum_{r}\biggl[A\biggl(
G^{2}(r)G^{2}(r+1)+G^{5}(r)G^{5}(r+1)+G^{7}(r)G^{7}(r+1)
\biggr) \\
+C\biggl(G^{1}(r)G^{1}(r+1)+G^{3}(r)G^{3}(r+1)+
G^{4}(r)G^{4}(r+1)\\ 
+G^{6}(r)G^{6}(r+1) +G^{8}(r)G^{8}(r+1)
+G_{\text{U(1)}}(r)G_{\text{U(1)}}(r+1) 
\biggr)\biggr] \; .
\end{multline}
From which we observe that 
the model with $A=C$ has an SU(3) symmetry.  

Interestingly, the remaining six SU(4) generators 
\[
 \bolS_{1}-\bolS_{2} \quad , \quad 
 2\left(  \bolS_{1}\times \bolS_{2}\right)
\]
are order parameters of the above SU(3)$\times$U(1).  
In fact, 
\[
(\bolS_{1}-\bolS_{2})
+i\, 2\left( \bolS_{1}\!\times \!\bolS_{2}\right) \quad 
\text{and}\quad 
(\bolS_{1}-\bolS_{2})
- i\, 2\left( \bolS_{1}\!\times \!\bolS_{2}\right)
\]
respectively transform like {\bf 3} and $\bar{\text{\bf 3}}$ under 
SU(3).  The spin-chirality U(1) acts like
\[
(\bolS_{1}-\bolS_{2})
\pm i\, 2\left( \bolS_{1}\!\times \!\bolS_{2}\right) 
\mapsto \be^{\pm i\theta}\left[
(\bolS_{1}-\bolS_{2})
\pm i\, 2\left( \bolS_{1}\!\times \!\bolS_{2}\right)
\right]  \; .
\]
%%%%%%%%%%%%%%%%%%%%%%%%%%%%%%%%%%%%%%%%%%%%%%%%%%%%%%%%%%%%%%
\section{Bosonization dictionary}
%%%%%%%%%%%%%%%%%%%%%%%%%%%%%%%%%%%%%%%%%%%%%%%%%%%%%%%%%%%%%%
\label{sec:dictionary}
%%%%%%%%%%%%%%%%%%%%%%%%%%%%%%%%%%%%%%%%%%%%%%%%%%%%%%%%%%%%%%
To establish the notations, we briefly summarize 
the main formulas used in our bosonization analysis.  
Consider $N$ free Dirac fermions defined by the following 
Hamiltonian:
\begin{equation}
{\cal H}_{\text{Dirac}} = 
-i\, v\, \int\! dx \sum_{a=1}^{N} \left(
\Psi^{\dagger}_{a,\text{R}}\partial_{x}\Psi_{a,\text{R}} 
-\Psi^{\dagger}_{a,\text{L}}\partial_{x}\Psi_{a,\text{L}} 
\right) 
%+m \, \int\!dx \left(\psi^{\dagger}_{\text{R}}\psi_{\text{L}}+
%\psi^{\dagger}_{\text{L}}\psi_{\text{R}}\right)  
\; .
\label{eqn:N_Dirac}
\end{equation}
Let us first introduce $N$ bosonic fields to 
bosonize $N$ Dirac fermions as follows \cite{Gogolin-N-T-book}:
\begin{equation}
\begin{split}
& \Psi_{a,\text{R}} = \frac{{\tilde \kappa}_{a}}{\sqrt{2\pi a_0}}
\exp\left(i \sqrt{4\pi} \phi_{a,\text{R}}\right)  
\\
& \Psi_{a,\text{L}} = \frac{{\tilde \kappa}_{a}}{\sqrt{2\pi a_0}}
\exp\left(-i \sqrt{4\pi} \phi_{a,\text{L}}\right) \\
& \qquad (a=1,\ldots,N) \; ,
\end{split}
\label{eqn:boson2fermion}
\end{equation}
$a_0$ being some ultraviolet cut-off (typically the lattice spacing)
and we use the following normalization for the chiral bosonic fields:
\begin{equation}
\begin{split}
& \langle \phi_{a,\text{R}} \left( \bar z\right)
\phi_{b,\text{R}} \left( \bar w\right)  \rangle
= - \frac{\delta_{ab}}{4 \pi} \ln \left(\bar z -
\bar w \right)   \\
& \langle \phi_{a,\text{L}} \left(z\right)
\phi_{b,\text{L}} \left( w\right)  \rangle
= - \frac{\delta_{ab}}{4 \pi} \ln \left(z - w \right) \; ,
\end{split}     
\label{eqn:normbos}
\end{equation}
where the complex coordinates $z$ and $\bar{z}$ are defined by 
$z=v\tau +i\, x$ and $\bar{z}=v\tau -i \, x$, respectively.  
The total bosonic fields $\phi_a$ and 
their dual fields $\theta_a$ are 
defined as:
$\Phi_a \equiv \phi_{a,\text{R}} + \phi_{a,\text{L}}$,
$\Theta_a \equiv -\phi_{a,\text{R}} + \phi_{a,\text{L}}$.
We impose the following commutation relation for the chiral 
bosonic fields
\begin{equation}
\left[\phi_{a,\text{R}}, \phi_{b,\text{L}}\right] = \frac{i}{4}
\delta_{ab}, 
\end{equation}
so that the left- and the right movers of the same species anti-commute: 
$\{\Psi_{a,\text{R}}(x), \Psi_{a,\text{L}}(y)\}=0$.
The anticommutation between fermions of  
{\em different} species is guaranteed by the presence of the 
Klein factors (here Majorana fermions) ${\tilde \kappa}_{a}$
in the definition (\ref{eqn:boson2fermion}), 
which obey the anticommutation rule:
\begin{equation}
\{{\tilde \kappa}_{a}, {\tilde \kappa}_{b}\} =
2 \delta_{ab} .
\end{equation} 
Using these bosons, ${\cal H}_{\text{Dirac}}$ can be written 
as a Hamiltonian of the so-called Tomonaga-Luttinger (TL) 
model \cite{Gogolin-N-T-book}:
\begin{subequations}
\begin{align}
{\cal H}_{\text{Dirac}} &=
\pi v\int\! dx \sum_{a=1}^{N}\left(
 J_{\text{L},a}^{2}+ J_{\text{R},a}^{2}
\right) \\
&= \frac{v}{2}\int\!dx \sum_{a=1}^{N} 
\left\{
(\partial_{x}\Phi_{a})^{2}
+ (\partial_{x}\Theta_{a})^{2} \right\}
={\cal H}_{\text{TL}} \; , 
\label{eqn:TL_ham}
\end{align}
\end{subequations}
where the operators of the form $A^{n}$ should always be understood as 
normal-ordered $\nord A^{n} \nord $\cite{Gogolin-N-T-book,Giamarchi-book}.
The above equivalence is established by the following 
bosonization formulas for 
fermion bilinears:
\begin{subequations}
\begin{align}
& J_{\text{L},a}=
\Psi_{a,\text{L}}^{\dagger}\Psi_{a,\text{L}}
= \frac{1}{\sqrt{\pi}}\partial_{x}\phi_{a,\text{L}} 
=\frac{1}{\sqrt{4\pi}} \partial_{x}(\Phi_{a}+\Theta_{a})
\label{eqn:current-L}
\\
& J_{\text{R},a}=
\Psi_{a,\text{R}}^{\dagger}\Psi_{a,\text{R}}
= \frac{1}{\sqrt{\pi}}\partial_{x}\phi_{a,\text{R}} 
=\frac{1}{\sqrt{4\pi}} \partial_{x}(\Phi_{a}-\Theta_{a}) \\
& J_{\text{L},a}^{2}
= +\frac{i}{\pi}\Psi_{a,\text{L}}^{\dagger}\partial_{x}\Psi_{a,\text{L}}
\, , \; 
J_{\text{R},a}^{2}
= -\frac{i}{\pi}\Psi_{a,\text{R}}^{\dagger}\partial_{x}\Psi_{a,\text{R}}
\\
\begin{split}
& \Psi_{a,\text{R}}^{\dagger}\Psi_{a,\text{L}}
= -\frac{i}{2\pi a_{0}}\, \be^{-i\sqrt{4\pi}\Phi_{a}} \\
& \Psi_{a,\text{L}}^{\dagger}\Psi_{a,\text{R}}
= +\frac{i}{2\pi a_{0}}\, \be^{+i\sqrt{4\pi}\Phi_{a}} 
\end{split}
\label{eqn:CDW-op}
\; .
\end{align}
\end{subequations}
%%%%%%%%%%%%%%%%%%%%
%%%%%%%%%%%%%%%%%%%%%%%%%%%%%%%%%%%%%%%%%%%%%%%%%%%%%%%%%%%%%%
\section{SU(4) effective action from Hubbard model}
%%%%%%%%%%%%%%%%%%%%%%%%%%%%%%%%%%%%%%%%%%%%%%%%%%%%%%%%%%%%%%
\label{sec:deriv_SU4}
%%%%%%%%%%%%%%%%%%%%%%%%%%%%%%%%%%%%%%%%%%%%%%%%%%%%%%%%%%%%%%
In this section, we sketch the derivation of the 
SU(4) fixed-point Hamiltonian from the two-band Hubbard model 
at quarter filling:
\begin{multline}
{\cal H}_{\text{Hubbard}}=
-t\, \sum_{i,a,\sigma}\left(
c^{\dagger}_{a,\sigma,i+1}\, c_{a,\sigma,i}+
c^{\dagger}_{a,\sigma,i}\, c_{a,\sigma,i+1}
\right) \\
+ \frac{U}{2}\sum_{\substack{i,a,b,\\
\sigma,\sigma^{\prime}}}
n_{a,\sigma,i}n_{b,\sigma^{\prime},i}\left(
1-\delta_{a,b}\delta_{\sigma,\sigma^{\prime}}
\right)  \; ,
\label{eqn:2-band-Hubbard}
\end{multline}
where $c^{\dagger}_{i,a,\sigma}$ creates an electron 
with spin $\sigma(=\uparrow,\downarrow)$ 
in orbital $a(=1,2)$ of the site-$i$ and 
$n_{a,\sigma,i} = c^{\dagger}_{a,\sigma,i} c_{a,\sigma,i}$.   
The key idea is the following.  
When the Hubbard interaction 
$U$ is sufficiently large, the system becomes 
a Mott insulator and the charge degrees of freedom gets decoupled 
from the low-energy physics \cite{Affleck-88,assaraf}.  However, we still have 
four states ($a=1,2$, $\sigma=\uparrow,\downarrow$) on each site and 
we identify them with four states of 
the fundamental representation {\bf 4} of SU(4).  
These states constitute the low-energy sector of the system.  
As in the usual single-band case, 
the effective Hamiltonian describing the low-energy physics 
is obtained by the second-order perturbation:
\begin{equation}
\begin{split}
{\cal H}_{\text{eff}} &=
\frac{2t^{2}}{U} P_{\text{G}}\sum_{\substack{%
a,\sigma\\b,\sigma^{\prime}
}}
\left(
c_{a,\sigma,i+1}^{\dagger}\, c_{a,\sigma,i} +
c_{a,\sigma,i}^{\dagger}\, c_{a,\sigma,i+1}\right) \\
& \phantom{\frac{2t^{2}}{U} P_{\text{G}}\sum_{\substack{%
a,\sigma\\b,\sigma^{\prime}
}}}\quad \times %% PHANTOM!
\left(
c_{b,\sigma^{\prime},i+1}^{\dagger}\, c_{b,\sigma^{\prime},i} +
c_{b,\sigma^{\prime},i}^{\dagger}\, c_{b,\sigma^{\prime},i+1}
\right)P_{\text{G}}  \\
&= \frac{4t^{2}}{U}\sum_{i} P_{i,i+1} 
= \frac{4t^{2}}{U}\sum_{i,a}X_{i}^{a}X_{i+1}^{a}+\text{const.} \; ,
\end{split}
\end{equation}
where the projection onto the subspace with 
one electron per site is enforced by 
$P_{\text{G}}$.  The operator $P_{i,i+1}$ appearing 
in the last line is the SU(4) permutation operator acting on 
{\bf 4}$\otimes${\bf 4} on neighboring sites.  
The Hamiltonian $\sum X_{i}^{a}X_{i+1}^{a}$ is nothing but 
the SU(4)-invariant model (\ref{eqn:SU4_Heisenberg}) in 
section \ref{sec:Model_and_Sym}.  

If we repeat the same steps starting from the continuum expression 
of the two-band Hubbard model (\ref{eqn:2-band-Hubbard}), 
we may obtain the desired continuum field theory for the SU(4) model 
(\ref{eqn:SU4_Heisenberg})\cite{Affleck-88,Azaria-G-L-N-99,%
Itoi-Q-A-00}.   

Let us begin with the first term of ${\cal H}_{\text{Hubbard}}$ 
Eq. (\ref{eqn:2-band-Hubbard}).  
As is well-known in the standard bosonization treatment of 
one-dimensional electron systems, the spectrum of the four 
electrons ($c_{1}\equiv c_{1,\uparrow}$, $c_{2}\equiv c_{1,\downarrow}$, 
$c_{3}\equiv c_{2,\uparrow}$, $c_{4}\equiv c_{2,\downarrow}$) 
near the Fermi points $k=\pm k_{\text{F}}=\pm \pi/4$ may be 
linearized to obtain the Hamiltonian of the Dirac fermions 
(\ref{eqn:N_Dirac}) with $N=4$ and $v =2t \sin k_{\text{F}}$. 
In order to obtain the continuum expression of the second term 
(Hubbard interaction ${\cal H}_{U}$), 
we need the expression of the density operator 
$n_{b,i}$.  
Since we have two species of Dirac fermions (L and R) for each fermion, 
the electron density is made up of four pieces:
\begin{equation}
\begin{split}
\frac{1}{a_0}n_{i,b}&\approx n_{b}(x) \\
= & \Psi^{\dagger}_{b,\text{R}}\Psi_{b,\text{R}} 
+ \Psi^{\dagger}_{b,\text{L}}\Psi_{b,\text{L}} \\
& \qquad + \be^{-2ik_{\text{F}}x}\,
\Psi^{\dagger}_{b,\text{R}}\Psi_{b,\text{L}}
+ \be^{+2ik_{\text{F}}x}\,\Psi^{\dagger}_{b,\text{L}}\Psi_{b,\text{R}} \\
& \qquad\qquad \qquad 
 (b=1,2,3,4)\; .
\end{split}
\end{equation}
With the help of Eqs. (\ref{eqn:current-L})-(\ref{eqn:CDW-op}), 
the right-hand side can be readily bosonized as:
\begin{multline}
n_{b}(x)=\frac{1}{\sqrt{\pi}}\partial_{x}\Phi_{b}
- \frac{i}{2\pi a_{0}}\be^{-2ik_{\text{F}}x}\, 
\be^{-i\sqrt{4\pi}\Phi_{b}} \\
+\frac{i}{2\pi a_{0}}\be^{+2ik_{\text{F}}x}\, 
\be^{+i\sqrt{4\pi}\Phi_{b}} \; .
\label{eqn:bosonized_density}
\end{multline}

Using this bosonized expression,
one finds that the effective Hamiltonian which 
describes the low-energy properties of the two-band Hubbard model
at quarter-filling consists of four terms:
\begin{equation}
{\cal H}_{U} = {\cal H}_{\text{c}}+{\cal H}_{\text{SU(4)}} 
+ {\cal H}_{2k_{\text{F}}}+ {\cal H}_{4k_{\text{F}}} \; .
\end{equation}
The two parts ${\cal H}_{2k_{\text{F}}}$ and ${\cal H}_{4k_{\text{F}}}$  
contains terms oscillating with wave vectors $q=\pm 2k_{\text{F}}$ 
and $q=4k_{\text{F}}$ respectively and may be neglected in the 
first approximation.    Before presenting the explicit forms of 
${\cal H}_{\text{c}}$ and ${\cal H}_{\text{SU(4)}}$, it is convenient 
to move to another set of boson basis:
\begin{equation}
\begin{pmatrix}
\Phi_{\text{c}} \\ \Phi_{\text{s}}\\ \Phi_{\text{f}}\\ 
\Phi_{\text{sf}} 
\end{pmatrix} = \frac{1}{2}
\begin{pmatrix}
1 &  1 &  1 &  1 \\
1 & -1 &  1 & -1 \\
1 &  1 & -1 & -1 \\
1 & -1 & -1 &  1 
\end{pmatrix}
\begin{pmatrix}
\Phi_{1} \\ \Phi_{2}\\ \Phi_{3}\\ 
\Phi_{4}  
\end{pmatrix} \; .
\label{eqn:SU4_bosons}
\end{equation}
Since the above is an orthogonal transformation, 
the Dirac part is transformed to 
\begin{equation}
{\cal H}_{\text{Dirac}} =
\frac{v}{2}\int\!dx \sum_{a=\text{c,s,f,sf}} 
\left\{
(\partial_{x}\Phi_{a})^{2}
+ (\partial_{x}\Theta_{a})^{2} \right\} \; .
\label{eqn:free-Dirac2}
\end{equation}
The $U$-dependent part (${\cal H}_{\text{c}}+{\cal H}_{\text{SU(4)}}$) 
can be recasted as:
\begin{equation}
\begin{split}
& \frac{U}{2\pi}\sum_{a,b=1}^{4} \partial_{x}\Phi_{a}
(\mathbb{M}-\text{\bf 1})_{a,b}\partial_{x}\Phi_{b} \\
& \qquad + \frac{2U}{(2\pi a_{0})^{2}}\sum_{a<b}^{4}
\cos\left[\sqrt{4\pi}(\Phi_{a}-\Phi_{b})\right]  \\
& =\frac{U}{2\pi}\left\{
3 (\partial_{x}\Phi_{\text{c}})^{2}  
-\sum_{a=\text{s,f,sf}} (\partial_{x}\Phi_{\text{a}})^{2}  
\right\} \\
& \qquad + \frac{U}{\pi^{2}a_{0}^{2}}\biggl\{
 \cos(\sqrt{4\pi}\Phi_{\text{s}})\cos(\sqrt{4\pi}\Phi_{\text{f}}) \\
& \phantom{\frac{U}{\pi^{2}}} \qquad \qquad +  
 \cos(\sqrt{4\pi}\Phi_{\text{f}})\cos(\sqrt{4\pi}\Phi_{\text{sf}}) \\
& \phantom{\frac{U}{\pi^{2}}} \qquad \qquad \qquad + 
 \cos(\sqrt{4\pi}\Phi_{\text{sf}})\cos(\sqrt{4\pi}\Phi_{\text{s}}) 
\biggr\}
\; ,
\label{eqn:uniform-U}
\end{split}
\end{equation}
where $\mathbb{M}$ is a 4$\times$4 matrix with all matrix 
elements equal to unity.    From Eqs. (\ref{eqn:free-Dirac2}) and 
(\ref{eqn:uniform-U}), we can read off 
\begin{subequations}
%%%%%%%%%%%%%%%%%%%%%%%%%
\begin{multline}
{\cal H}_{\text{c}} = \\ 
\int\!dx \left\{
\frac{v}{2} \left(
(\partial_{x}\Phi_{\text{c}})^{2}  
+ (\partial_{x}\Theta_{\text{c}})^{2} 
\right) + \frac{3U}{2\pi} (\partial_{x}\Phi_{\text{c}})^{2} 
\right\}
\end{multline}
and 
\begin{equation}
\begin{split}
& {\cal H}_{\text{SU(4)}} \\
& = \int\!dx \sum_{a=\text{s,f,sf}}\biggl\{
\frac{v}{2} \left(
(\partial_{x}\Phi_{\text{a}})^{2}  
+  (\partial_{x}\Theta_{\text{a}})^{2}  
\right) 
- \frac{U}{2\pi} (\partial_{x}\Phi_{\text{a}})^{2}
\biggr\} \\
& \qquad + \frac{U}{\pi^{2}a_{0}^{2}}\int\!dx \biggl\{
 \cos(\sqrt{4\pi}\Phi_{\text{s}})\cos(\sqrt{4\pi}\Phi_{\text{f}}) \\
& \phantom{\frac{U}{\pi^{2}}\int\!dx}
\qquad \qquad + 
 \cos(\sqrt{4\pi}\Phi_{\text{f}})\cos(\sqrt{4\pi}\Phi_{\text{sf}}) \\
& \phantom{\frac{U}{\pi^{2}}\int\!dx} \qquad \qquad \qquad + 
 \cos(\sqrt{4\pi}\Phi_{\text{sf}})\cos(\sqrt{4\pi}\Phi_{\text{s}}) 
\biggr\} \; ,
\end{split}
\label{eqn:SU4_step1}
\end{equation}
%%%%%%%%%%%%%%%%%%%%%%%%
\end{subequations}
%%%%%%%%%%%%%%%%%%%%%%%%
which is written only in terms of 
three bosons $\Phi_{\text{s}}$, $\Phi_{\text{f}}$, and
$\Phi_{\text{sf}}$ and dictates the physics of the SU(4)-sector. 
Comparing with the bosonization treatment of the single-band 
Hubbard model, one finds that the ($4k_{\text{F}}$) umklapp 
scattering present in the single-band case is absent here 
while the SU(4)-part (`SU(4)' should be replaced with `spin' in the 
single-band case) contains the marginally irrelevant backscattering 
as in the case of the single band.  
This may seem contradicting since we know the two-band Hubbard model 
will become insulating (i.e. charge-gapped) for 
sufficiently large $U$ ($U \sim 2.8 t$) \cite{assaraf}.    
This paradox is remedied by taking into account the second-order 
($U^{2}$) perturbation coming from the OPE 
${\cal H}_{4k_{\text{F}}}{\cal H}_{4k_{\text{F}}}$.  
In fact, the above OPE yields the contribution like
\[
 -\frac{3U^{2}}{8\pi^{4}a_{0}^{2}}\, \cos (\sqrt{16\pi}\Phi_{\text{c}})  
\]  
which will pin the charge $\Phi_{\text{c}}$-field for 
large enough $U$.  
Therefore, we may fix the value of the $\Phi_{\text{c}}$-field 
and drop ${\cal H}_{\text{c}}$ in the following analysis.  

The final step is to rewrite the SU(4) Hamiltonian 
${\cal H}_{\text{SU(4)}}$ in terms of six Majorana fermions.  
The mapping is provided by the following correspondence:
\begin{equation}
\begin{split}
& \xi_{\text{R/L}}^{2}+i\, \xi_{\text{R/L}}^{1} = 
\frac{\eta_{1}}{\sqrt{\pi a_{0}}}
\exp(-ip \sqrt{4\pi}\varphi_{\text{s,R/L}})  \\
& \chi_{\text{R/L}}^{1}+i\, \chi_{\text{R/L}}^{2} = 
\frac{\eta_{2}}{\sqrt{\pi a_{0}}}
 \exp(+ip \sqrt{4\pi}\varphi_{\text{f,R/L}})  \\
& \chi_{\text{R/L}}^{3}+i\, \xi_{\text{R/L}}^{3} = 
\frac{\eta_{3}}{\sqrt{\pi a_{0}}}
 \exp(-ip \sqrt{4\pi}\varphi_{\text{sf,R/L}}) \; ,
\end{split}
\end{equation}
where the integer $p$ is defined by $p=+1$(L) and $p=-1$(R). 
The Majorana fermions $\eta_{a}$ $(a=1,2,3)$ have been introduced 
to guarantee the anticommutation.  
The following formulas proved by taking OPEs are useful 
in rewriting the SU(4) Hamiltonian (\ref{eqn:SU4_step1}):
\begin{align}
\begin{split}
& \partial_{x}\varphi_{\text{s,R/L}} = i\sqrt{\pi}\, 
\xi_{\text{R/L}}^{1}\xi_{\text{R/L}}^{2} \\
& \partial_{x}\varphi_{\text{f,R/L}} = i\sqrt{\pi}\, 
\chi_{\text{R/L}}^{1}\chi_{\text{R/L}}^{2} \\
& \partial_{x}\varphi_{\text{sf,R/L}} = i\sqrt{\pi}\, 
\xi_{\text{R/L}}^{3}\chi_{\text{R/L}}^{3} 
\end{split}
\label{eqn:boson_to_Majorana1}
\\
\begin{split}
&  \cos(\sqrt{4\pi}\, \Phi_{\text{s}}) 
= i\pi a_{0}(\kappa_{1}+\kappa_{2}) \\
& \cos(\sqrt{4\pi}\, \Phi_{\text{f}}) 
= i\pi a_{0}(\kappa_{4}+\kappa_{5}) \\
& \cos(\sqrt{4\pi}\, \Phi_{\text{sf}})
= i\pi a_{0}(\kappa_{3}+\kappa_{6}) \; .
\end{split}
\label{eqn:boson_to_Majorana2}
\end{align}
The Majorana fermions are normalized so that 
\begin{equation}
\begin{split}
& \VEV{\xi^{a}(z)\xi^{b}(w)}=\VEV{\chi^{a}(z)\chi^{b}(w)}
=\frac{\delta_{a,b}}{2\pi(z-w)} \\
& \VEV{\xi^{a}(z)\chi^{b}(w)}=0 ,
\end{split}
\end{equation}
and the Majorana bilinears $\kappa_{i}$ are defined as 
\begin{equation}
\kappa_{a}\equiv \xi_{\text{R}}^{a}\xi_{\text{L}}^{a} 
\quad , \quad 
\kappa_{a+3} \equiv \chi_{\text{R}}^{a}\chi_{\text{L}}^{a} 
\quad (a=1,2,3) \; .  
\end{equation} 
By using Eqs. (\ref{eqn:boson_to_Majorana1}) and 
(\ref{eqn:boson_to_Majorana2}), 
the first term in Eq. (\ref{eqn:SU4_step1}) can be written as: 
\begin{equation}
\begin{split}
& - \frac{i}{2}\tilde{v}
\sum_{a=1}^{3}\int\! dx 
\left( \xi_{\text{R}}^a \partial_x \xi_\text{R}^a 
- \xi_\text{L}^a \partial_x \xi_\text{L}^a
+ \chi_\text{R}^a \partial_x \chi_\text{R}^a
- \chi_\text{L}^a \partial_x \chi_\text{L}^a \right) \\
& \qquad -U\int\! dx \left(\kappa_{1}\kappa_{2}+\kappa_{4}\kappa_{5}
+\kappa_{3}\kappa_{6} \right)   \; ,
\end{split}
\end{equation}
where we have introduced the renormalized velocity 
$\tilde{v} =(v-U/(2\pi))$.   
Similarly, Eq. (\ref{eqn:boson_to_Majorana2}) enables us to 
recast the second term of ${\cal H}_{\text{SU(4)}}$ as:
\begin{equation}
-\frac{U}{2} \int\!dx \, \left(\sum_{a=1}^{6}\kappa_{a}\right)^{2}
+U\int\!dx \left( \kappa_{1}\kappa_{2}+\kappa_{4}\kappa_{5}
+\kappa_{3}\kappa_{6} \right) \; .
\end{equation}
From these equations, we finally obtain the desired result:
\begin{equation}
\begin{split}
& {\cal H}_{\text{SU(4)}}= \\
& - \frac{i}{2}\tilde{v}
\sum_{a=1}^{3}\int\! dx 
\left( \xi_{\text{R}}^a \partial_x \xi_\text{R}^a 
- \xi_\text{L}^a \partial_x \xi_\text{L}^a
+ \chi_\text{R}^a \partial_x \chi_\text{R}^a
- \chi_\text{L}^a \partial_x \chi_\text{L}^a \right) \\
& \qquad 
-\frac{U}{2} \int\!dx \, \left(\sum_{a=1}^{6}\kappa_{a}\right)^{2} 
\; .
\end{split}
\label{eqn:SU4_step2}
\end{equation}
For repulsive interaction $U>0$, the interaction (the last term) 
is marginally irrelevant and known to yield logarithmic 
corrections\cite{Affleck-G-S-Z-89,majumdar} to physical quantities.  

%%%%%%%%%%%%%%%%%%%%%%%%%%%%%%%%%%%%%%%%%%%%%%%%%%%%%%%%%%%%%%
%%%%%%%%%%%%%%%%%%%%%%%%%%%%%%%%%%%%%%%%%%%%%%%%%%%%%%%%%%%%%%
\section{Derivation of the duality transformation in the continuum 
limit}
%%%%%%%%%%%%%%%%%%%%%%%%%%%%%%%%%%%%%%%%%%%%%%%%%%%%%%%%%%%%%%
\label{sec:derivation_duality}
%%%%%%%%%%%%%%%%%%%%%%%%%%%%%%%%%%%%%%%%%%%%%%%%%%%%%%%%%%%%%%
If the duality transformation (\ref{eqn:duality2}) works 
not only on a lattice but also in a low-energy field theory, 
the uniform part of the generators derived above should obey the {\em same} 
transformation rule (since the duality transformation does 
not change momentum of the system).  
Plugging Eqs. (\ref{eqn:SC_Majorana2}),(\ref{eqn:SU4byMajorana1}) into 
(\ref{eqn:O2_doublet}), we obtain the following 
set of equations for the uniform part of $\bolS_{1}$:
\begin{multline}
\tilde{\xi}^{a}\tilde{\xi}^{b} =\\
\cos^{2}\left(\frac{\theta}{2}\right)\xi^{a}\xi^{b}
+\sin^{2}\left(\frac{\theta}{2}\right)\chi^{a}\chi^{b}
-\frac{1}{2} \sin\theta (\xi^{a}\chi^{b}-\xi^{b}\chi^{a}) \; ,
\label{eqn:duality_eq}
\end{multline}
where we have dropped the indices L/R and
$(a,b)=(2,3),(3,1),(1,2)$.  
Since all three equations are satisfied simultaneously 
for a transformation which is independent both of color ($a=1,2,3$) and 
of the chirality (L/R), we try the following:
\begin{equation}
\begin{pmatrix} \tilde{\xi}^{a}_{\text{L/R}} \\ 
\tilde{\chi}^{a}_{\text{L/R}} \end{pmatrix}
 = 
\begin{pmatrix} 
\cos \phi & -\sin \phi \\ \sin \phi & \cos \phi 
\end{pmatrix} 
\begin{pmatrix} \xi^{a}_{\text{L/R}} \\ 
\chi^{a}_{\text{L/R}} \end{pmatrix}  \; .
\end{equation}
After some algebra, we see that the above equations are satisfied 
if we choose the parameter $\phi$ as 
\[
 \phi = \frac{\theta}{2} \; .
\] 
Moreover, it is not difficult to verify that the above choice 
works for the equations for $\bolS_{2}$ as well:
\begin{multline}
\tilde{\chi}^{a}\tilde{\chi}^{b} = \\
\sin^{2}\left(\frac{\theta}{2}\right)\xi^{a}\xi^{b}
+\cos^{2}\left(\frac{\theta}{2}\right)\chi^{a}\chi^{b}
+\frac{1}{2}\sin\theta (\xi^{a}\chi^{b}-\xi^{b}\chi^{a})
\; .
\end{multline}
Therefore, we may conclude that the following color- and chirality (L/R) 
independent SO(2) transformation $R(\theta)$ generates the `duality' in the 
continuum limit:
\begin{equation}
\begin{split}
\tilde{\xi}_{\text{L/R}}^{a} &= 
\xi_{\text{L/R}}^{a}\cos \frac{\theta}{2} 
- \chi_{\text{L/R}}^{a}\sin \frac{\theta}{2} \; , \\
\tilde{\chi}_{\text{L/R}}^{a} &= 
\xi_{\text{L/R}}^{a}\sin \frac{\theta}{2} 
+ \chi_{\text{L/R}}^{a}\cos \frac{\theta}{2} \; .
\label{eqn:duality_cont}
\end{split}
\end{equation}

This is not the end of the story. 
Besides the uniform part discussed above, the continuum expression 
of the SU(4) generators 
contain two kinds of oscillating parts; one has 
momentum $\pi(=4k_{\text{F}})$ and oscillates in a staggered manner, 
and the others exhibits $\pm \pi/2(=2k_{\text{F}})$ oscillation.  
The argument goes similarly for the $4k_{\text{F}}$-part and we can 
verify that the choice (\ref{eqn:duality_cont}) works.   

The easiest way to write down the transformation of 
the $2k_{\text{F}}$-part would be to use the Dirac fermions instead of 
Majorana fermions $\bolxi$ and $\bolchi$ and look for the correct 
transformation rule for the Dirac quartet.    
Again, as in Eq. (\ref{eqn:duality_eq}), we require that 
the $2k_{\text{F}}$-part of the {\em transformed} SU(4) generators 
\[
 \hat{\tilde{G}} = \tilde{L}^{\dagger}X^{a} \tilde{R} \;
(2k_{\text{F}}) \quad \text{ or } \quad  
\tilde{R}^{\dagger}X^{a} \tilde{L} \; (-2k_{\text{F}})
\]
be correctly reproduced by linear combinations of the original 
generators as in Eqs. (\ref{eqn:transf1}) and (\ref{eqn:transf2}). 
The answer is quite simple and is given as follows:
\begin{equation}
\tilde{R} = U(\theta)R \quad , \quad 
\tilde{L} = U(\theta)L \; ,
\end{equation}
where 
\[
\begin{split}
& R = (\Psi_{\text{R},1\uparrow},
\Psi_{\text{R},1\downarrow},
\Psi_{\text{R},2\uparrow},
\Psi_{\text{R},2\downarrow}) \; , \\
& L = (\Psi_{\text{L},1\uparrow},
\Psi_{\text{L},1\downarrow},
\Psi_{\text{L},2\uparrow},
\Psi_{\text{L},2\downarrow}) \; 
\end{split}
\]
and $U(\theta)$ is given by Eq. (\ref{eqn:duality2}).  
%%%%%%%%%%%%%%%%%%%%%%%%%%%%%%%%%%%%%%%%%%%%%%%%%%%%%%%%%
% Non-zero OPE coefficients
%%%%%%%%%%%%%%%%%%%%%%%%%%%%%%%%%%%%%%%%%%%%%%%%%%%%%%%%%
\section{Non-zero OPE coefficients}
\label{sec:non-zero_OPE}
%%%%%%%%%%%%%%%%%%%%%%%%%%%%%%%%%%%%%%%%%%%%%%%%%%%%%%%%%
The operator-product expansion of interaction 
operators is defined as:
\[
 {\cal H}_{i}(z,\bar{z}){\cal H}_{j}(w,\bar{w}) 
\sim \frac{C_{ij}^{k}}{|z-w|^{2}}{\cal H}_{k}(w,\bar{w}) 
\quad (i,j=1,2,3,4,7) \; .
\]
Since ${\cal H}_{i}$ are written in terms of free fermions 
$\xi^{a}$ and $\chi^{a}$, it is straightforward to 
compute the right-hand side.  
The only non-zero OPE coefficients are listed below:
\begin{align*}
&C_{11}^{1}=C_{22}^{1}=C_{44}^{1}=-\frac{1}{\pi^{2}}\, , \; 
C_{33}^{1}= -\frac{5}{\pi^{2}}\, , \\
&C_{12}^{2}=C_{21}^{2}=-\frac{1}{\pi^{2}} \, , \;
C_{34}^{2}=C_{43}^{2}=-\frac{3}{\pi^{2}} \\
&C_{47}^{2}=C_{74}^{2}=-\frac{1}{2\pi^{2}} \, , \; 
C_{13}^{3}=C_{31}^{3}=-\frac{3}{\pi^{2}} \, , \\
&C_{24}^{3}=C_{42}^{3}=-\frac{1}{\pi^{2}} \, ,  \;
C_{14}^{4}=C_{41}^{4}=-\frac{1}{\pi^{2}} \, \\
&C_{23}^{4}=C_{32}^{4}=-\frac{3}{\pi^{2}} \, , \;
C_{27}^{4}=C_{72}^{4}=-\frac{1}{2\pi^{2}} \, , \\
&C_{13}^{7}=C_{31}^{7}=\frac{8}{\pi^{2}} \, ,\;
C_{24}^{7}=C_{42}^{7}=-\frac{8}{\pi^{2}} \, .
\end{align*}

%%%%%%%%%%%%%%%%%%%%%%%%%%%%%%%%%%%%%%%%%%%%%%%%%%%%%%%%%
% Ground-state degeneracy
%%%%%%%%%%%%%%%%%%%%%%%%%%%%%%%%%%%%%%%%%%%%%%%%%%%%%%%%%
\section{Ground-state degeneracy}
\label{sec:GS_degeneracy}
%%%%%%%%%%%%%%%%%%%%%%%%%%%%%%%%%%%%%%%%%%%%%%%%%%%%%%%%%
In this appendix, we describe how to obtain ground-state 
degeneracy by inspecting the classical ground states
within the bosonization approach.  
As has been mentioned before, bosonic expressions of 
the physical operators have a kind of `gauge redundancy' 
and the ground states should be counted modulo the gauge 
redundancy.  

%%%%
To this end, we have first to identify physical operators 
for which we define `gauge transformations'.  
One obvious choice may be U(3) Dirac fermions in Eq. (\ref{eqn:Diracfer}). 
However, the expansion of physical (lattice) operators 
in terms of the continuum ones contains the $2k_{\text{F}}$-terms 
which cannot be expressed by these Dirac fermions.  
This obscures how to impose the gauge-equivalence for 
three bosons $\varphi$, $\varphi_{\text{s}}$, and $\varphi_{\text{f}}$.  
Instead, we start from the {\em four} Dirac fermions used 
in Appendix \ref{sec:deriv_SU4} to 
obtain the effective Hamiltonian of the SU(4)-model. 
Throughout this section, we shall use the same notations 
as in Appendix \ref{sec:deriv_SU4}.
%%%%%%%%%%%%%%%%%%%%%%%%%%%%%%%%%%%%%%%%%%%%%%%%%%%%%%%%%
By using the SU(4) bosons $\Phi_{\text{s}}$, $\Phi_{\text{f}}$, 
and $\Phi_{\text{sf}}$ (see Eq. (\ref{eqn:SU4_bosons})), the interactions 
${\cal V}_{A}=-\lambda_{A} \left({\cal O}^{\pi}_{A}\right)^{2}$ 
($A=$Q, SD, SC, and RQ) 
read: 
\begin{subequations}
%%%%%%%%%%%%%%%%%%%%%%%%%%%%%%%%
\begin{equation}
\begin{split}
{\cal V}_{\text{Q}} &= -\frac{2\lambda_{\text{Q}}}{\pi^{2}a_{0}^{2}}
\biggl\{
\cos(\sqrt{4\pi}\Phi_{\text{s}})
 \cos(\sqrt{4\pi}\Phi_{\text{f}}) \\
& \phantom{-\frac{2\lambda_{\text{Q}}}{\pi^{2}}
\biggl(} \qquad
+ \cos(\sqrt{4\pi}\Phi_{\text{s}}) 
 \cos(\sqrt{4\pi}\Phi_{\text{sf}})   \\
& 
\phantom{-\frac{2\lambda_{\text{Q}}}{\pi^{2}}
\biggl(} \qquad\qquad 
+ \cos(\sqrt{4\pi}\Phi_{\text{f}})
 \cos(\sqrt{4\pi}\Phi_{\text{sf}}) 
\biggr\} 
\end{split}
\end{equation}
%%%%%%%%%%%%%%%%%%%%%
\begin{equation}
\begin{split}
{\cal V}_{\text{SD}} =& -\frac{2\lambda_{\text{SD}}}{\pi^{2}a_{0}^{2}}
\biggl\{
-\cos(\sqrt{4\pi}\Phi_{\text{s}})
 \cos(\sqrt{4\pi}\Phi_{\text{f}}) \\
& \phantom{-\frac{2\lambda_{\text{Q}}}{\pi^{2}}
\biggl(} \qquad
+ \cos(\sqrt{4\pi}\Phi_{\text{s}})
\cos(\sqrt{4\pi}\Theta_{\text{sf}})  \\
& \phantom{-\frac{2\lambda_{\text{Q}}}{\pi^{2}}
\biggl(} \qquad\quad
- \cos(\sqrt{4\pi}\Phi_{\text{f}})
 \cos(\sqrt{4\pi}\Theta_{\text{sf}})   
\biggr\} 
\end{split}
\end{equation}
%%%%%%%%%%%%%%
\begin{equation}
\begin{split}
{\cal V}_{\text{SC}} =& -\frac{2\lambda_{\text{SC}}}{\pi^{2}a_{0}^{2}}
\biggl\{
\sin\left[\sqrt{\pi}\left(\Phi_{\text{s}}+\Theta_{\text{s}}
+\Phi_{\text{f}}-\Theta_{\text{f}}\right)\right] \\
&\qquad \qquad \qquad 
\times \sin\left[\sqrt{\pi}\left(\Phi_{\text{s}}-\Theta_{\text{s}}
+\Phi_{\text{f}}+\Theta_{\text{f}}\right)\right]  \\
& + \sigma^{z}
\sin\left[\sqrt{\pi}\left(\Phi_{\text{s}}+\Theta_{\text{s}}
+\Phi_{\text{f}}-\Theta_{\text{f}}\right)\right]
 \sin(\sqrt{4\pi}\Theta_{\text{sf}}) \\
& + \sigma^{z}
 \sin(\left[\sqrt{\pi}\left(\Phi_{\text{s}}-\Theta_{\text{s}}
+\Phi_{\text{f}}+\Theta_{\text{f}}\right)\right]
 \sin(\sqrt{4\pi}\Theta_{\text{sf}})
\biggr\}
\end{split}
\end{equation}
%%%%%%%%%%%%%%
\begin{equation}
\begin{split}
{\cal V}_{\text{RQ}} =& -\frac{2\lambda_{\text{RQ}}}{\pi^{2}a_{0}^{2}}
\biggl\{
- \sin\left[\sqrt{\pi}\left(\Phi_{\text{s}}+\Theta_{\text{s}}
+\Phi_{\text{f}}-\Theta_{\text{f}}\right)\right] \\
& \qquad \qquad \qquad 
\times \sin\left[\sqrt{\pi}\left(\Phi_{\text{s}}-\Theta_{\text{s}}
+\Phi_{\text{f}}+\Theta_{\text{f}}\right)\right]  \\
& -\sigma^{z}
 \sin(\left[\sqrt{\pi}\left(\Phi_{\text{s}}+\Theta_{\text{s}}
+\Phi_{\text{f}}-\Theta_{\text{f}}\right)\right]
 \sin(\sqrt{4\pi}\Phi_{\text{sf}})  \\
& +\sigma^{z}
 \sin\left[\sqrt{\pi}\left(\Phi_{\text{s}}-\Theta_{\text{s}}
+\Phi_{\text{f}}+\Theta_{\text{f}}\right)\right]
 \sin(\sqrt{4\pi}\Phi_{\text{sf}})
\biggr\}  \; .
\end{split}
\end{equation}
These expressions can be readily obtained from Eqs. (\ref{orderparadfer}).
The $2\times 2$ Hamiltonian is diagonal and we can freely choose one of 
the eigenvalues (say, +1) of the Pauli matrix $\sigma^{z}$ throughout 
the calculation.  
\end{subequations}
%%%%%%%%%%%%%%%%%%%%%%%%%%%%%%%%%%%%%%%%%%%%
In principle, our system may be described only by these three bosons. 
However, the relationship between three bosons 
$\Phi_{\text{s}}$($\Theta_{\text{s}}$), 
$\Phi_{\text{f}}$($\Theta_{\text{f}}$),  
$\Phi_{\text{sf}}$($\Theta_{\text{sf}}$) and 
physical operators is not so obvious and it is convenient to 
go back to the original (2-band) Hubbard model 
(see Appendix \ref{sec:deriv_SU4}).  For this reason, we recover the charge 
boson $\Phi_{\text{c}}$($\Theta_{\text{c}}$) and 
add the following umklapp term to find the semiclassical 
vacua of our problem:
\begin{equation}
-\lambda_{\text{umklapp}} \; \cos \left(
\sqrt{16\pi}\, \Phi_{\text{c}}\right) \; .
\end{equation}

The procedure is as follows.  First of all, 
the gauge redundancy\cite{Lin-B-F-98} of the original four Dirac fermions 
(see Appendix \ref{sec:deriv_SU4} for the definition of these fermions)
\begin{align*}
& \Psi_{a,\text{L}}= \frac{\kappa_{a}}{\sqrt{2\pi a_{0}}}
\exp\left(-i \sqrt{4\pi}\, \Phi_{a,\text{L}}\right) \\
& \Psi_{a,\text{R}}= \frac{\kappa_{a}}{\sqrt{2\pi a_{0}}}
 \exp\left(+i \sqrt{4\pi}\, \Phi_{a,\text{R}}\right) \\
& \qquad \qquad (a=1,\ldots,4)
\end{align*}
reads:
\begin{equation}
\Phi_{a,\text{L/R}} \mapsto 
\Phi_{a,\text{L/R}} + \sqrt{\pi}\; N_{a,\text{L/R}}
 \qquad (N_{a,\text{L/R}}\in \mathbb{Z} )\; .
\end{equation}
That is, physical quantities should be unchanged even if we make 
the above shift.   Therefore, the semiclassical ground states 
which are `gauge-equivalent' by this shift should be treated as one 
and the true ground states are the equivalence classes 
of this gauge transformation.  

For the two phases Q and SD, a straightforward semiclassical 
analysis (for either $\Phi$ or $\Theta$) is applicable 
and we can proceed in essentially the same 
manner as in Ref.\onlinecite{Lin-B-F-98}.  
For the cases of SC and RQ phases, however, the situation 
is slightly tricky. 
Since both $\Phi$ and 
$\Theta$ appear in a single sine/cosine interactions, 
we should introduce new fields 
\begin{equation}
\begin{split}
& \Phi^{\prime}_{\text{s}}=\frac{1}{\sqrt{2}}\left(
\Phi_{\text{s}}+\Phi_{\text{f}} \right) \, , \, 
\Phi^{\prime}_{\text{f}}=\frac{1}{\sqrt{2}}\left(
\Phi_{\text{s}}-\Phi_{\text{f}} \right) \\
& \Theta^{\prime}_{\text{s}}=\frac{1}{\sqrt{2}}\left(
\Theta_{\text{s}}+\Theta_{\text{f}} \right) \, , \, 
\Theta^{\prime}_{\text{f}}=\frac{1}{\sqrt{2}}\left(
\Theta_{\text{s}}-\Theta_{\text{f}} \right) 
\end{split}
\end{equation}
before applying a semiclassical argument.  
Then we can readily find the semiclassical ground states of 
${\cal V}_{\text{SC,RQ}}$ to apply the method 
of Ref.\onlinecite{Lin-B-F-98}.   

%%%%%%%%%%%%%%%%%%%%%%%%%%%%%%%%%%%%%%%%%%%%%%%%%%%%%%%%%%%%%%
%     BIBLIOGRAPHY 
%%%%%%%%%%%%%%%%%%%%%%%%%%%%%%%%%%%%%%%%%%%%%%%%%%%%%%%%%%%%%%

%%%%%%%%%%%%%%%%%%%%%%%%%%%%%%%%%%%%%%%%%%%%%%%%%%%%%%%%%

\begin{thebibliography}{101}
\bibitem{comment1}
To be precise, since the scalar can be constructed by taking a trace
of $Q_{ab}$, we have to subtract
$2 \delta_{ab} \bolS_{1,r}{\cdot}\bolS_{2,r}/3 $
from them in order to have correct quadrupolar order parameters.  
Nevertheless, we use a simplified expression $Q_{ab}$ hereafter.
\bibitem{comment3}
The model with $B(=D)=0$ and $G\gg 1$ may have a pathological
ground state with a macroscopic number of degenerate ground states.
\bibitem{commentunivconv}
This reminds us of the correspondence
between SU(2) and SO(3).
More precisely, as Lie groups, SO(6) is not 
equivalent to SU(4): SO(6) $\sim$ SU(4)$/\mathbb{Z}_{2}$ .
\bibitem{comment4}
One important exception would be a class of self-dual models which
is solvable by the so-called matrix-product ansatz\cite{Kolezhuk-M-98}.
For these models, we can explicitly show that we have a 1-parameter
family of the (exact) ground states and that the above doublet can
take {\em arbitrary finite} values characterized by a single parameter.
\bibitem{comment5}
By an explicit RG calculation,
we can show that ${\cal H}_{5}$ is not generated in the course
of renormalization. Therefore, we may safely suppress $g_{5}$.
\end{thebibliography}
\end{document}